\documentclass[11pt]{report}
\usepackage{upennstyle} 

\hypersetup{
    colorlinks=true,
    linkcolor=blue,
    filecolor=magenta,      
    urlcolor=cyan,
    citecolor=blue,
    pdftitle={Heisenberg Limit beyond Quantum Fisher Information},
    }

\usepackage{dsfont}
\usepackage{revsymb}

\usepackage[T1]{fontenc}
\usepackage{slashed}
\usepackage[makeroom]{cancel}
\usepackage{xfrac}
\usepackage{xcolor}
\usepackage{mathrsfs}
\usepackage{mathtools}
\usepackage[width=.9\textwidth]{caption}

\renewcommand{\t}[1]{\textrm{#1}}

\definecolor{brightmaroon}{rgb}{0.76, 0.13, 0.28}
\definecolor{carmine}{rgb}{0.59, 0.0, 0.09}

\newcommand{\ket}[1]{|#1\rangle}
\newcommand{\kett}[1]{|#1 \rangle \! \rangle}
\newcommand{\bbra}[1]{\langle \! \langle #1|}
\newcommand{\bra}[1]{\langle#1|}
\newcommand{\braket}[1]{\langle #1 \rangle}

\usepackage{dsfont}
\newcommand{\up}{\uparrow}
\newcommand{\down}{\downarrow}
\newcommand{\var}{\theta}
\newcommand{\bvar}{{\boldsymbol{\var}}}
\newcommand{\cost}{\mathcal C}

\newcommand{\M}{M}
\newcommand{\tr}{\t{Tr}}

\newcommand{\bx}{{\boldsymbol{x}}}
\newcommand{\ml}{{\tilde\var_{\t{ML}}}}
\newcommand{\mlk}{{\tilde\var_{\t{ML}}}}
\newcommand{\E}{\mathds{E}}

\newcommand{\mm}{\t{minimax}}
\newcommand{\bay}{\t{Bayesian}}
\newcommand{\vm}{\widehat{\Delta^2\tilde\var}}
\newcommand{\vb}{\overline{\Delta^2\tilde\var}}
\newcommand{\cm}{\widehat{\cost}}
\newcommand{\cb}{\overline{\cost}}
\newcommand{\bX}{{\bold{X}}}
\newcommand{\bL}{{\bold{L}}}

\newcommand{\mH}{\mathcal H}
\newcommand{\C}{C}
\newcommand{\bF}{{\bold{F}}}
\newcommand{\ptr}{\t{tr}}

\newcommand{\HH}{H}
\newcommand{\G}{\Lambda}
\newcommand{\oG}{\Lambda}
\newcommand{\bG}{\bold{\G}}
\newcommand{\ch}{{\mathcal E}}
\newcommand{\figref}[1]{\hyperref[#1]{Fig.~\ref{#1}}}
\renewcommand{\eqref}[1]{\hyperref[#1]{Eq.~(\ref{#1})}}
\newcommand{\chref}[1]{\hyperref[#1]{Ch.~\ref{#1}}}
\newcommand{\secref}[1]{\hyperref[#1]{Sec.~\ref{#1}}}
\newcommand{\appref}[1]{\hyperref[#1]{App.~\ref{#1}}}
\newcommand{\NN}{\bar n}
\newcommand{\h}{\hat f}
\newcommand{\est}{\t{est}}
\newcommand{\lu}{\t{l.u.}}
\newcommand{\lin}{\t{Lin}}
\newcommand{\EV}[1]{{V_{#1}^{\kett{\Omega}}}}

\newcommand{\sepcr}{\Delta^2\tilde\bvar^{\t{CR}}_{\t{SEP}}}
\newcommand{\jntcr}{\Delta^2\tilde\bvar^{\t{CR}}_{\t{JNT}}}
\newcommand{\sepmm}{\widehat{\Delta^2\tilde\bvar}_{\t{SEP}}}
\newcommand{\jntmm}{\widehat{\Delta^2\tilde\bvar}_{\t{JNT}}}
\newcommand{\sepcrp}{\Delta^2\tilde\bvar^{\t{CR}}_{\t{SEP+}}}
\newcommand{\sepmmp}{\widehat{\Delta^2\tilde\bvar}_{\t{SEP+}}}

\newcommand{\subsectionnull}[1]{\section*{#1}
\addcontentsline{toc}{subsection}{\protect\numberline{}#1}}

\usepackage[backend=biber,style=authoryear,maxbibnames=20,giveninits=true,uniquename=init,date=year,doi=false,isbn=false,url=false,uniquelist=false,maxcitenames=2]{biblatex}

\DeclareFieldFormat{date}{\bfseries{#1}} 
\DeclareFieldFormat{author}{\bfseries{#1}}

\renewbibmacro{in:}{}
\AtEveryBibitem{%
	\clearfield{issue}%
	\clearfield{number}}

\DeclareFieldFormat[article,inbook,inproceedings,online]{title}{\mkbibitalic{,,#1''}}
\DeclareFieldFormat[book]{title}{\mkbibitalic{#1}}
\DeclareFieldFormat[inbook]{booktitle}{\mkbibitalic{\href{http://dx.doi.org/\thefield{doi}}{#1}}}
\DeclareFieldFormat[article]{journaltitle}{\href{http://dx.doi.org/\thefield{doi}}{#1}}
\DeclareFieldFormat[article]{volume}{\mkbibbold{\href{http://dx.doi.org/\thefield{doi}}{#1}}}
\DeclareFieldFormat[article]{pages}{\href{http://dx.doi.org/\thefield{doi}}{\mkfirstpage[{\mkpageprefix[bookpagination]}]{#1}}}

\DeclareFieldFormat[incollection]{booktitle}{\mkbibitalic{\href{http://dx.doi.org/\thefield{doi}}{#1}}}
\DeclareFieldFormat[incollection]{pages}{\href{http://dx.doi.org/\thefield{doi}}{\mkfirstpage[{\mkpageprefix[bookpagination]}]{#1}}}
\DeclareFieldFormat[incollection]{publisher}{\href{http://dx.doi.org/\thefield{doi}}{#1}}
\DeclareFieldFormat[incollection]{title}{\mkbibitalic{#1}}

\DeclareCiteCommand{\parencite}
{\usebibmacro{prenote}}
{\usebibmacro{citeindex}%
	\printtext[bibhyperref]{\usebibmacro{cite}}}
{\multicitedelim}
{\usebibmacro{postnote}}

\DeclareCiteCommand*{\parencite}
{\usebibmacro{prenote}}
{\usebibmacro{citeindex}%
	\printtext[bibhyperref]{\usebibmacro{citeyear}}}
{\multicitedelim}
{\usebibmacro{postnote}}

\DeclareCiteCommand{\parencite}[\mkbibparens]
{\usebibmacro{prenote}}
{\usebibmacro{citeindex}%
	\printtext[bibhyperref]{\usebibmacro{cite}}}
{\multicitedelim}
{\usebibmacro{postnote}}

\DeclareCiteCommand*{\parencite}[\mkbibparens]
{\usebibmacro{prenote}}
{\usebibmacro{citeindex}%
	\printtext[bibhyperref]{\usebibmacro{citeyear}}}
{\multicitedelim}
{\usebibmacro{postnote}}

\DeclareCiteCommand{\footcite}[\mkbibfootnote]
{\usebibmacro{prenote}}
{\usebibmacro{citeindex}%
	\printtext[bibhyperref]{ \usebibmacro{cite}}}
{\multicitedelim}
{\usebibmacro{postnote}}

\DeclareCiteCommand{\footcitetext}[\mkbibfootnotetext]
{\usebibmacro{prenote}}
{\usebibmacro{citeindex}%
	\printtext[bibhyperref]{\usebibmacro{cite}}}
{\multicitedelim}
{\usebibmacro{postnote}}

\DeclareCiteCommand{\textcite}
{\boolfalse{cbx:parens}}
{\usebibmacro{citeindex}%
	\printtext[bibhyperref]{\usebibmacro{textcite}}}
{\ifbool{cbx:parens}
	{\bibcloseparen\global\boolfalse{cbx:parens}}
	{}%
	\multicitedelim}
{\usebibmacro{textcite:postnote}}

\makeatletter

\newrobustcmd*{\parentexttrack}[1]{%
	\begingroup
	\blx@blxinit
	\blx@setsfcodes
	\blx@bibopenparen#1\blx@bibcloseparen
	\endgroup}

\AtEveryCite{%
	\let\bibcloseparen=\bibclosebracket}

\makeatother

\title{Heisenberg Limit beyond Quantum Fisher Information}
\author{Wojciech G\'orecki}
\date{2022} 
\supervisor{dr hab. Rafał Demkowicz-Dobrzański}
\committee{prof. dr hab. Dariusz Chruściński, Nicolaus Copernicus University in Toruń}
\committee{prof. dr hab. Michał Horodecki, University of Gdańsk}
\committee{dr hab. Emilia Witkowska, Polish Academy of Sciences} 

\authorlegal{Wojciech G\'orecki} 

\dedication{Write your dedication text in italics here.} 

\acknowledgement{

First and foremost, I would like to sincerely thank my promoter Rafal Demkowicz-Dobrzanski for his unlimited support throughout my doctoral studies, combined at the same time with leaving great freedom for development and strong motivation to achieve results.

My sincere thanks to Dominic Berry and Howard Wiseman, who, long before me, began to address the Heisenberg limit beyond quantum Fisher information, whose publications inspired me and in collaboration with whom I wrote the first of the articles that form the basis of this thesis. I would also like to thank Sisi Zhou and Liang Jiang, with whom I also had the pleasure of working in the early years of my Ph.D. on an independent project. Contact with these four at such an outstanding level so early in my career made a huge impression on me and influenced the quality of my future work.

I would also like to thank Stanislaw Kurdziałek, Francesco Albarelli, and Lorenzo Maccone, with whom I worked directly in Warsaw and Pavia, for all the substantive discussions and a great atmosphere of cooperation, which fueled each other.

I thank Jacek Krajczok, Tomasz Cheda, Albert Rico, and Paweł Kurzyński for all the conversations at the intersection of quantum information, algebra, and statistics, which allowed me to become convinced that the field of my research and the knowledge gained during my Ph.D. is not as hermetic as it might seem.

I also want to thank everyone who supported me early in my scientific career and helped me find the direction I want to go in theoretical physics. In particular, I would like to thank my high school teacher, Elżbieta Zawistowska, as well as the supervisors of my internships during my early studies, Kazimierz Rzążewski and Krzysztof Pawłowski.

I thank the reviewers Dariusz Chruściński, Michał Horodecki, and Emilia Witkowska for their useful comments on this thesis, which allowed me to improve its clarity and accessibility.

I would also like to thank everyone who has supported me in my personal life over the years.

During my doctorate, I was financially supported by the National Science Center (Poland) grant No. 2016/22/E/ST2/00559 and the National Science
Center (Poland) Grant No. 2020/37/B/ST2/02134. I was also supported by the Foundation for Polish Science (FNP) via the START scholarship.

} 

\abstract{The Heisenberg limit provides a fundamental bound on the achievable estimation precision with a limited number of $N$ resources used (e.g., atoms, photons, etc.). Using entangled quantum states makes it possible to scale the precision with $N$ better than when resources would be used independently. Consequently, the optimal use of all resources involves accumulating them in a single execution of the experiment. Unfortunately, that implies that the most common theoretical tool used to analyze metrological protocols - quantum Fisher information (QFI) -- does not allow for a reliable description of this problem, as it becomes operationally meaningful only with multiple repetitions of the experiment.

In this thesis, using the formalism of Bayesian estimation and the minimax estimator, I derive asymptotically saturable bounds on the precision of the estimation for the case of noiseless unitary evolution. For the case where the number of resources $N$ is strictly constrained, I show that the final measurement uncertainty is $\pi$ times larger than would be implied by a naive use of QFI. I also analyze the case where a constraint is imposed only on the average amount of resources, the exact value of which may fluctuate (in which case QFI does not provide any universal bound for precision). In both cases, I study the asymptotic saturability and the rate of convergence of these bounds.

In the following part, I analyze the problem of the Heisenberg limit when multiple parameters are measured simultaneously on the same physical system. In particular, I investigate the existence of a gain from measuring all parameters simultaneously compared to distributing the same amount of resources to measure them independently. Using two examples -- the measurement of multiple phase shifts in a multi-arm interferometer and the measurement of three magnetic field components -- I show the existence of qualitative differences between the results obtained using Bayesian estimation/minimax estimator and those resulting from the use of QFI. I also derive a lower bound on the achievable precision of the measurement in the general case (not always saturable).
} 
\publications{

\section*{Publications this thesis is based on:}

\textbf{W. Górecki}, R Demkowicz-Dobrzański,
\textit{Multiparameter quantum metrology in the Heisenberg limit regime: Many-repetition scenario versus full optimization},
\href{https://journals.aps.org/pra/abstract/10.1103/PhysRevA.106.012424}{Physical Review A 106, 012424 (2022)}

\textbf{W. Górecki}, R. Demkowicz-Dobrzański,
\textit{Multiple-Phase Quantum Interferometry: Real and Apparent Gains of Measuring All the Phases Simultaneously},
\href{https://journals.aps.org/prl/abstract/10.1103/PhysRevLett.128.040504}{Physical Review Letters 128, 040504 (2022)}

\textbf{W. Górecki}, R. Demkowicz-Dobrzański, H.M. Wiseman, D.W. Berry,
\textit{$\pi$-Corrected Heisenberg Limit},
\href{https://journals.aps.org/prl/abstract/10.1103/PhysRevLett.124.030501}{Physical Review Letters 124, 030501 (2020)}

\vspace{1cm}
\section*{Publications related to quantum metrology cited in this thesis:}

S. Kurdziałek, \textbf{W. Górecki}, F. Albarelli, R. Demkowicz-Dobrzański,
\textit{Using adaptiveness and causal superpositions against noise in quantum metrology},
\href{https://arxiv.org/abs/2212.08106}{arXiv:2212.08106 (2022)}

\textbf{W. Górecki}, A. Riccardi, L. Maccone,
\textit{Quantum metrology of noisy spreading channels},
\href{https://journals.aps.org/prl/abstract/10.1103/PhysRevLett.129.240503}{Physical Review Letters 129, 240503 (2022)}

R. Demkowicz-Dobrzański, \textbf{W. Górecki}, M. Guţă,
\textit{Multi-parameter estimation beyond quantum Fisher information},
\href{https://iopscience.iop.org/article/10.1088/1751-8121/ab8ef3}{Journal of Physics A: Mathematical and Theoretical 53, 363001 (2020)}
	
\textbf{W. Górecki}, S. Zhou, L. Jiang, R. Demkowicz-Dobrzański,
\textit{Optimal probes and error-correction schemes in multi-parameter quantum metrology},
\href{https://quantum-journal.org/papers/q-2020-07-02-288/}{Quantum 4, 288 (2020)}

\vspace{1cm}
\section*{Other publications:}

J. Kopyciński, M. Łebek, \textbf{W. Górecki}, K. Pawłowski,
\textit{Ultrawide dark solitons and droplet-soliton coexistence in a dipolar Bose gas with strong contact interactions},
\href{https://journals.aps.org/prl/abstract/10.1103/PhysRevLett.130.043401}{Physical Review Letters 130, 043401 (2023)}

J. Kopyciński, M. Łebek, M. Marciniak, R. Ołdziejewski, \textbf{W. Górecki}, K. Pawłowski,
\textit{Beyond Gross-Pitaevskii equation for 1D gas: quasiparticles and solitons},
\href{https://www.scipost.org/SciPostPhys.12.1.023}{SciPost Physics 12, 023 (2022)}

W. Golletz, \textbf{W. Górecki}, R. Ołdziejewski, K. Pawłowski,
\textit{Dark solitons revealed in Lieb-Liniger eigenstates},
\href{https://journals.aps.org/prresearch/abstract/10.1103/PhysRevResearch.2.033368}{Physical Review Research 2, 033368 (2020)}

R. Ołdziejewski, \textbf{W. Górecki}, K. Pawłowski, K. Rzążewski,
\textit{Strongly correlated quantum droplets in quasi-1D dipolar Bose gas},
\href{https://journals.aps.org/prl/abstract/10.1103/PhysRevLett.124.090401}{Physical Review Letters 124, 090401 (2020)}

R. Ołdziejewski, \textbf{W. Górecki}, K. Pawłowski, K. Rzążewski,
\textit{Roton in a few-body dipolar system},
\href{https://iopscience.iop.org/article/10.1088/1367-2630/aaf295/meta}{New Journal of Physics 20, 123006 (2018)}

R. Ołdziejewski, \textbf{W. Górecki}, K. Pawłowski, K. Rzążewski,
\textit{Many-body solitonlike states of the bosonic ideal gas},
\href{https://journals.aps.org/pra/abstract/10.1103/PhysRevA.97.063617}{Physical Review A 97, 063617 (2018)}

\textbf{W. Górecki}, K. Rzążewski,
\textit{Electric dipoles vs. magnetic dipoles-for two molecules in a harmonic trap},
\href{https://iopscience.iop.org/article/10.1209/0295-5075/118/66002/meta}{EPL (Europhysics Letters) 118, 66002 (2017)}

\textbf{W. Górecki}, K. Rzążewski,
\textit{Making two dysprosium atoms rotate—Einstein-de Haas effect revisited},
\href{https://iopscience.iop.org/article/10.1209/0295-5075/116/26004/meta}{EPL (Europhysics Letters) 116, 26004 (2016)}

R. Ołdziejewski, \textbf{W. Górecki}, K. Rzążewski,
\textit{Two dipolar atoms in a harmonic trap},
\href{https://iopscience.iop.org/article/10.1209/0295-5075/114/46003/meta}{EPL (Europhysics Letters) 114, 46003 (2016)}
}

\abbreviations{

\begin{center}
\begin{tabular}{ l l }
MSE & mean square error,\\
RMSE & root mean square error,\\
FI & Fisher information,\\
QFI & quantum Fisher information\\
CR & Cram\'er-Rao inequality,\\
$N$ & the total number of resources used to estimate the parameter.\\
$n,k$ & $n$ is number of resources used in single trial, $k$ is number of trials.\\
HS & Heisenberg scaling (MSE $\sim 1/N^2$),\\
POVM & positive operator-valued measure,\\
SLD & symmetric logarithmic derivative,\\
ML estimator & maximum likelihood estimator,\\
$\var$ & parameter to be measured,\\
$\tilde\var(\cdot)$ & estimator,\\
$\HH$ & generator of the evolution of the whole system,\\
$\oG$ & generator of the evolution of a single particle,\\
$\ch(\cdot)$ & single particle quantum channel,\\
$\mH_S$ & single particle Hilbert space,\\
$\mH_A$ & ancillary system Hilbert space,\\  
\end{tabular}
\end{center}

\vfill
All formulas cited from other papers are translated to this notation, with additional footnote comments, where needed.
}

\begin{document}

\maketitle 

\pagenumbering{arabic}
\setcounter{page}{2}


\makeabstract
\makeacknowledgement 

\makepublications

\makeabbreviations

\setcounter{tocdepth}{2}
\tableofcontents
\vfill

{\setstretch{1.8}
\listoffigures
}


\begin{mainf} 
\setcounter{page}{9}

\chpt{Introduction}
\section{Heisenberg limit in quantum metrology}
In classical mechanics, all parameters characterizing a physical system can, in principle, be measured to any accuracy, but, in practice, the imperfections of the measurement apparatus introduce certain fluctuations in the results. Repeated measurement reduces the uncertainty of the estimation of a parameter, such that the mean square error of the estimator (MSE) scales inversely to the number of these repetitions (the so-called shot noise limit).

In contrast to the above, the formalism of quantum mechanics rigorously defines achievable precision, as the act of measurement itself is not deterministic but probabilistic. As a result, the question of the fundamental limit on the achievable precision of the measurement of a given parameter becomes a well-posed problem without the need to specify a particular measurement procedure. Moreover, quantum mechanics opens up the prospect of achieving a better scaling of measurement accuracy with the number of resources used in the experiment (which can be understood as the number of atoms, number of photons, total time, total energy, etc.). Using appropriate entangled quantum states \parencite{mitchell2004super,horodecki2009quantum} (or sequential multiple passes \parencite{higgins2007entanglement}) makes it possible to decrease MSE quadratically with the number of resources. Such scaling has been named after Werner Heisenberg by Holland and Burnett \parencite{holland1993interferometric}, where they referred to number-phase
uncertainty relation in \parencite{heitler1954quantum}.

The issue of the feasibility of reaching the Heisenberg limit is intensively discussed in the current literature \parencite{giovannetti2006quantum,Paris2009,giovannetti2011advances,Toth2014,demkowicz2015optical,Schnabel2016,degen2017quantum,Pezze2018,Pirandola2018,niezgoda2019optimal}. At the same time, in some cases, there are ambiguities related to the interpretation of results obtained by different methods.

\section{Major problems discussed in the thesis}
The existence of Heisenberg scaling implies that when the only constraint is the total number of resources, the optimal strategy is to accumulate them all in a single execution of the experiment. However, the results obtained with most of the theoretical tools commonly used in quantum metrology, such as Fisher's quantum information (QFI) and the related Cram\'er-Rao (CR) inequality, have an unambiguous interpretation only in the limit of many repetitions. As a result, the actual limit on achievable precision is not very well explored, and there are inconsistent (sometimes even contradictory) statements on this topic in the current literature.

The most widely discussed example in the quantum metrology literature is the issue of phase shift estimation in an interferometer. Back in the 1990s, it was noted that when the value of the shift is initially completely unknown (it can take any value in the interval $[0,2\pi[$), the minimum uncertainty, obtainable using $N$ photons, is $\pi$ times larger than would result from naive use of QFI ~\parencite{luis1996,buzek1999,Berry2000}. In 2011, an argument was put forward ~\parencite{hayashi2011} justifying that, in the limit of $N$ going to infinity, 
even if it is known from the beginning that the parameter belongs to a smaller interval, the rate remains equal to $\pi$, independently of the size of this interval. 
In this thesis, I generalize this observation to the case of any unitary evolution generator with a bounded spectrum. Furthermore, I show that the limit on precision is also valid for more advanced adaptive metrology protocols. Finally, I do not restrict myself to the limit of $N$ going to infinity but also discuss the case of finite $N$. These results have already been published in \parencite{Gorecki2020pi}.

A second major issue related to the misunderstandings arising from the misuse of quantum Fisher information in the context of the Heisenberg limit is the precision achievable when only the expectation value of the total energy is bounded. In such a situation, the QFI can be unconstrainedly large. Some authors have postulated the existence of protocols that allow for better scaling than Heisenberg scaling~~\parencite{Anisimov2010quantum,rivas2012sub}; however, these works were criticized rather quickly \parencite{Pezze_2013}. In 2012, it was shown for the general case that the MSE cannot decrease faster than quadratically with the mean energy \parencite{Giovannetti2012beyond}, but the derived bound was not saturable.
In the thesis, I analyze this problem by establishing the saturable bound and investigating the rate of convergence. These results are based on the methodology and observations from my papers \parencite{Gorecki2020pi,gorecki2021multiple}, but they were not published before in explicit form.

Finally, I examine the issue of the Heisenberg limit in the context of multi-parameter metrology. One of the aspects addressed is the comparison of the efficiency of the combined measurement of all parameters simultaneously with the situation when they are measured separately. The analysis of such an issue necessitates a well-defined procedure for resource allocation between the parameters, which has been treated radically differently in various publications ~\parencite{humphreys2013quantum,yousefjani2017estimating} leading to contradictory results regarding the existence of a gain from performing a joint measurement.
This thesis discusses the issue of resource allocation and makes a comprehensive comparison of the results obtained using the two paradigms (QFI and Bayesian estimation theory/minimax approach). The presented results has been published in \parencite{gorecki2021multiple,gorecki2022multiparameter}.

\section{Outline of the thesis}
In Chapter Two, I present, by way of example, the fundamental problems that arise when attempting to analyze the Heisenberg limit using concepts restricted to local estimation, such as QFI or the error propagation formula. I also discuss the state of the art of the precision bound for average energy constraint. I formulate the problem of a quantum channel estimation, and I introduce a distinction between entangled-parallel and adaptive-sequential metrological schemes.

In Chapter Three, I discuss the Fisher quantum information formalism, highlighting its strengths (such as the extensively developed theory concerning metrology in the presence of noise) while emphasizing its limitations in the context of Heisenberg limit analysis.

In Chapter Four, I analyze the operationally achievable Heisenberg limit under constraints on the total amount of resources available in the system or on the average energy. In doing so, I use the formalism of Bayesian estimation and minimax estimator. In this chapter, I present the results from \parencite{Gorecki2020pi,gorecki2021multiple}.

In Chapter Five, I go further to the multiparameter case, introducing basic concepts in both QFI formalism and Bayesian/minimax one. 
I reproduce the proof of optimality of entangled-parallel strategy for the covariant problems~\parencite{chiribella2008memory}, which I will need in the next chapter.

Finally, in Chapter Six, I discuss the Heisenberg limit in multiparameter case, presenting the results from \parencite{gorecki2021multiple,gorecki2022multiparameter}.

\chpt{Quantum metrology and the Heisenberg limit}

"Everything should be made as simple as possible, but not simpler," Albert Einstein was once supposed to have said. Keeping in mind this rule, before introducing broader mathematical formalism, I would like to briefly explain the essence of the main problem I will discuss in this thesis by introducing a simple but expressive example.

\section{Heisenberg limit by an example}
\label{sec:hle}

Consider the problem of estimating the strength of a weak magnetic field oriented in vertical direction $z$, sensed by atoms with internal magnetic moment -- spin $\sfrac{1}{2}$. Since action of the field on the atoms results in rotating them around the $z$ axis via unitary operation $U_\var=e^{-i\var J_z}$ (where $J_z$ is the spin $z$ component\footnote{For simplicity of further formulas, dividing by the constant factor $\hbar$ is incorporated into the definition of $J_z$, in such way, that $J_z$ is dimensionless.}), the problem is equivalent to estimating the angle of the rotation $\var$, see \figref{fig:bals}(a). The aim is to perform the whole procedure so that the mean squared error of the estimator $\Delta^2\tilde\var$ is minimized.

\begin{figure}[h!]
\includegraphics[width=1.\textwidth]{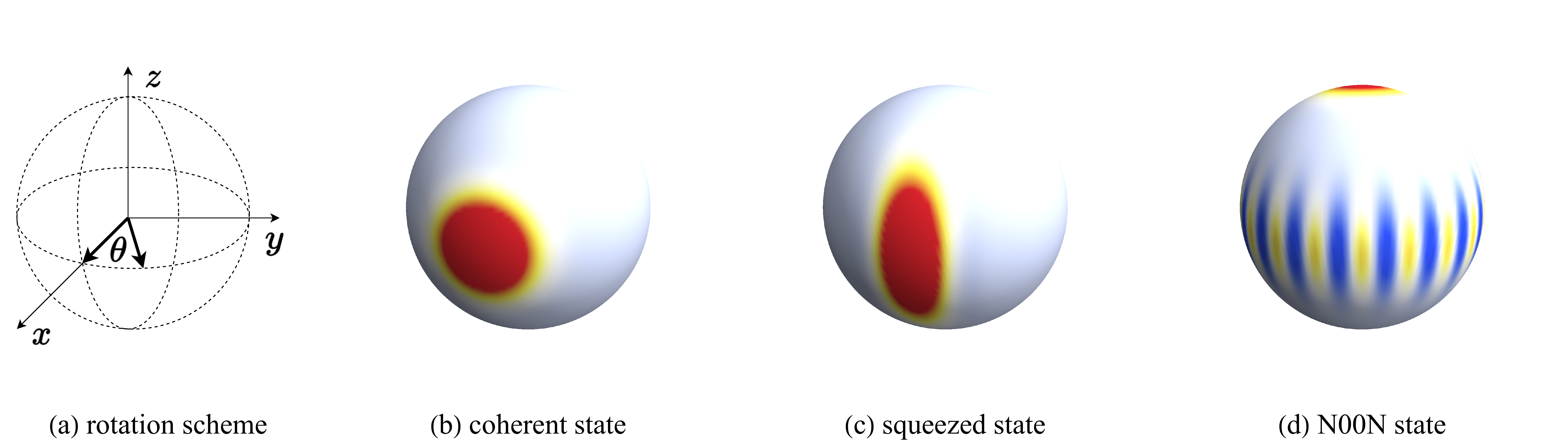}
\caption[Wigner function for the estimation of the spin's rotation]{(a) The vector of internal magnetic moment (spin) is rotated by an angle $\var$ around vertical axis $z$. (b-d) Due to fundamental laws of quantum mechanics, all three components of the spin vector cannot be simultaneously well-defined. Wigner function~\parencite{Jing2019split,davis2021wigner} expressing these uncertainties are plotted for three different quantum states of $N=20$ atoms}\label{fig:bals}
\end{figure}

The simplest and most intuitive strategy is to orient the atom's spin in direction $x$ and, after rotation, measure the $y$ component of the spin $J_y$. As for small rotation the value of $\braket{J_y}\approx \frac{1}{2}\theta$, one can simply estimate the value of parameter as $\tilde\var=2\braket{J_y}$. From the error propagation formula:
 \begin{equation}
 \label{errpro}
    \Delta^2\tilde\theta=\frac{\Delta^2 J_y}{\left|\frac{d\braket{J_y}}{d\theta}\right|^2}.
\end{equation}
As for a single particle, angular momentum operator $J_y$ is simply Pauli matrix multiplied by factor one-half $\frac{1}{2}\sigma_y$,
while the initial state pointing the $x$ direction is given as $\ket{+}=\frac{1}{\sqrt{2}}(\ket{\up}+\ket{\down})$, the variance of such measurement's result is finite and equal $\Delta^2 J_y=\frac{1}{4}$, which leads to $\Delta^2\tilde\theta=1$ (in radians). If one repeats measurement on $N$ independent atoms and then estimates the angle value from the averaged result, the variance of the measurement results will decrease by a factor $\frac{1}{N}$. At the same time, the denominator will remain unchanged, leading at the end to $\Delta^2\tilde\var=\frac{1}{N}$. Such scaling with the amount of resources used is called standard scaling or shot noise limit.

The question is whether one may obtain better scaling by collectively using all $N$ atoms. The first guess would be to coherently orient all atoms to the same direction $x$, such that it may be effectively seen as a single particle of spin $N/2$, see \figref{fig:bals}(b).
Then even a small rotation by an angle $\var$ will lead to $N$ times bigger changes of the value of the collective spin $\braket{J_y}\approx \frac{1}{2}N\theta$.
Unfortunately, it would also increase the fluctuation of the $y$ component itself $\Delta^2 J_y=N$. As a result, these two effects will cancel each other, leading again to the shot noise limit, i.e., the same precision could be obtained by measuring each atom separately and averaging the result.

May this problem be overcome in quantum mechanics by choosing another state for sensing? The laws of quantum mechanics do not allow us to cancel the uncertainties completely. 
The famous Heisenberg uncertainty principle, which, in its most canonical form, establishes the relationship between momentum and position variances $\Delta x\Delta p\geq \frac{\hbar}{2}$ (where $\Delta x:=\sqrt{\Delta^2 x}$ is a root mean squared error (RMSE)), in general bounds the product of variances of any two observables by the expectation value of their commutator~\parencite{schrodinger1930heisenbergschen,robertson1934indeterminancy}:
\begin{equation}
\label{heisen}
    \Delta J_y\Delta J_z\geq \frac{1}{2}|\braket{[J_y,J_z]}|,
\end{equation}
which, in our case, is simply equal to the value of the spin component in $x$ direction $|\braket{J_x}|$. Therefore, even keeping $J_x$ still large, we have some freedom to decrease the variance $\Delta^2 J_y$, by squeezing the state along $y$ axes.
It simultaneously increases the uncertainty in $z$ direction $\Delta^2 J_z$, but this effect does not affect our estimation procedure, see \figref{fig:bals}(b).

Is there any fundamental bound for the precision of estimating angle $\theta$ using $N$ spins $\sfrac{1}{2}$? To answer this question, note that as the evolution of the system is simply rotating around $z$ axis $e^{-i\theta J_z}$, the speed of change of expectation value $\braket{J_y}$ is also determined by the expectation value of the commutator:
\begin{equation}
\label{evol}
    \frac{d\braket{J_y}}{d\var}=i\braket{[J_y,J_z]}.
\end{equation}
Substituting $\frac{1}{i}\frac{d\braket{J_y}}{d\theta}$ for $\braket{[J_y,J_z]}$ in \eqref{heisen} and dividing both sides by $\Delta J_z$ and $\frac{d\braket{J_y}}{d\theta}$ one obtain following bound~\parencite{Mandelstam1991}:
\begin{equation}
\label{bound}
    \Delta^2 \tilde\var\geq\frac{1}{4\Delta^2 J_z}.
\end{equation}
Quite surprisingly, even though in our procedure we decided to obtain information about the rotation angle by measuring the spin component in the $y$-direction, neither $\braket{J_y}$ nor $\Delta^2 J_y$ appears in the final formula. Indeed, the whole reasoning remains valid for any other measurement that extracts information about $\theta$.

It strongly suggests that looking for the state optimal for measuring rotation, we should choose the one which maximizes the variance $\Delta^2 J_z$. It would be the famous Greenberger–Horne–Zeilinger (GHZ) state (in the context of quantum optics also called $\ket{N00N}$ state) -- a quantum superposition of situation, where all $N$ atoms' spins are oriented up and $0$ are oriented down, with the situation, where $0$ is oriented up, and $N$ is oriented down -- $\ket{N00N}\propto\frac{1}{\sqrt{2}}(\ket{\up}^{\otimes N}+e^{iN\var}\ket{\down}^{\otimes N})$, for which $\Delta^2 J_z=N^2$.  We see a significant improvement compared to the initially proposed strategy -- the bound \eqref{bound} indicate that precision could scale quadratically with $N$, which we will call Heisenberg scaling (HS). However, looking at the graphical representation of this state \figref{fig:bals}(d), one may find an interesting property -- in addition to the two strongly outlined areas at opposite poles (corresponding directly to the downward and upward spins independently), one can also see dense fringes at the equator, resulting from the quantum interference of the two states. The fringes are very narrow so that even when rotated by a small angle, the arrangement of the fringes changes significantly, making it possible to observe even a minimal change precisely. On the other hand, the cyclic nature of the distribution of the fringes means that when rotated through an angle $\theta=\frac{2\pi}{N}$ (or multiple thereof), their distribution becomes identical again (which corresponds to the phase factor $e^{iN\var}$ in the formula above). As the results in such a scenario, we can estimate the value of the parameter only up to the term $+\frac{m 2\pi}{N}$ ($m\in\mathbb Z$). It means that to use $\ket{N00N}$ state (potentially leading to $\Delta\tilde\var=\frac{1}{N}$), from the very beginning, we need to know the exact value of the parameter with accuracy $~2\pi/N$.
This fact makes the single usage of $\ket{N00N}$ extremely poor strategy, as it does not significantly increase initial knowledge.

Another way to see how the $\ket{N00N}$ cannot be effective for estimating general unknown angle $\var$ is information theory. Note that even if $\ket{N00N}$ consist $N$ atoms, independently on the exact value of $\var$ it remain into two-dimensional Hilbert space spanned by $\{\frac{1}{\sqrt{2}}(\ket{\up}^{\otimes N}+\ket{\down}^{\otimes N}),\frac{1}{\sqrt{2}}(\ket{\up}^{\otimes N}-\ket{\down}^{\otimes N})\}$. Therefore, from the point of view of the information capacity, it should be treated as a simple qubit and hence can transmit at most a single bit of information. That shows that even if it can record information on a tiny angle change in a readable way, it is impossible to record information on an angle change over a wide range. 

When the true parameter value is guaranteed to lay in a small neighborhood of size $\sim2\pi/N$ of some fixed value $\var_0$, the $\ket{N00N}$ is extremely useful for detecting arbitrary small changes of $\var$. If one is then able to repeat procedure $k$ times to estimate the angle from the averaged result, they will obtain $\Delta^2\tilde\var\approx\frac{1}{kN^2}$ (which would lead to a significant update of the initial knowledge).

However, such a strategy became suboptimal from the view of total resources $k\cdot N$ used, as precision scales only linearly with $k$,
while one still may ask, if the precision $\approx \frac{1}{k^2N^2}$ is obtainable. Consequently, the inequality \eqref{bound} constitutes the valid bound for the estimation precision, but it is not necessarily saturable. Therefore, it does not answer the question about fundamentally optimal precision of estimation in a situation where only the total amount of resources is restricted. 

The problem may be formulated more abstractly in the general form -- for any unitary evolution $U_\var=e^{i\var \HH}$ governed by the hermitian generator of transformation $\HH$ and arbitrary measured observable $A$, one may directly repeat the steps \eqref{heisen}, \eqref{evol}, and \eqref{bound} by replacing $J_z\to H$, $J_y\to A$. As the result, for any peculiar state used for sensing the parameter $\var$, the variance of the estimator will be bounded by the inverse of the variance of the $\HH$ calculated on this state (independently on the choice of $A$):
\begin{equation}
\label{bound2}
    \Delta^2 \tilde\var\geq\frac{1}{4\Delta^2 \HH},
\end{equation}
while the problem of its saturability remains unsolved. This issue cannot be solved solely using an error propagation formula or any other tool based on local formalism, i.e., the one which only considers the first derivatives of the functions around the working point (as Fisher information formalism). This thesis aims to analyze it going beyond this methodology.

\section{General formulation of the channel estimation problem}
\label{sec:gf}

Given unitary quantum channel acting on the system, depending on the unknown parameter $\var$: $\rho_\var=U_\var \rho U^\dagger_\var$.
The aim is to get an information about the parameter $\var$ by preparing a proper input state $\rho$, acting the channel and performing a general measurement -- positive operator-valued measure (POVM) $\{\M_x\}$ ($\forall_x \M_x\geq 0,\,\sum_x \M_x=\openone$) on the output state $\rho_\var$. The probability of obtaining the result $x$ when the parameter value is equal $\var$ is given by the Born rule $p(x|\var)=\tr(\M_x \rho_\var)$. The measurement may be repeated many times, resulting in the sequence of outcomes $\bx=\{x_1,...,x_k\}$. At last, one needs to choose the estimator $\tilde\var(\bx)$
assigning an estimated parameter value to a sequence of results. The measure of the efficiency of the estimator is the MSE:
\begin{equation}
\label{MSE}
    \Delta^2\tilde\var=\sum_\bx p(\bx|\var)(\tilde\var(x)-\var)^2,
\end{equation}
where $p(\bx|\var)=\prod_{i=1}^k p(x_i|\var)$. In the case when the mean value of the estimator indicates the correct value of the parameter (the estimator is unbiased):
\begin{equation}
\label{unbi}
   \sum_\bx p(\bx|\var)\tilde\var(x)=\var,
\end{equation}
the MSE is equal to the variance of the estimator. In such cases, I will use these words interchangeably if it does not lead to ambiguity.

One should notice that for a given measurement and estimator, the value of MSE \eqref{MSE} depends on the exact value of $\var$, so the problem of finding optimal metrology strategy is not well defined until we define a range of parameter value for which the estimator is to work correctly. In principle, we would like the estimator to return actual value for all possible $\var$s. However, this condition in many situations may be too strict or even impossible to satisfy for a finite number of repetitions. Moreover, we often have some approximated knowledge about the parameter value, and we are interested in local estimation around a specific point $\var\approx\var_0$, so it is not needed for the estimator to work well very far from this value. Still, if one demands from estimator to work correctly only in single point $\var_0$, then trivial constant estimator $\forall_\bx\tilde\var(\bx)=\var_0$ would lead to zero cost, while it extracts no information from the measurement. To avoid such pathological constructions, heuristically, we would say we require that the estimator works correctly, at least in a certain immediate neighborhood of $\var_0$. As further analysis will show, different approaches to the mathematical formalization of this condition will lead to different results.

One way is to impose the condition that estimator is \textbf{locally unbiased} around point $\var_0$, i.e., \eqref{unbi}, as well as the equation resulting from taking the derivative of both sides, 
are satisfied at point $\var_0$:
\begin{equation}
\begin{split}
\label{lunbi}
   &\sum_\bx p(\bx|\var_0)\tilde\var(x)=\var_0,\\
   &\sum_\bx \frac{dp(\bx|\var)}{d\var}\Big|_{\var=\var_0}\tilde\var(x)=1.
   \end{split}
\end{equation}
Note that the above condition allows for an exact rederivation of the error propagation formula \eqref{errpro} for the situation, where the parameter $\var$ is estimated with the use of some observable $A=\sum_a a\ket{a}\bra{a}$. Indeed, for measurement $\{\M_a\}=\{\ket{a}\bra{a}\}$ direct minimization of the estimator variance with locally unbiased condition results in $\tilde\var(a)=\var_0+a\left(\frac{d\braket{A}}{d\var}|_{\var=\var_0}\right)^{-1}$ and $\Delta^2\tilde\var=\Delta^2 A/\left(\frac{d\braket{A}}{d\var}|_{\var=\var_0}\right)^2$.

As shown in the discussed example, the analysis performed only with local unbiasedness conditions may not be meaningful when the experiment is performed only once. However, the concept of local unbiasedness is a very efficient tool when the experiment is repeated many times. The relation between the minimal cost obtainable with local unbiasedness condition for single measurement realization vs. the cost with global unbiasedness condition in the limit of many repetitions will be discussed in \chref{fisher}. Still, as argued at the beginning, we are interested in the optimal usage of whole resources, which demands accumulating all of them in a single realization so we will need another theoretical tool.

The problem of full minimization of the MSE of the estimator, with restriction only on the total amount of resources $N$ will be discussed in \chref{ch:baymin} with the usage of two alternative approaches. In the Bayesian approach, one introduces a priori distribution of the value of parameter $p(\var)$ and as the figure of merit, consider the \eqref{MSE} averaged over this probability. Alternatively, in the minimax approach, the finite size set of possible values of parameter $\Theta$ is under consideration, and the figure of merit is \eqref{MSE} maximized over $\var\in \Theta$.

To make the results from \chref{fisher} and \chref{ch:baymin} somehow comparable and to avoid confusion, I introduce a proper notation for the number of used resources. By $N$ I will always understand the total amount of resources used. In the scenario with many experiment's repetitions $k$, I will denote the number of resources used in a single trial as $n$, such that $N=k\cdot n$. In this context, one may think about the results from \chref{fisher} as the ones obtained with optimization over protocols using in total $N$ resources, with additional constraints for the maximal amount of resources used in a single trial $n$ (so the measurement is repeated $k=N/n$ times).

\section{The role of observable and optimality of projective measurements}
\label{sec:projective}

The attentive reader may note that in \secref{sec:hle} I discussed the direct estimation of a parameter from the mean value of the observables, whereas in \secref{sec:gf} we consider independent optimization over both the measurement and the estimator itself. Therefore, one may ask whether the bound \eqref{bound2} is still valid with such an extension.

In the following, I will show that, in fact, for the case where the measurement is not repeated many times (which corresponds to optimal use of resources), optimization over the observable itself is equivalent to optimization over the general measurement and the estimator.

In the literature, this fact has been derived independently in different formalisms (see \parencite{Macieszczak_2014} for the Bayesian approach). Below I want to show that for all introduced approaches (locally unbiased estimators, Bayesian or minimax approach), it may be seen as a direct consequence of a single matrix inequality.

For a given measurement $\{M_x\}$ and estimator $\tilde\var(x)$, the mean value of the estimator for any true value of the parameter $\var$ may be written as the expectation value of the observable $A$ defined as follow:
\begin{equation}
  \braket{\tilde\var}=\sum_x \tr(\rho_\var \M_x)\tilde\var(x)=\tr\Big(\rho_\var\underbrace{\sum_x\M_x\tilde\var(x)}_{A}\Big).
\end{equation}
Now I will argue that such a procedure may only be improved if one measures the observable directly (i.e., performs the projection onto its eigenvectors $\{\ket{a}\bra{a}\}$) and then estimates the value of $\var$ from the average value of $A$ (formally: attributes to each of $\ket{a}\bra{a}$  the estimator $\tilde\var'(a)$, such that $A=\sum_a \tilde\var'(a)\ket{a}\bra{a})$.

To show that, I construct the positive operator of the following form (its positivity comes from $\forall_x M_x\geq 0)$):
\begin{equation}
  0\leq \sum_x\left((\tilde\var(x)-\var)\openone-(A-\openone\var)\right)M_x\left((\tilde\var(x)-\var)\openone-(A-\openone\var)\right)=\sum_x M_x(\tilde\var(x)-\var)^2-(A-\openone\var)^2
\end{equation}
(note that no assumptions about unbiasedness have been used here; the last equality comes directly 
from $\sum_x\M_x=\openone$ and $A=\sum_x\M_x\tilde\var(x)$). From that, we have:
\begin{equation}
\label{rspr}
  \forall_\var\quad \Delta^2\tilde\var=\sum_x \tr(\rho_\var \M_x)(\tilde\var(x)-\var)^2\geq \sum_x \tr(\rho_\var (A-\openone\var)^2),
\end{equation}
where the last inequality is tight if all $\M_x$ are the projections onto eigenstates of $A$.


As \eqref{rspr} holds for any strategy $\{\M_x\},\,\tilde\var(x)$ at any $\var$, it remains valid if one restricts to locally unbiased estimators, as well as if one averages both sides with arbitrary $p(\var)$ or take the maximum over $\var\in\Theta$.

Still, it should be stressed that above we have considered the single parameter estimation, with single measurement realization, where the figure of merit is MSE. In more general cases, separating estimator form measurement may still be useful and necessary, i.e., for the situations where:
\begin{itemize}
    \item one considers many repetitions of the experiment, so the measurement is local, but the estimator may depend on all measurement results jointly.
    \item the cost function other than MSE is under consideration.
    \item more than one parameter is to be estimated.
\end{itemize}
All these cases will also appear in further chapters.

\section{Average energy case and the inability to overcome the Heisenberg scaling}
\label{sec:meanen}

\eqref{bound2} establishes a fundamental lower bound (not necessarily saturable) for the estimator's variance concerning the variance of the evolution generator for the sensing state. As shown in the discussed example, if the total number of involved resources is $N$, the generator variance may scale like $N^2$, leading to quadratic improvement to the classical strategy. The natural question is, what happens if one imposes constraints only on the average number of resources, not the maximum one?

\begin{figure}[h!]
\begin{center}
\includegraphics[width=0.6\textwidth]{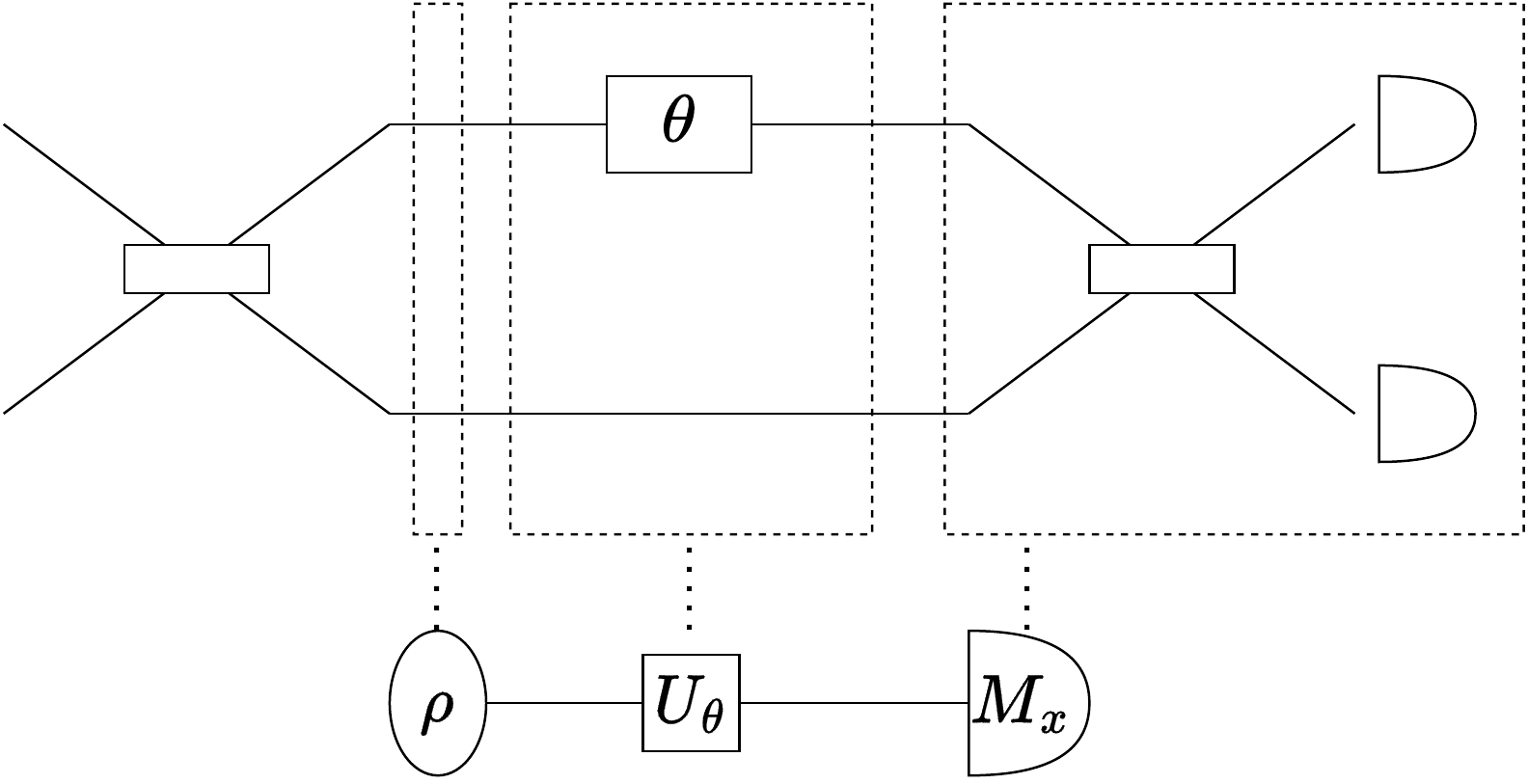}
\caption[Interferometer]{The simplest scheme of the optical interferometer. The incoming state of light is divided between the sensing (upper) and the reference (lower) arm. After going through the interferometer, the light again passes through a beam splitter. In the end, photon-counting detectors are applied to each output arm, where the statistics of this photon counting depend on the phase shift $\var$ in the sensing arm. Generally, the procedure may be optimized over preparing the input state $\rho$ and applying more advance measurement.}\label{fig:interf}
\end{center}
\end{figure}
The problem was extensively discussed in 2010s in the context of optical interferometry. Consider two-arms interferometer, with unknown phaseshift $\var$ in the upper arm, see \figref{fig:interf}. Let us label the modes connected to both arms by $\hat a, \hat a^\dagger$ and $\hat b,\hat b^\dagger$; the corresponding generator of the evolution is then $\HH=\hat a^\dagger \hat a$. Note that for the case when the number of photons $N$ is fixed the problem is mathematically exactly equivalent to the one discussed in \ref{sec:hle}, if in the latter one we consider only the fully symmetric subspace of states. Indeed, if we appropriately identify the up $\ket{\up}$ and down $\ket{\down}$ spin states with lower and upper arms in the interferometer, the single-particle evolution generator is the same (up to, physically irrelevant, global phase). 

However, an interesting effect may be observed if the constraint is imposed only on the average number of photons $\braket{\HH}=\overline N$.
Then one may easily construct the state with arbitrary large $\Delta^2 \HH$ -- for example, if the probability of finding exactly $k$ photons in the upper arm decreases like $\propto \frac{1}{k^3}$, the average converges to constant, while the variance converges to infinity. In some papers, this fact was interpreted as the opportunity to overcome this Heisenberg scaling $\frac{1}{N^2}$, and a few concrete potential protocols have been proposed. Further analysis showed that they lead to strongly biased results or require a large number of repetitions, such that at the end, the scaling with a whole amount of used resources is not better than $\frac{1}{N^2}$
(see also \secref{sec:infinite} for the discussion by the example within Fisher information formalism).

The universal solution for such a class of problems has been proposed in \parencite{Giovannetti2012beyond}, where the authors presented the bound based on quantum speed limit\footnote{Similar reasoning has been proposed earlier in \parencite{zwierz2010general}, however, the derivation of the bound was invalid \parencite{zwierz2011erratum}.}
(e.i. how fast the fidelity between states $\ket{\psi}$ and $e^{i\var \HH}\ket{\psi}$ may decrease with increasing $\var$)
\parencite{margolus1998maximum,Giovannetti2003evolution}. They have shown that for any reasonable estimator
\begin{equation}
\label{boundlorenzo}
    \Delta^2\tilde\var\geq \max\left[\frac{\kappa^2}{(\braket{\HH}-E_g)^2},\frac{\gamma^2}{\Delta^2 \HH} \right],
\end{equation}
where $E_g$ is the ground state energy (the minimal eigenvalue of $\HH$), while $\kappa$ and $\gamma$ are constant (not given analytically). By "reasonable," they only assume that the region where the estimator works well should be at least two times bigger than resulting $\Delta\tilde\var$.

At a similar time, the bound for the particular case of this problem (the estimation of completely random phase shift in interferometer) has been derived \parencite{hall2012universality,hall2012does}, based on entropic uncertainty relations \parencite{bialynicki1975uncertainty}. This bound is tighter but less general, while qualitative consequences are the same.

The above works close the discussion on the optimal scaling with the average amount of resources and the impossibility of beating the Heisenberg scaling. However, due to the not-saturability of mentioned bounds, they do not provide an answer to the question regarding the multplicative constant in the minimal obtainable cost.

\section{Entangled-parallel and sequential-adaptive metrological schemes}
\label{sec:schemes}

So far, we have discussed the problem of unitary evolution governed by generator $\HH$, which allows for formulating the bound in terms of its average value or its variance. Especially for the system with non-interactive particles or linear optics, $\braket{\HH}$ typically scales linearly with a number of particles (or photons), leading to Heisenberg scaling of the precision of the form $\sim \frac{1}{N^2}$. Let us focus on such a case to introduce a notation for the rest of the thesis.

For the sake of simplicity and homogeneity of notation, let me now ignore the indistinguishability of particles (bosons or fermions) and treat them as distinguishable (the problem will be discussed in more detail in specific cases, if necessary).
It is obvious that for distinguishable particles, the estimation would be easier (because a larger states' space would be available), so all the bounds derived for such a case are still valid for bosons and fermions (while saturation may depends on which case is considered).

Let me name the Hilbert space of the single particle in the discussed system by $\mH_S$ and the corresponding single particle evolution generator by $\G\in \lin(\mH_S)$ (where $\lin(\mH)$ denotes the space of linear operators acting on $\mH$). The $N$-particle evolution is therefore governed by:
\begin{equation}
    H=\sum_{i=1}^N \openone^{\otimes (i-1)}\otimes\G\otimes \openone^{\otimes (N-i)},\quad e^{i\var \HH}=\left(e^{i\var \G}\right)^{\otimes N}.
\end{equation}
In most of the analyses presented in this thesis, I will not significantly go beyond the problem of unitary evolution. However, to provide a more realistic context, at some point, I will also discuss the recent results about the noisy case (obtained typically within QFI formalism). To have a unified framework, therefore, I denote a single particle quantum gate as $\ch_\var(\rho)=\rho_\var$, which for unitary evolution is given as $\ch_\var(\rho)=e^{i\var\G}\rho e^{-i\var\G}$ (the more general case of arbitrary completely positive trace preserving (CPTP) map \parencite{chruscinski2009spectral} will be discussed later \secref{sec:singlenoisy}). Let us now consider a different way of using $N$ quantum gates.

In the simplest scenario, the experimentalist uses $N$ uncorrelated particles, lets them evolve, and makes the measurement on each of them (see \figref{fig:schemes}(a)): 
\begin{equation}
\label{separable}
    \rho_\var=\ch_\var(\rho),\quad p(\{x_1,...,x_N\}|\var)=\prod_{i=1}^N\tr(M_{x_i}\rho_\var).
\end{equation}
In such an approach, the MSE will scale as $\sim1/N$, due to the standard statistical properties. Alternatively, one may use general entangled $N$-particle state $\rho\in\lin(\mH_S^{\otimes N})$, for which the evolution is given by acting the product of single-particle channels $\ch_\var^{\otimes N}(\rho)$. In general, such state may be additionally entangled with some external ancillary system $\mH_A$, and then for $\rho\in\lin(\mH_S^{\otimes N}\otimes \mH_A)$ we have (see \figref{fig:schemes}(b)):
\begin{equation}
\label{parall}
    \rho_\var=(\ch_\var^{\otimes N}\otimes\openone)(\rho)
\end{equation}
which in principle may allow for $\sim1/N^2$ scaling of MSE. 

Keeping this formalism, we may consider an even more general scenario, where one can use the channel $\ch_\var(\cdot)$ $N$ times in a completely arbitrary way, including multi-times acting on a single particle (or group of entangled particles), with arbitrary unitary control and/or partial measurement between. The most general adaptive scheme\footnote{In modern literature, even more general schemes appear, where indefinite causal order of acting of gates is considered \parencite{zhao2020quantum,liu2022strict}. However, its physical interpretation is not fully clear, and therefore this case will not be discussed in this thesis.}
may be written as (see also \figref{fig:schemes}(c)):
\begin{equation}
\label{adapt}
\rho_\var=V_N\circ(\ch_\var\otimes\openone)\circ...V_1\circ(\ch_\var\otimes\openone)\circ\rho,
\end{equation}
where $\rho$ is a single-particle state entangled with the ancilla $\rho\in\lin(\mH_S\otimes\mH_A)$, $V_i\circ \rho$ is a shortcut for $V_i\rho V_i^\dagger$, while $V_i$ are unitary controls acting jointly on the system and ancilla.
Note that this equation covers all mentioned cases. Indeed, any actions of the channel on multi-particle states may be simulated with the above equation by applying proper SWAP operators in unitary control $V_i$. Furthermore, the partial measurement may also be simulated, assuming that the ancillary system is sufficiently large.

\begin{figure}[h!]
\begin{center}
\includegraphics[width=1\textwidth]{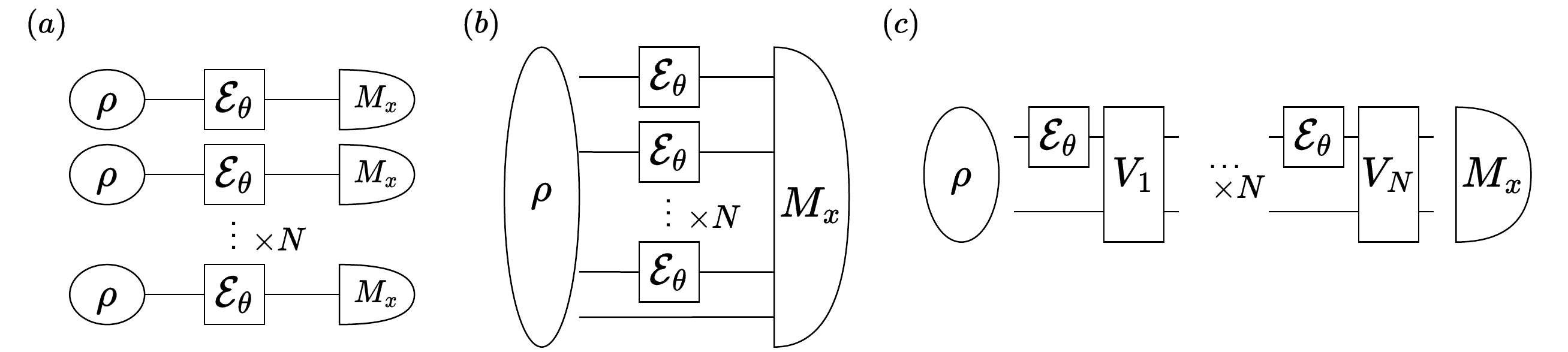}
\caption[Metrological schemes]{Three metrological schemes of usage $N$ quanum gates to estimate the parameter $\var$. a) gates are used independently, which may lead only to standard scaling of the MSE $\sim1/N$. b) entangled-parallel scheme: all $N$ gates act jointly in a parallel way on some entangled stated (extended by ancilla if needed). c) sequential-adaptive scheme: most general  scheme, including sequential acting of the gate $N$ times with unitary control between and entanglement with arbitrary large ancilla.}\label{fig:schemes}
\end{center}
\end{figure}

Formulating the number of resources in terms of quantum gates used in the protocol is helpful, as it allows for generalizing the problem for arbitrary channel estimation and easily includes more sophisticated sequential adaptive strategies. Moreover, by taking proper limits, it may also be applied to the problems of continuous time evolution \parencite{Gorecki2020}.

However, it should be noted that it does not cover all potential metrological issues. For example, for the systems with two-body interaction, the evolution of the state of $N$ particles cannot be written as \eqref{parall}. Moreover, in such case, the energy may scale as $N^2$ with the number of particles, so the bound \eqref{boundlorenzo} allows for quadratic scaling $\Delta^2\tilde\var\sim \frac{1}{N^4}$ \parencite{beltran2005breaking,boixo2007generalized}. Another example may be amplitude estimation of trapped ions \parencite{wolf2019motional} or light \parencite{gorecki2022quantum}, where in the latter case, the channel may increase the number of photons. Finally, it does not provide an accurate description of the problem where the noise acting of different probes is mutually correlated~\parencite{chabuda2020tensor}. Such cases will not be this thesis's topic; we should still be aware of their existence.

Another point should be highlighted here. During reformulating the problem from scaling with the energy to scaling with the number of gates used, we have restricted ourselves to the problems of the well-defined number of gates (representing photons etc.). On the other hand, previously cited bound \eqref{boundlorenzo} allows for non-well-defined energy (unbounded from above). 

In \parencite[Section 5.3.]{demkowicz2015optical} it was pointed out that for the standard problems in optical interferometry, it is much more reasonable to formulate the bounds in terms of the maximal amount (not average) of the resources. The reason is that, in practice, we cannot measure the superposition of the states with a different number of photons, so they will be no more useful than corresponding mixtures
(or, in more abstract way, lacking a phase reference implies a photon-number superselection rule \parencite{bartlett2007reference}). Then we can use the argument that for most reasonable models, the minimal variance obtainable with a fixed amount of resources $N$ is a convex function of $N$ (typically $\sim\frac{1}{N}$ or $\sim\frac{1}{N^2}$), so the optimal strategy with the mean amount of resources will be the one with the well-defined amount of resources. Still, the bound for optimal precision with indefinite number of photons (with fixed average) may be relevant in optical interferometry, if someone distinguishes between sensing beam and reference one and the constrain is applied only to the sensing one (i.e., if the limitation comes from the strength of the sample and not from the capacity of the photon counters). Moreover, this bound will be helpful in analyzing a specific problem of multiparameter estimation in \secref{sec:multiphases}.

The problem of optimal strategy with constraining only on the average amount of resources may be also relevant in analyzing any kind of quantum clocks, understood as the problem of estimation of time by performing the measurement on the state with not well defined energy. In this context, the bounds like \eqref{boundlorenzo} and analysis performed later in this thesis allow going beyond standard Mandelstam-Tamm inequality \parencite{tamm1991uncertainty,busch2008time}.

Recently, some approaches to formulating the problem of sequential adaptive schemes with energy restriction appear in literature \parencite{zhao2020quantum}, but the general formulation is not yet established; moreover, for some problems commonly understood "energy" may refer to something different expectation value of the generator $\braket{H}$ \parencite{zhao2020quantum,gorecki2022quantum}. This cases, however, will not be explored in this thesis. While talking about the bound for the estimator's variance with the constraint on mean energy, I will always understand it as expectation value of the generator and I will restrict myself only to the entangled-parallel scheme; that will be discussed in \secref{sec:airy}, \secref{sec:meanbound}.

\chpt{Fisher information and Cram\'er-Rao bound}
\label{fisher}

In this chapter, I introduce the formalism based on Fisher information and Cram\'er-Rao inequality, stating the fundamental bounds on estimation precision when the experiment is repeated many times. My emphasis is on pointing out how the results obtained by this method should be interpreted and what their operational significance is. At the end of the chapter, I also present current developments in analyzing the estimation problem in the presence of noise.

\section{Classical Cram\'er-Rao bound}
Given probability distribution $p(x|\var)$ of the measurement result $x$ depended on the value of unknown parameter $\var\in\Theta$. The aim is to estimate the value of the parameter from the measurement outcomes. Then for any estimator, locally unbiased at $\var=\var_0$, i.e., the one satisfying:
\begin{equation}
    \braket{\tilde\var}\Big|_{\var=\var_0}=\sum_{x}p(x|\var_0)\tilde\var(x)=\var_0
\end{equation}
\begin{equation}
    \frac{d\braket{\tilde\var}}{d\var}\Big|_{\var=\var_0}=\sum_{x}\frac{dp(x|\var)}{d\var}\Big|_{\var=\var_0}\tilde\var(x)=1,
\end{equation}
its variance is bounded from below by Cram\'er-Rao bound (CR):
\begin{equation}
\label{clasfish}
    \Delta^2\tilde\var\geq\frac{1}{F(p)},\quad F(p)=\sum_{x}\frac{1}{p(x|\var)}\left(\frac{dp(x|\var)}{d\var}\right)^2,
\end{equation}
where $F(p)$ is classical Fisher information (FI) of the probability distribution $p(x|\var)$. 

CR bound is typically derived using Cauchy Schwarz inequality; it may also be obtained by direct minimization over estimators with locally unbiased constraint using the Lagrange multipliers method. In the latter method, one would obtain:
\begin{equation}
\label{lue}
    \tilde\var^{\lu}(x)=\frac{1}{F(p)}\frac{1}{p(x|\var)}\frac{dp(x|\var)}{d\var}.
\end{equation}
Note that while the above estimator is by construction always locally unbiased at point $\var_0$, we have no knowledge about the size of the neighborhood around $\var_0$, where it works well, which in general makes FI and CR bound not very useful for the case when the experiment is performed only once.

Directly from the definition, FI is convex, i.e., for any two probability distributions $p_1(x|\var)$, $p_2(x|\var)$
\begin{equation}
\label{clasconvex}
    F(c_1 p_1+c_2 p_2)\leq c_1 F(p_1) +c_2 F(p_2),
\end{equation}
and additive, i.e., for the joint probability of two variables of the product form $p_{12}(x_1,x_2|\var)=p_1(x_2|\var)p_2(x_2|\var)$ we have
\begin{equation}
\label{clasaddit}
    F(p_{12})= F(p_1) +F(p_2).
\end{equation}
Especially for the $k$-length sequence of measurement results $\bx=\{x_1,...,x_k\}$, wherein each repetition's outcome is drawn independently, so $p(\bx)=p(\{x_1,...,x_k\})=p(x_1)p(x_2)...p(x_k)$, the FI is simply $k$ times bigger than FI for single realization, so:
\begin{equation}
\label{clasfish2}
    \Delta^2\tilde\var\geq\frac{1}{k\cdot F(p)}.
\end{equation}
 It is consistent with a simple observation that one may always choose as the collective estimator the average of single-result estimators $\tilde\var(\bx)=\frac{1}{k}\sum_i\tilde\var(x_i)$, without violating locally unbiasedness condition. However, such a procedure would be a poor choice in the limit of large repetitions number $k$, as it would not solve the problem of biasedness far from the point $\var_0$.

It turns out that in the limit of large $k$, \eqref{clasfish2} may be saturated in all points $\var\in\Theta$ by the same estimator. That is the maximum likelihood estimator (ML estimator), i.e., the one associating to any sequence of the measurement outcomes the value of $\var$, for which the probability of obtaining this sequence is the biggest possible:
\begin{equation}
    \ml(\bx)=\t{arg}\max\limits_\var p(\bx|\var).
\end{equation}
For such estimator, \parencite[5.3. Asymptotic Normality]{vaart_1998}
\parencite[CHAPTER 7]{nagaoka2005asymptotic}
\begin{equation}
\begin{split}
\label{optML}
\forall_\var\quad \lim_{k\to\infty}\braket{\ml(\bx)}&=\var\\
    \lim_{k\to\infty}k\Delta^2\ml(\bx)&=\frac{1}{F[p(\cdot|\var)]},
\end{split}
\end{equation}
where in principle the value of $F[p(\cdot|\var)]$ may depend on the value of parameter.

As the precise formulation of converging ML estimator to the normal distribution is crucial for understanding ambiguities regarding the interpretation of FI in the context of HL, for completeness, in \secref{sec:ML}, I attach the proof of optimality of ML estimator. Before, let me discuss it with a simple but expressive example.

\section{Example -- two arms interferometer}
\label{sec:inter}
Let us go back to the problem of estimating the phase shift in two arms interferometer \figref{fig:interf} and assume that in each trial, only one photon is put into the system. The probabilities of clicking of each detector are given as follows:
\begin{equation}
    p(0|\var)=(1-\sin(\var))/2,\quad p(1|\var)=(1+\sin(\var))/2.
\end{equation}
The problem of estimating $\var$ is therefore equivalent to the analyzing the problem the unfair coin of probability $q=(1+\sin(\var))/2$. To find the maximum of $p(\{x_1,...,x_k\}|\var)=q^{\sum x_i}(1-q)^{k-\sum x_i}$ over $\var$, it is easier to deal with its logarithm:
\begin{equation}
    \log\left(p(\{x_1,...,x_k\}|\var)\right)=\left(\sum x_i\right)\log(q)+\left(k-\sum x_i\right)\log(1-q).
\end{equation}
Then we have:
\begin{equation}
    \frac{dp(\{x_1,...,x_k\}|\var)}{dq}=\frac{\sum x_i}{q}-\frac{k-\sum x_i}{1-q}=\frac{\sum x_i-kq}{q(1-q)}=0\Rightarrow \frac{\sum x_i}{k}=q,
\end{equation}
which leads to the exact formula for the ML estimator:
\begin{equation}
    \ml(\{x_1,...,x_k\})=\arcsin\left(2\frac{\sum x_i}{k}-1\right).
\end{equation}
The expected value of this estimator's indication and MSE (depending on the number of performed trials and the actual value of the parameter $\var$) are presented in \figref{fig:mse}. For comparison, the performances of locally unbiased estimators \eqref{lue} at point $\var_0=0$ and $\var_0=1$ (for $k=100$) are also presented. 

\begin{figure}[h!]
\includegraphics[width=1.\textwidth]{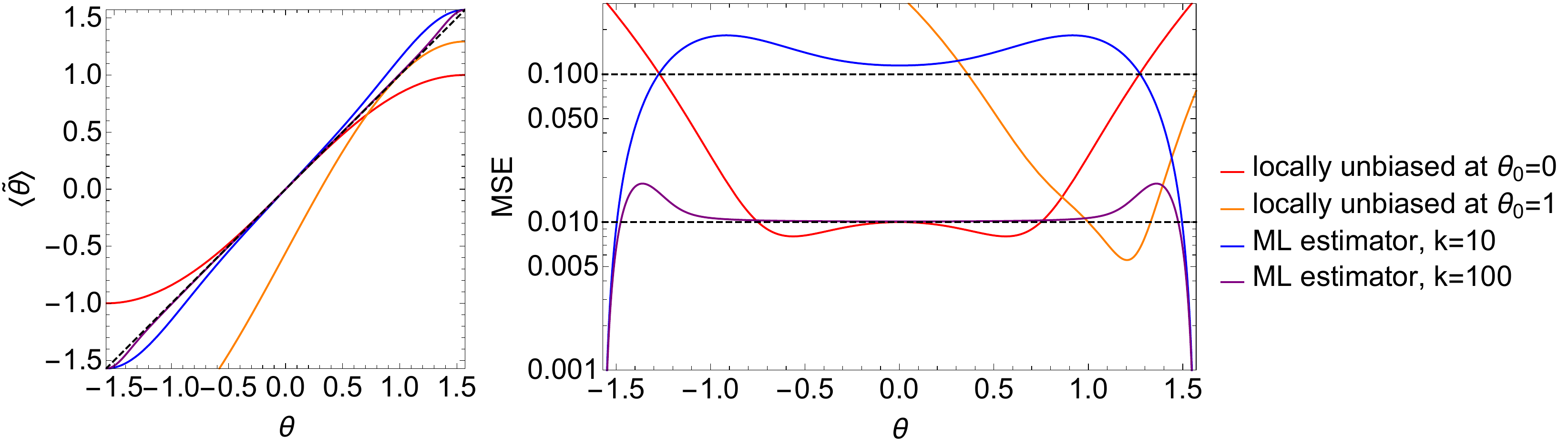}
\caption[Maximum likelihood estimator]{Performance of ML estimator and locally unbiased estimator for the problem of estimation of the phase shift in the interferometer. (Left panel) the expectation value of the estimator. (Right panel) MSE dependence on the actual value of $\var$. One may see that a locally unbiased estimator works well only in the closest neighborhood of $\var_0$, while ML already for $100$ trials returns reasonable values for all $\var$ and it is close to converging CR bound (dashed horizontal lines).}\label{fig:mse}
\end{figure}

\section{Asymptotic normality of maximum likelihood estimator}
\label{sec:ML}
To formulate theorem precisely, let me remind two types of convergence of random variables.

The sequence of random variables $Y_k$ converge in distribution to random variable $Y$ iff for any $y$  cumulative distribution function of $Y_k$ converge to  cumulative distribution functions of $Y$:
\begin{equation}
    Y_k\overset{d}{\to}Y \Leftrightarrow \forall_y\lim_{k\to\infty}p(Y_k\leq y)=p(Y\leq y).
\end{equation}
The sequence of random variables $Y_k$ converge in probability to a constant $c$ iff:
\begin{equation}
    Y_k\overset{p}{\to}c \Leftrightarrow \forall_\epsilon\lim_{k\to\infty}p(|Y_k-c|\geq\epsilon)=0.
\end{equation}
\textbf{Theorem 1.} The sequence of random variables $\sqrt{k}(\ml(\bx)-\var_0)$ converge in distribution to normal distribution of variance $1/F(\var_0)$ for any value of $\var_0$:
\begin{equation}
    \sqrt{k}(\ml(\bx)-\var_0)\overset{d}{\to}\mathcal N(0,F(\var_0)^{-1}).
\end{equation}

\textit{Proof.} (based on \parencite{newey1994chapter,gundersen2019asymptotic})

Let us introduce the normalized log-likelihood function:
\begin{equation}
    L_k(\var|\bx)=\frac{1}{k}\log p(\bx|\var),
\end{equation}
with $\bx=\{x_1,...,x_k\}$. From the definition of the ML estimator
\begin{equation}
\label{null}
    \mlk(\bx)=\t{arg}\max\limits_{\var} p(\{x_1,...,x_k\}|\var) \Rightarrow \forall_{\bx}\dot L_k(\mlk(\bx)|\bx)=0.
\end{equation}
where sign dot denotes derivative $\frac{d}{d\var}$.
Assume without a loss that $\mlk(\bx)\geq\var_0$ (analogous reasoning may be performed for the opposite inequality). From mean value theorem there exists $\bar\var(\bx)$ laying between $\var_0$ and $\mlk(\bx)$ such that:
\begin{equation}
    \dot L_k(\mlk(\bx)|\bx)=\dot L_k(\var_0|\bx)+\ddot L_k(\bar\var(\bx)|\bx)(\mlk(\bx)-\var_0).
\end{equation}
From \eqref{null} LHS is equal $0$; after rearranging terms and multiplying both sides by $\frac{\sqrt{k}}{\ddot L(\bar\var(\bx)|\bx)}$ we obtain:
\begin{equation}
\label{convv}
    \sqrt{k}(\mlk(\bx)-\var_0)=-\frac{\sqrt{k}\dot L(\var_0|\bx)}{\ddot L(\bar\var(\bx)|\bx)}.
\end{equation}
Next, we will show the distribution convergence of the nominator and probability convergence of the denominator and next apply Slutsky's theorem. Let us start with the denominator.

For any $\var$ from the weak law of large numbers (WLLN):
\begin{equation}
    \ddot L_k(\var|\bx)=\frac{1}{k}\sum_{i=1}^k\frac{d^2}{d\var^2}\log p(x_i|\var)\overset{p}{\to}\E\left(\frac{d^2}{d\var^2}\log p(x_1|\var)\right)=-F(\var),
\end{equation}
where $\E(\cdot)$ denotes averaging. Next, as $\var_0\leq \bar\var(\bx)\leq \ml(\bx)$ and $\ml(\bx)\overset{p}{\to}\var_0$, also $\bar\var(\bx)\overset{p}{\to}\var_0$, so combining with above (for sufficiently regular $L(\var|\bx)$):
\begin{equation}
\label{tof}
    \ddot L_k(\bar\var(\bx)|\bx)\overset{p}{\to}-F(\var_0).
\end{equation}

The nominator is given as:

\begin{equation}
    \sqrt{k}\dot L(\var_0|\bx)=\sqrt{k}\left( \frac{1}{k}\sum_{i=1}^k\frac{d}{d\var}\log p(x_i|\var)\Big|_{\var=\var_0}
    -\underbrace{\E\left(\frac{d}{d\var}\log p(x_1|\var)\right)\Big|_{\var=\var_0}}_{=0})
        \right)
\end{equation}
therefore, from the central limit theorem, it converges to the normal distribution with variance:
\begin{equation}
    \E\left(\left(\frac{d}{d\var}\log p(x_1|\var)\right)\Big|_{\var=\var_0}^2\right)=F(\var_0).
\end{equation}
Applying above, together with \eqref{tof}, to \eqref{convv} we finally get:
\begin{equation}
    \forall_{\var_0}\quad\sqrt{k}(\ml(\bx)-\var_0)\overset{d}{\to}\mathcal N\left(0,\frac{F(\var_0)}{F(\var_0)^2}\right)=\mathcal N\left(0,\frac{1}{F(\var_0)}\right),
\end{equation}
what was to be proven. $\square$

\section{Quantum Cram\'er-Rao bound}
\label{sec:qcr}
Previously we have asked how well the parameter $\var$ value may be estimated by probing probability distribution $p(x|\var)$, i.e., we have minimized MSE over all locally unbiased estimators. Here we generalize the problem slightly: there is a given family of states $\rho_\var$ dependent on the unknown parameter $\var$, and we want to optimize the whole procedure over both choices of measurement $\{M_x\}$ as well as an estimator. 

Using the result obtained for the classical case, we may see the problem as optimizing the classical FI \eqref{clasfish} over the choice of measurement. After direct calculation, one would obtain \parencite{Braunstein1994statistical}
\begin{equation}
\label{qfi}
    \max_{\{\M_x\}}F\big[p(x|\var)=\tr(\rho_\var\M_x)\big]=:F_Q(\rho_\var)=\tr(\rho_\var L^2),
\end{equation}
where $L$ is symmetric logarithmic derivative (SLD) $\frac{d\rho_\var}{d\var}=\frac{1}{2}\left(L\rho_\var+\rho_\var L\right)$.

As for the tensor product of matrices $\rho_\var\otimes\sigma_\var$ correspodning SLD operators has the form $L_\rho\otimes\openone+\openone\otimes L_\sigma$, the QFI is additive:
\begin{equation}
F_Q(\rho_\var\otimes\sigma_\var)=F_Q(\rho_\var)+F_Q(\sigma_\var).
\end{equation}
From the fact that QFI is FI maximized over measurements, it is also monotonic, i.e., for any $\var$-independent completely positive trace preserving map $\ch(\cdot)$
\begin{equation}
\label{monotonic}
F_Q(\rho_\var)\geq F_Q(\ch(\rho_\var)),
\end{equation}
as acting $\ch(\cdot)$ followed by POVM on $\ch(\rho_\var)$ clearly defined equivalent POVM on $\rho_\var$ directly. 

Moreover, from above, it is also convex, i.e., for any $\var$-independent parameters $c_\rho, c_\sigma$
\begin{equation}
c_\rho F_Q(\rho_\var)+c_\sigma F_Q(\rho_\sigma)\geq F_Q(c_\rho \rho_\var+c_\sigma \sigma_\var),
\end{equation}
as one may consider the mapping $\mathcal E: c_\rho \rho_\var\otimes \ket{1}\bra{1}+c_\sigma \rho_\sigma\otimes \ket{2}\bra{2}\mapsto c_\rho \rho_\var+c_\sigma \rho_\sigma$. For the situation where $c_\rho,\,c_\sigma$ depend on $\var$, the extended convexity may be defined~\parencite{alipour2015extended}, but it will not be discussed in this thesis.

As stated before, QFI may be derived by direct maximization of classical FI over the measurement. Here I want to present an alternative derivation based on joint optimization over both measurement and estimator (such an approach will be beneficial in multiparameter case \chref{ch:multi}). From the saturability of classical CR bound, we know that:
\begin{equation}
    F_Q^{-1}=\min_{M_x,\t{l.u.} \tilde\var(x)}\Delta^2\tilde\var.
\end{equation}
From the discussion performed in \secref{sec:projective}, looking for the optimal strategy, we may restrict to projective measurements. For given $\var_0$ let me define:
\begin{equation}
   X=\sum_{x}(\tilde\var(x)-\var_0)\ket{x}\bra{x}.
\end{equation}
Therefore QFI is given as:
\begin{equation}
    F_Q^{-1}=\min_{M_x,l.u. \tilde\var(x)}\Delta^2\tilde\var=\min_{X}\left[\tr(\rho_\var X^2)\quad \t{with}\quad \tr(\dot\rho_\var X)=1\right],
\end{equation}
where the minimization is made over all hermitian $X$. It may be performed with the Lagrange multipliers' method, from which:
\begin{equation}
    \forall_{\delta X}\tr((\rho_\var X+X\rho_\var-\lambda \dot\rho_\var)\delta X)=0\Rightarrow \dot\rho_\var=\lambda(\rho_\var X+X\rho_\var),
\end{equation}
where the last equality is equivalent to the definition of SLD (up to constant factor $\lambda$). Local unbiasedness condition implies $\lambda=\tr(\rho_\var L^2)$, and therefore
\begin{equation}
    \tr(\rho_\var X^2)=\frac{\tr(\rho_\var L^2)}{(\tr(\rho_\var L^2))^2}=\frac{1}{\tr(\rho_\var L^2)}\Rightarrow F_Q=\tr(\rho_\var L^2).
\end{equation}

Here we should stress that even if we have derived the value of QFI with only local unbiasedness condition, in the limit of many repetitions, it has similar consequences as the classical one. Indeed, the corresponding measurement, i.e., the projection onto eigenvectors of $L$, results in the probability distribution maximizing classical FI. That indicates that this measurement will be optimal also from the global perspective and allow for effective usage of the ML estimator.

One issue should be mentioned here. In general, the measurement maximizing classical FI may depend on the exact value of the parameter, so the above reasoning cannot be applied directly. Still, in such case, one may asymptotically saturate quantum CR globally, using a two-step procedure \parencite{gill2000state,kahn2008local,yang2019attaining}.
First, use $\sim\sqrt{k}$ copies performing full tomography of the state \parencite{Haah2017sample} to obtain some approximate value of $\var_{\est}$ with the risk of any finite error exponentially decreasing with $\sqrt{k}$\footnote{See \parencite[Section 2.1]{guta2008optimal} for broader discussion and direct calculation for the qubit case, where Hoeffding's inequality \parencite{hoeffding1963probability} has been used.}. Then use the remaining $k-\sqrt{k}$ copies to perform the measurement in the basis of $L$ for $\var=\var_{\est}$. Consequently
 \begin{equation}
\forall_{\var\in\Theta}\lim_{k\to\infty}k\Delta^2\tilde\var=\frac{1}{F_Q(\rho_\var)}.
\end{equation}

It is worth stressing that for pure states, the formula for QFI takes an especially simple form\footnote{From \eqref{qfipure}, one may notice, that Fisher information is related to natural metric on Hilbert space induced by a scalar product, namely $|\braket{\psi_{\var}|\psi_{\var+d\var}}|^2=1-\frac{1}{4}F_Q d\var^2$~\parencite{braunstein1996generalized}. In a general mixed state case, it may be related to the Bures metric, induced by fidelity formula \parencite{hubner1992explicit,sommers2003bures} (see also \parencite{zhou2019exact} for discussion about the ambiguities at points, where the rank of matrices changes).}:
\begin{equation}
\label{qfipure}
F_Q=4(\braket{\dot\psi_\var|\dot\psi_\var}-|\braket{\dot\psi_\var|\psi_\var}|^2).
\end{equation}

\section{Quantum metrology -- unitary evolution}
\label{sec:unitary}
Now we can move to a more general case, i.e., the problem of channel estimation, where the procedure may be optimized not only over measurement and estimator but also the input state or even additional features, as discussed in \secref{sec:schemes}. For simplicity, let us start with the problem of unitary evolution, i.e., the one given as $U=e^{i\var \G}$. From the convexity of the QFI, for any channel, the optimal input state will be the pure state. Therefore, for a single usage of the channel, from \eqref{qfipure}, we have:
\begin{equation}
\label{qfipure2}
F_Q=4(\braket{\psi_\var|\G^2|\psi_\var}-|\braket{\psi_\var|\G|\psi_\var}|^2)=
4(\braket{\psi|\G^2|\psi}-|\braket{\psi|\G|\psi}|^2)=
4\Delta^2\G.
\end{equation}
In the case when generator $\G$ has a bounded spectrum with minimal and maximal eigenvalues $\lambda_-,\lambda_+$, the state maximizing QFI will be an equal superposition of corresponding eigenstates $\ket{\psi_\var}=\frac{1}{\sqrt{2}}(e^{i\lambda_-\var}\ket{\lambda_-}+e^{i\lambda_+\var}\ket{\lambda_+})$, for which
\begin{equation}
    F_Q=(\lambda_+-\lambda_-)^2.
\end{equation}
The reasoning does not change significantly for $n$ parallel usage of the gate \figref{fig:schemes}(b). We simply need to focus on the states maximizing $\Delta^2 \G^{\otimes n}$, which be the corresponding $n00n$ state $\ket{\psi^n_\var}=\frac{1}{\sqrt{2}}(e^{in\lambda_-\var}\ket{\lambda_-}^{\otimes n}+e^{i\lambda_+\var}\ket{\lambda_+}^{\otimes n})$ with:
\begin{equation}
    F_Q=n^2(\lambda_+-\lambda_-)^2.
\end{equation}
Finally, one may ask if applying a more advanced adaptive strategy \figref{fig:schemes}(c) improves this result. The answer is no, which was shown \parencite{giovannetti2006quantum} in the following way. For general adaptive evolution with $n$ usage of gates, the output state is given by:
\begin{equation}
    W_\var \ket{\psi}, \quad \t{with} \quad W_\var=V_n (U_\var\otimes\openone) V_{n-1} (U_\var\otimes\openone)\cdots V_1 (U_\var\otimes\openone).
\end{equation}
so the FQI is equal 
\begin{equation}
\label{unitarybound}
    F_Q=4(\braket{\dot\psi_\var|\dot\psi_\var}-|\braket{\psi_\var|\dot\psi_\var}|^2)=
    4\left(\braket{\psi|\dot W_\var^\dagger\dot W_\var|\psi}-|\braket{\psi|W_\var^\dagger\dot W_\var|\psi}|^2\right)=4\Delta^2 \left[-iW_\var^\dagger\dot W_\var\right],
   \end{equation}
where in the last step, identity $\openone=W_\var W_\var^\dagger$ was added inside the first bracket, and factor $-i$ has been added to make the connection with the original evolution generator more visible. Indeed, $-iW_\var^\dagger\dot W_\var$ is simply sum of $n$ properly rotated generators $\G$
\begin{multline}
-iW_\var^\dagger\dot W_\var=\sum_{j=1}^n\G'_j(\var),\quad \t{where}\quad \\ \G'_j(\var)=\left(V_{j-1} (U_\var\otimes\openone) \cdots V_1 (U_\var\otimes\openone)\right)^\dagger (\G\otimes \openone) \left(V_{j-1} (U_\var\otimes\openone) \cdots V_1 (U_\var\otimes\openone)\right),
\end{multline}
so its both minimal and maximal eigenvalues may be at most $n\lambda_-,n\lambda_+$, which is obtained with $\forall_j V_j=\openone$ (as $U_\var$ commutes with $\G$). It is also clear that no additional entanglement with ancilla is needed. Therefore, optimal state may be simply obtain by $n$ acting of the gates on single system copy: $\ket{\psi_\var^n}=\frac{1}{\sqrt{2}}(e^{in\lambda_-\var}\ket{\lambda_-}+e^{i\lambda_+\var}\ket{\lambda_+})$.

Note that states optimal in both parallel and adaptive schemes has the same problem as the one discussed in \secref{sec:hle} -- they became identical for parameters differ by $\frac{k 2\pi}{n(\lambda_+-\lambda_-)}$ (up to the irrelevant global phase).

When the experimentalist has a large number of trials $k$ to perform, it does not lead to a problem, as one may easily repeat the reasoning from the end of the previous section -- first, perform the general tomography (for example, by using simple $\ket{\psi_\var}=\frac{1}{\sqrt{2}}(e^{i\lambda_-\var}\ket{\lambda_-}+e^{i\lambda_+\var}\ket{\lambda_+})$ state (with a single action of the gate in each trial), and only after that focus on optimal local estimation. That would finally lead to:
\begin{equation}
 \lim_{k\to\infty}k\Delta^2\tilde\var=\frac{1}{n^2(\lambda_+-\lambda_-)^2}.
\end{equation}

Still, more is needed to answer the general problem of \textbf{optimal usage of all $N$ resources}. Intuitively one can see that starting from finite size $\Theta$, to use all resources with at least close to optimal way, one could divide this $N$ between different $n00n$ states (with different $n$), each devoted to estimating $\var$ with a precision of another order. It was shown formally \parencite{higgins2009demonstrating}, that such an approach (assuming that the resources have been divided up in such a way as to allow sufficient repetition of the different orders of precision), indeed allows for obtaining quadratic scaling of the precision with $N$ for the problem of phase shift in the interferometer:
\begin{equation}
\label{unihs}
 \Delta^2\tilde\var\sim\frac{1}{N^2}.
\end{equation}
However, a quantitative discussion of these results demands tools beyond QFI so that it will be performed later in \chref{ch:baymin}.

\section{Infinite QFI for average energy constraints}
\label{sec:infinite}

In \secref{sec:meanen}, I have discussed the problem of the optimization of metrology protocol with constraint only on the average (not total) value of the energy, pointing at the fact that in such case $\Delta^2\HH$ may be arbitrarily large. Let me go back here to this problem. Note that the above comment, together with \eqref{qfipure2}, implies the possibility of obtaining the infinite value of QFI for finite average energy. How should it be interpreted in the context of the asymptotical saturability of CR? Let me discuss a simple example.

Let us consider the interferometer, where the input state is the mixture of $n00n$ states with different $n$, where the weight of the next state decreases like the inverse of the third power of $n$:
\begin{equation}
\label{infin}
    \rho=\frac{1}{\zeta(3)}\sum_{m=1}^{+\infty}\frac{1}{m^{3}}\ket{m00m}\bra{m00m},
\end{equation}
where $\zeta(\cdot)$ is Rieman zeta function and $\zeta(3)\approx 1.20$ is normalization factor. The mean energy is finite:
\begin{equation}
 \braket{\HH}=\frac{1}{\zeta(3)}\sum_{m=1}^{+\infty}\frac{1}{2m^2}=
 \frac{\pi^2}{12\zeta(3)}\approx 0.64,
\end{equation}
while its variance goes to infinity:
\begin{equation}
 \braket{\Delta^2\HH}=\frac{1}{\zeta(3)}\sum_{m=1}^{+\infty}\frac{1}{4m}=+\infty,
\end{equation}
so does QFI. Knowing the measurement optimal for each $n00n$ state, consider the general measurement:
\begin{equation}
 \{\M_{m,\pm}\}=\{ \tfrac{1}{2}(\ket{0}^{\otimes m}\pm\ket{1}^{\otimes m})(\bra{0}^{\otimes m}\pm\bra{1}^{\otimes m})\},
\end{equation}
for which $p_{m,+}(\var)=\frac{1}{\zeta(3)m^3}\cos^2(m\var/2)$, $p_{m,-}(\var)=\frac{1}{\zeta(3)m^3}\sin^2(m\var/2)$. This leads also to infinite classical FI. Moreover, such a formulated example is free from the problems found for a single $n00n$ state, discussed in \secref{sec:hle}, i.e., the probability distribution distinct between any two values of $\var$. All these imply that for the ML estimator from equation \eqref{optML}, we have:
\begin{equation}
\label{tozero}
 \lim_{k\to\infty}k\Delta^2\tilde\var_{\t{ML}}=0.
\end{equation}
How such a result should be understood? To understand it better, let us cut the sum in \eqref{infin} at some large $M$ to get:
\begin{equation}
    \rho_M=\frac{1}{\mathcal N_M}\sum_{m=1}^{M}\frac{1}{m^{3}}\ket{m00m}\bra{m00m},
\end{equation}
where $\mathcal N_M=\sum_{m=1}^M 1/m^3\leq \zeta(3)$ is normalization factor. For any finite $M$:
\begin{equation}
\label{finiteK}
 p_{m,+}(\var)=\frac{\cos^2(m\var/2)}{\mathcal N_Mm^3},\, p_{m,-}(\var)=\frac{\sin^2(m\var/2)}{\mathcal N_Mm^3}\quad \Longrightarrow\quad F(M)=\left(\sum_{m=1}^M m^2/m^3\right)/\mathcal N_M,
\end{equation}
so classical FI is an unlimited growing function of $M$.

The MSE of ML estimator for different $M$ has been plotted in figure \figref{fig:conv}. The higher $M$, the bigger number of repetitions $k$ is needed for saturating CR bound. Especially before saturating CR, we observe the scaling of MSE faster than $1/k$. Therefore for $M=+\infty$, we expect such scaling for all $k$, which is consistent with \eqref{tozero}.

\begin{figure}[h!]
\begin{center}
\includegraphics[width=0.6\textwidth]{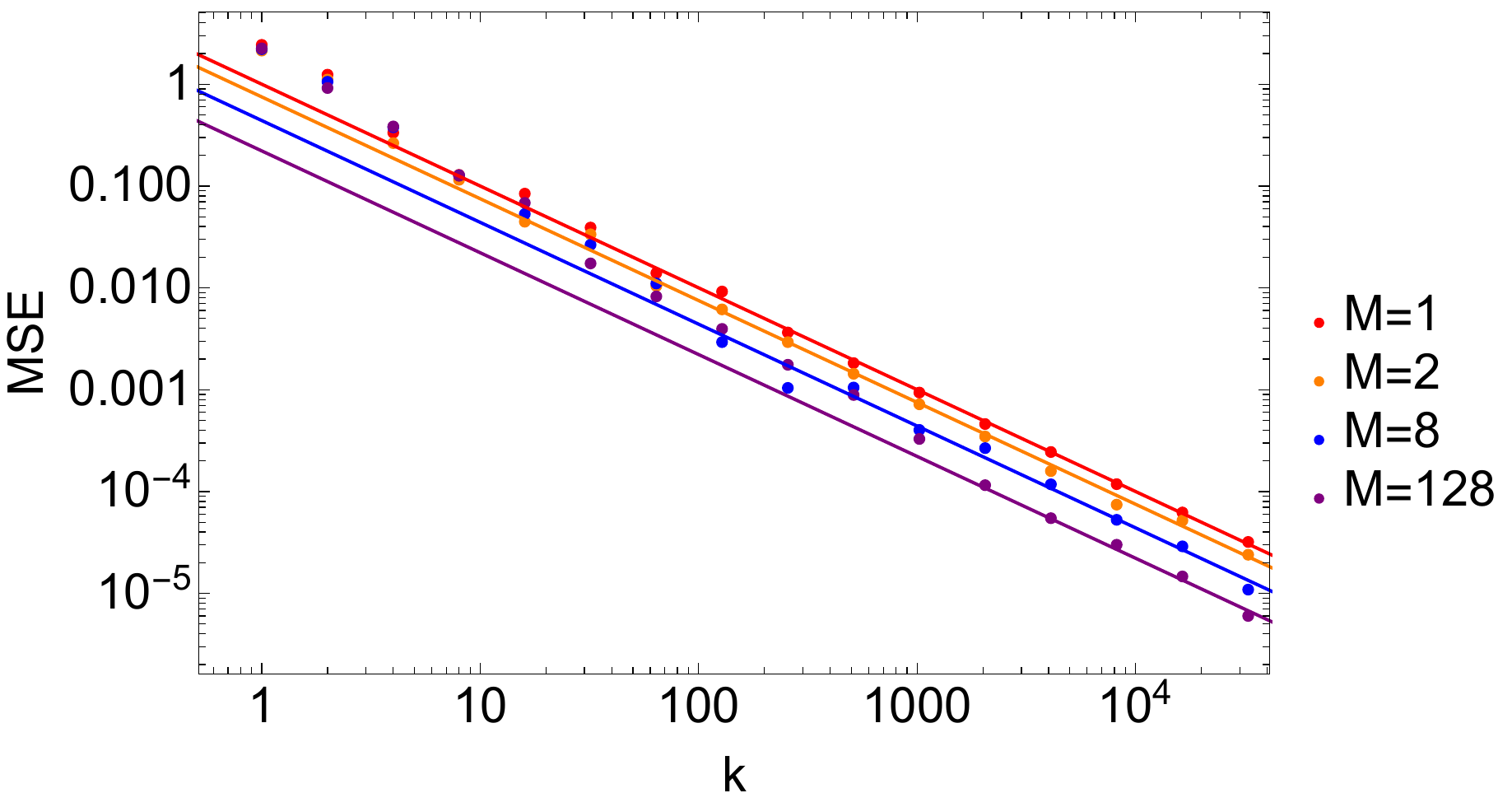}
\caption[Saturating QFI with average energy constrain]{MSE of ML estimator vs. trials number $k$ for the model \eqref{finiteK} is plotted. Results are presented for $M=1,2,8,128$. The dots are the MSE drawn from $p_{m,+}(\var), p_{m,-}(\var)$ probabilities at $\var=1.78072$ (to reduce statistical noise, for each $k$, the MSE was calculated $100$ times and averaged). Solid lines are corresponding CR bounds.} 
\label{fig:conv}
\end{center}
\end{figure}

As the average energy used in a single trial is finite end equal $\approx 0.64$, the above indicates that for such a model, the final MSE will decrease faster than the inverse of the total energy used $\approx 0.64k$. What exactly will this scaling be? The answer for this equation is far from the above analysis obtained by QFI. However, from \eqref{boundlorenzo}, we can immediately say that it cannot decrease faster than quadratically with total average energy used $\sim\frac{1}{k^2}$.

\section{General theorem for a noisy estimation}
\label{sec:singlenoisy}

As stated in the introduction, the main object of interest in this thesis will be the unitary noiseless evolution, therefore from a strictly theoretical point of view, the discussion performed so far is enough for understanding the context of the results presented later in \chref{ch:baymin}. However, this is the idealized situation, as in practice, some noise occurs in any actual metrological situation. Therefore, for completeness, I present the state-of-the-art general theorem about the possibility of attending the Heisenberg scaling in noisy metrology, formulated only within QFI formalism. 
This presentation is only to give a broader context to the results, and the noise case will not be analyzed quantitatively later in this work.

For physical system $\mH_S$, consider general channel $\ch_\var:\lin(\mH_S)\to\lin(\mH_S)$, i.e., completely positive trace preserving (CPTP) map. Any such map may be defined by the set of (not unique) Kraus operators $\{K_k(\var)\}$
\begin{equation}
 \ch_\var(\rho)=\sum_{k=0}^r K_k(\var) \rho K_k^\dagger(\var),
\end{equation}
where $\{K_k(\var)\}_k$ satisfies $\sum_{k=1}^r K_k(\var)^\dagger K_k(\var)=\openone$ (an alternative way of characterizing quantum channels via Choi matrices will be given in \secref{sec:covadapt}). Further, for more compact notation, the parameter in the bracket will be omitted $K_k:=K_k(\var)$, but one should keep in mind that they depend on $\var$.

Then the QFI of the output state maximized over the input state is bounded by \parencite{fujiwara2008fibre,demkowicz2012elusive}
\begin{equation}
\label{singlebound0}
 \max_{\rho}F_Q\leq 4\left\| \sum_{k=0}^r \dot K_k^\dagger \dot K_k \right\|, 
\end{equation}
which comes from the fact that QFI of any mixed state $\rho_\var$ is smaller or equal to QFI of its purification (as the consequence of \eqref{monotonic}). Next, as the set of Krauss operators is defined only up to unitary transformation: \parencite[Proposition 2.]{fujiwara2008fibre}
\begin{equation}
 \tilde K_k(\var)=\sum_{j}u_{kj}(\var)K_j(\var),
\end{equation}
(where in principle the unitary $u_{kj}(\var)$ mixing different Krauss operator may also depend on $\var$), the inequality may be tightened by performing minimization over all Krauss representation. Around $\var_0$, denoting $u_{kj}(\var)=u_{kj}(\var_0)+ih_{kj}(\var-\var_0)+\mathcal O((\var-\var_0)^2)$ (where $h$ is hermitian), by direct subjection we get:
\begin{equation}
 \max_{\rho}F_Q\leq \min_h  4\left\| \sum_{k=0}^r \dot {\tilde K}_k^\dagger \dot {\tilde K}_k \right\|   =\min_h 4\left\| \sum_{k=0}^r \left(\dot K_k^\dagger+i\sum_j h_{kj}K^\dagger_j\right) \left(\dot K_k-i\sum_jh_{kj}K_j\right) \right\|, 
\end{equation}
 Moreover, the above bound is proved to be saturable if ancilla allowed $\rho\in\mH_S\otimes\mH_A$. Starting from this bound, one may derive practical bounds for parallel and adaptive strategies discussed in \figref{fig:schemes}(b,c) using $n$ quantum gates, in terms of the Krauss operators of a single gate. 

For the parallel scheme \parencite{fujiwara2008fibre}:
\begin{equation}
\label{parallel}
 F_Q\leq \min_{h}4\left(n\|\alpha\| +n(n-1)\|\beta\|^2\right)
\end{equation}
where 
\begin{equation}
\begin{split}
\alpha=&\sum_{k=0}^r \dot {\tilde K}_k^\dagger\dot {\tilde K}_k=
\sum_{k=0}^r \left(\dot K_k^\dagger+i\sum_jh_{kj}K^\dagger_j\right) \left(\dot K_k-i\sum_jh_{kj}K_j\right),\\
\beta=&\sum_{k=0}^r {\tilde K}_k^\dagger\dot {\tilde K}_k=\sum_{k=0}^r K_k^\dagger\left(\dot K_k-i\sum_jh_{kj}K_j\right).
\end{split}
\end{equation}
Note especially that $\|\alpha\|\geq \|\beta\|^2$.

For a general sequential adaptive scheme, the bound may be formulated in the iterative form \parencite{kurdzialek2022using} (which is an improved version of the bound form \parencite{demkowicz2014using}):
\begin{equation}
F_Q\leq 4 a^{(n)},\quad \t{where}\quad a^{(i+1)}=\min_{h}\left[a^{(i)}+\|\alpha\|+2\|\beta\|\sqrt{a^{(i)}}\right],\quad \t{with}\quad a^{(0)}=0,
\end{equation}
which may be further bounded by two alternative closed formulas:
\begin{equation}
\label{adaptive}
 F_Q\leq \min_h
 \begin{cases}
 4\left(n\|\alpha\| +n(n-1)\|\beta\|\sqrt{\|\alpha\|}\right),\\
 4\left(n\|\alpha\| +n(n-1)\|\beta\|^2+(\|\alpha\|-\|\beta\|^2)n\log n\right),
 \end{cases}
\end{equation}
where the first one is more usefull in regime $n\approx 1,\,\|\beta\|/\sqrt{\|\alpha\|}\ll 1$, while the second one in the regime $n\gg 1,\, \|\beta\|/\sqrt{\|\alpha\|}\approx 1$.

Looking at both \eqref{parallel} and \eqref{adaptive} one can see that the necessary condition for occurring HS is that $\forall_{h}\|\beta\|\neq 0$. It has been identified as a "Hamiltonian-not-in-Kraus-span" (HNKS) condition. Namely, define:
\begin{equation}
\label{HNKS}
\tilde \G=i\sum_k K_k^\dagger \dot K_k,\quad \mathcal S=\t{span}_{\mathbb H}\{K_i^\dagger K_j\, \t{for all},\ i,j\},
\end{equation}
where  $\t{span}_{\mathbb H}\{\cdot\}$ represents all Hermitian operators that are linear combinations of operators in $\{\cdot\}$. Then HNKS is satisfied iff $\tilde \G\notin \mathcal S$.

Otherwise, in the limit of large $n$, both bounds reduce to:
\begin{equation}
\label{nohs}
 F_Q\lesssim \min_{h:\beta=0}4n\|\alpha\|,
\end{equation}
or, more formally, $\lim_{n\to\infty} \frac{1}{n} F_Q=\min_{h:\beta=0}\|\alpha\|$.
In such case, the bound has been proven to be asymptotically tight \parencite[Theorem 2]{zhou2021asymptotic} :
\begin{equation}
 \lim_{n\to\infty}nF_Q= 4\|\alpha\|.
\end{equation}

The most prominent example of practical importance, when the above bound has been saturated, revealing quantum advantage, was using squeezed vacuum state \parencite{caves1981quantum} in the interferometer in the LIGO gravitational wave detector \parencite{ligo2011gravitational,aasi2013enhanced}, where the experimentalists obtain the MSE improved by factor respectively $3.5\t{dB}$ and $2.15\t{dB}$ compared to the optimal "classical" strategy with the same output energy. The value of MSE is in perfect accord with theoretical model \parencite{aasi2013enhanced}, see also \appref{examplelosses} for exact formula and relation with the bound.

In the case where $\min_{h}\|\beta\|>0$, in the limit of large $n$ the bounds  \eqref{parallel}, \eqref{adaptive} (the second line in the latter one) reduce to:
\begin{equation}
\label{parallas}
 \lim_{n\to\infty}\frac{1}{n^2}F_Q\leq\min_{h}4\|\beta\|^2
\end{equation}
and it may be saturated \parencite{zhou2021asymptotic}, up to this leading term. The authors of \parencite{zhou2021asymptotic} propose the following procedure.

They show that there exists two dimensional subspace of joint system and ancilla $\t{span}\{\ket{0_L},\ket{1_L}\}\in\mH_S\otimes \mH_A$ and the recovery operation $\mathcal R:\mathcal \lin(\mH_S\otimes \mH_A)\to \lin(\t{span}\{\ket{0_L},\ket{1_L}\})$ such that for any $\rho\in \lin(\t{span}\{\ket{0_L},\ket{1_L}\})$ in the closest neighborhood of $\var_0$, effectively unitary evolution rotating by the angle $2\|\beta\|(\var-\var_0)$ is obtained:
\begin{equation}
 \mathcal R(\ch_\var\otimes \openone_A(\rho))\overset{\var\approx \var_0}{\approx} e^{i\|\beta\|(\var-\var_0)\sigma_z}\rho e^{-i\|\beta\|(\var-\var_0)\sigma_z}.
\end{equation}
More precisely, exactly at point $\var_0$ this procedure satisfies\footnote{In original paper \parencite{zhou2021asymptotic} the additional mapping between 2-dimensional logical space to physical space $\t{span}\{\ket{0_L},\ket{1_L}\}\in\mH_S\otimes \mH_A$ appears, which I omitted for simplicity of the formulas}:
\begin{equation}
\label{localqec}
\begin{split}
 \mathcal R(\ch_{\var_0}\otimes \openone_A(\rho))&=\rho\\
 \frac{d}{d\var}\mathcal R(\ch_\var\otimes \openone_A(\rho))\Big|_{\var=\var_0}&=i\|\beta\|(\sigma_z\rho-\rho\sigma_z)
\end{split}
\end{equation}
To get some intuition about the applicability of this protocol, let us consider the simplest non-trivial example.

\subsectionnull{Example}
 Consider a qubit sensing the phase-signal in $z$ direction, subjected to dephasing noise in $x$ direction. What is important, the dephasing operator act before the unitary operator. The exemplary Kraus representation for this model is:
\begin{equation}
 K_1=p e^{i\var \sigma_z},\quad K_2=\sqrt{1-p}e^{i\var \sigma_z}\sigma_x
\end{equation}
for which:
\begin{equation}
\alpha=\openone \Rightarrow \|\alpha\|=1,
\end{equation}
\begin{equation}
 \beta=-i(p\sigma_z+(1-p)\sigma_x\sigma_z\sigma_x)
 =-i(1-2p)\sigma_z\Rightarrow \|\beta\|=|1-2p|.
\end{equation}
Instead of showing that this representation minimizes $\|\beta\|$, we just show the exemplary protocol saturating \eqref{parallas} (so optimal $h=0$),

Let $\ket{0_L}=\ket{00}\in \mH_S\otimes \mH_A$ and $\ket{1_L}=\ket{11}\in \mH_S\otimes \mH_A$. Then for the state $\ket{\psi}=a\ket{0_L}+b\ket{0_L}=\ket{\psi}=a\ket{00}+b\ket{11}$ we have:
\begin{equation}
\begin{split}
 \ch_\var(\ket{\psi}\bra{\psi})=&\\
 p&(ae^{-i\var}\ket{00}+be^{+i\var}\ket{11})(ae^{+i\var}\bra{00}+be^{-i\var}\bra{11})\\
 +(1-p)&(ae^{+i\var}\ket{10}+be^{-i\var}\ket{01})(ae^{-i\var}\bra{10}+be^{+i\var}\bra{01})
 \end{split}
\end{equation}

Consider recovery channel $\mathcal R$ defined by Krauses $R_k$:
\begin{equation}
 R_1=e^{+i\var_0}\ket{00}\bra{00}+e^{-i\var_0}\ket{11}\bra{11},\quad R_2=e^{-i\var_0}\ket{00}\bra{10}+e^{+i\var_0}\ket{11}\bra{01}
\end{equation}
(as $\mathcal R$ is CPTP map, it may be simulated by acting of unitaries, if sufficiently large ancilla allowed so that it may be seen as the special case of \figref{fig:schemes}(c)). Then (restricted to subspace $\ket{0_L},\ket{1_L}$): 
\begin{equation}
 \mathcal R(\ch_\var(\ket{\psi}\bra{\psi}))=
 \begin{bmatrix}
 |a|^2 & ab^*\left(pe^{-i(\var-\var_0)}+(1-p)e^{+i(\var-\var_0)}\right)  \\
a^*b\left(pe^{+i(\var-\var_0)}+(1-p)e^{-i(\var-\var_0)}\right) & |b|^2 
\end{bmatrix}
 \end{equation}

Exactly at point $\var_0$ above satisfies \eqref{localqec}, and the bound \eqref{parallas} may by obtain by choosing as the input state $\ket{\psi}=\tfrac{1}{\sqrt{2}}(\ket{0_L}^{\otimes n}+\ket{1_L}^{\otimes n})$. Note, however, that for finite difference $\var-\var_0$ the factor multiplying off-diagonal elements has the absolute value
$|p e^{-i(\var-\var_0)}+(1-p)e^{+i(\var-\var_0)}|=\sqrt{1-4p(1-p)\sin^2(\var-\var_0)}$, so to make it negligible in analyzing the evolution of the state of the form $\ket{\psi}$ it is necessary $\sqrt{1-4p^2(1-p^2)\sin^2(\var-\var_0)}^n\approx 1\Rightarrow (\var-\var_0)\ll \frac{1}{\sqrt{n}}$.

Again, this is not a problem when the number of gates that can be used in a single implementation $n$ is limited, while the experiment may be repeated a large number of times $k$. For such case, one may obtain:
\begin{equation}
 \lim_{k\to\infty}k\Delta^2\tilde\var=\frac{1}{n^2\|\beta\|^2}.
\end{equation}

However, the possibility of using such a protocol in a situation where one wants to use whole $N$ resources optimally in analogous to the procedure discussed at the and of \secref{sec:unitary} is highly nontrivial. Namely, it is not clear if, for sequential usage of $n00n$ state with increasing $n$, up-to-date knowledge of $\var$ will be sufficient for effectively applying QEC procedure in the next step in such a way to provide finally 
\begin{equation}
\Delta^2\tilde\var\overset{?}{\sim}\frac{1}{N^2},
\end{equation}
in analogous to \eqref{unihs}, proven for unitary noisless estimation in ~\parencite{higgins2009demonstrating}. This, however, remain an open question outside of the topic of this thesis.

\chpt{Obtainable Heisenberg limit in single parameter unitary estimation}
\label{ch:baymin}

In this chapter, I will discuss the fundamental achievable bounds for the precision obtainable in single parameter unitary estimation with a total amount of resources limited. As mentioned before, this problem cannot be analyzed with the usage of QFI or any other formalism based solely on the concept of the local unbiasedness of the estimator. Therefore to perform quantitative analysis, I will use the formalism of minimax estimation and Bayesian estimation.

\section{Bayesian and minimax costs}
Let me start again with the "classical" theory of estimation, i.e., where the probability distribution of the measure outcomes dependent on the value of the parameter $p(x|\var)$ is fixed.

The most popular alternative for FI formalism is the Bayesian one. It postulates that the unknown parameter $\var$ may take a given value according to some known a priori distribution $p(\var)$. In this sense, both $\var$ and $x$ are treated as random variables with joint probability distribution $p(x,\var)=p(x|\var)p(\var)$. That implies marginal distribution $p(x)=\int d\var p(x|\var)p(\var)$, and also conditional probability:
\begin{equation}
   p(\var|x)=\frac{p(x|\var)p(\var)}{\int d\var p(x|\var)p(\var)},
\end{equation}
also called a posteriori distribution of $\var$ after obtaining the result $x$. By Baysiesan MSE, we define the standard MSE, averaged with a priori distribution:
\begin{equation}
\label{bayesformula}
   \vb=\int p(\var) d\var \int dx p(x|\var)(\tilde\var(x)-\var)^2.
\end{equation}

This approach may be applied to the problem when the parameter to be measured represent the quantity which may take different values and in any peculiar realization it has been randomly drawn from a known distribution. Then we may use the knowledge about this distribution to make the procedure more effective on average (i.e. we may focus to make it effective around the values of $\var$, which are more probable, accepting the loss of accuracy in the unlikely cases). For example, when designing a thermometer to measure the temperature inside a heated room, we can focus on having good accuracy in the $15^\circ C$-$30^\circ C$ range and allow very little accuracy outside of this range as we expect it to be used very rarely at other temperatures. To validate the efficiency of such a thermometer in practice, one should perform many repetitions in typical circumstances, comparing the results with the ones coming from a high-quality trusted device. After that, one would recover the formula \eqref{bayesformula}. On the other hand, Bayesian estimation may be also applied in measuring the parameter which is fundamentally postulated to be constant (like, for example, the transition frequency of an atom). In that case, the a priori probability $p(\var)$ will express our imperfect knowledge, coming from previous measurements, which (similarly as in the previous case) may be used to optimize the measurement protocol (the way, how $p(\var)$ may arise from previous measurements, will be discussed at the end of \secref{sec:intuitive} in subsection "Intuitive interpretation"). However, in this case, the formula \eqref{bayesformula} will not be recovered after performing many repetitions of the experiment in typical circumstances, as the parameter will take the same specific value in each realization. Still, using the Bayesian formula is reasonable to optimize the metrology strategy or to validate a peculiar strategy on the level of a mathematical model (i.e. without performing measurements).

The alternative for the Bayesian approach is the minimax one \parencite{hajek1972local,hayashi2011}. Here one only needs to assume that the actual value of the parameter belongs to some set $\var\in\Theta$, and then to validate the effectiveness of the estimation strategy, one assumes the most pessimistic scenario, i.e, maximizes the MSE over all possible values of parameter:
\begin{equation}
   \vm=\sup_{\var\in\Theta}\int dx p(x|\var)(\tilde\var(x)-\var)^2.
\end{equation}

Directly from the definition, one can see that if a priori distribution in the Bayesian approach has support inside of the set considered in the minimax approach $\t{supp}\,p(\var)\subseteq\Theta$, then for any fixed estimator, the Bayesian cost is smaller or equal than the minimax one:
\begin{equation}
\label{bayesmm}
   \forall_{\tilde\var(\cdot)}\quad\vb\leq\vm.
\end{equation}
Bayesian cost is also connected with FI, via the van Trees inequality \parencite{gill1995}:
\begin{equation}
   \vb\geq\frac{1}{\int d \var p(\var)F(\var)+I},\quad I=\int d\var \frac{1}{p(\var)}\left(\frac{dp(\var)}{d\var}\right)^2,
\end{equation}
where $I$ should be understood as the information included in a priori distribution.

From the asymptotic optimality of the ML estimator, in the limit of many repetitions:

\begin{equation}
   \min_{\tilde\var}\lim_{k\to\infty}k\vb
   =\lim_{k\to\infty}k\vb_{\t{ML}}
   =\int p(\var)d\var \frac{1}{F(\var)},
\end{equation}
\begin{equation}
   \min_{\tilde\var}\lim_{k\to\infty}k\vm=
   \lim_{k\to\infty}k\vm_{\t{ML}}
   =   \sup_{\var\in\Theta} \frac{1}{F(\var)}.
\end{equation}
Especially these two coincide if $F(\var)$ does not depend on $\var$.

Now we want to use these formalisms to analyze the problem of quantum metrology, i.e., $p(x|\var)=\tr(\M_x\rho_\var)$. However, before starting, two issues should be mentioned here.

First, we focus on the situation where all resources are used in the optimal way, so, as discussed before, all resources are accumulated in a single measurement realization. This implies that any single measurement outcome is connected directly with the estimator indication (unlike in the previous case, where the estimator could be the function of a sequence of outcomes). Therefore, to simplify notation, we may label the measurement directly by the value of estimator $(\tilde\var(x)-\var)^2\tr(\M_x\rho_\var)dx\to (\tilde\var-\var)^2\tr(\rho_\var\M_{\tilde\var})d\tilde\var$.

Second, as we are no longer restricted only to local estimation, we would like to consider a more general cost than the quadratic one. Therefore, in general, we will want to minimize the mean value of cost function $\cost(\var,\tilde\var)$, which we typically choose to satisfy $\cost(\var,\tilde\var)\overset{|\tilde\var-\var|\ll1}{\approx}(\tilde\var-\var)^2$, which allows us to maintain a relationship with the results obtained using Fisher information. The proper formulas in both formalisms are, respectively, in Bayesian formalism:
\begin{equation}
\label{baycost}
   \cb=\int p(\var) d\var \int d\tilde\var \tr(\rho_\var\M_{\tilde\var})\cost(\var,\tilde\var),
\end{equation}
and in minimax formalism:
\begin{equation}
\label{mincost}
   \cm=\sup_{\var\in\Theta}\int d\tilde\var \tr(\rho_\var\M_{\tilde\var})\cost(\var,\tilde\var).
\end{equation}

\section{Optimality of covariant measurements}
\label{sec:covariant}

Before discussing the fundamental bound derived in~\parencite{Gorecki2020pi}, I want to recall the theorem, 
which significantly simplifies the measurement optimization when the estimation problem satisfies proper symmetry conditions. While this theorem was not used in~\parencite{Gorecki2020pi}, and it is not necessary for understanding it, it allows for deriving very expressive examples providing the reader with good intuition.

Consider a problem where the family of channels is a unitary representation of a compact Lie group $G$, such that $\ch_g(\rho)=U_g\rho U^\dagger_g$ and the aim is to estimate the group element $g\in G$. We would say that the estimation problem is covariant (in both Bayesian or minimax approach), iff:
\begin{itemize}
    \item the cost function is invariant under the acting of the group $\cost(g_1,g_2)=\cost(hg_1,hg_2)$.
    \item (required in Bayesian formalism) a priori distribution is invariant under acting of the group $p(g)dg=p(hg)d(hg)$ -- the prior is uniform with respect to the Haar measure on the group. For simplicity of the notation, further, I assume that $dg$ is the normalized Haar measure, $\int_G dg=1$ (so covariant a priori distribution is $p(g)=1$).
\end{itemize}

Then for a given input state $\rho$ with single usage of the channel, the optimal cost is obtainable with covariant measurement, i.e., the one of the form \parencite[Chapter 4]{Holevo1982}:
\begin{equation}
\label{covpovm}
    \M_{\tilde g}=U_{\tilde g}\M_e U^\dagger_{\tilde g},\quad 
    \int dg U_{\tilde g}\M_e U^\dagger_{\tilde g}=\openone.
\end{equation}
After applying it to \eqref{baycost}, \eqref{mincost} it simplified the problem to the form:

\begin{equation}
\label{covariantbound}
\begin{drcases*}
\inf_{M_{\tilde g}}\cm=\inf_{M_{\tilde g}}\quad\,\,\sup_{\var\in\Theta}\quad\,\, \int d\tilde g \tr(U_g\rho U^\dagger_g\M_{\tilde g})\cost(g,\tilde g) \\
\inf_{M_{\tilde g}}\cb=\inf_{M_{\tilde g}}\int dg \int d\tilde g \tr(U_g\rho U^\dagger_g\M_{\tilde g})\cost(g,\tilde g)
\end{drcases*}=\inf_{\M_e}\int d\tilde g \tr(
    \rho U_{\tilde g}\M_e U^\dagger_{\tilde g})\cost(e,\tilde g)
\end{equation}

The proof is based on the observation that for any measurement $\{\M'_g\}$ one may construct corresponding covariant measurement by proper averaging: $\M_g:=U_g\left(\int dh U^\dagger_{h}M'_h U_{h}\right)U_g^\dagger$, while this procedure does not affect the Bayesian cost and may only decrease the minimax cost. Moreover, within minimax formalism, the statement may be generalized also for non-compact groups \parencite{bogomolow1982minimax}, where formal averaging may not be possible.

The above directly implies that the covariant measurement is optimal for any parallel strategy with the usage of $N$ gates -- as $U_g^{\otimes N}$ is still a group representation. In \chref{sec:covadapt}, I will reproduce the proof of an even stronger statement, saying that if ancilla is allowed, optimal parallel scheme with covariant measurement cannot be beaten by any adaptive strategy (the derivation, however, will demand more advanced formalism, which is not necessary at this point).

\section{Interferometer with a fixed total number of photons}
\label{sec:piinter}

The most canonical example of applying the above theorem is the problem of estimating the unknown phase in the interferometer. In such case the bound \eqref{bound2} gives:
\begin{equation}
\label{naive}
   \Delta^2\tilde\var\geq \frac{1}{N^2},
\end{equation}
without guarantee of saturability. It was shown~\parencite{luis1996,buzek1999,Berry2000}, that for estimation of a completely unknown phase the additional $\pi^2$ factor appears. 

First, note that the value $\var$ should be identified with $\var+2\pi$; therefore, we need to choose a proper cost function considering this fact. The common choice is
\begin{equation}
   \cost(\var,\tilde\var)=4\sin^2\left(\tfrac{\tilde\var-\var}{2}\right),
\end{equation}
which for small difference $\tilde\var-\var$ may be well approximated by $(\tilde\var-\var)^2$. The average cost for the covariant measurement is therefore equal:
\begin{equation}
  \cb=\frac{1}{2\pi}\int d\tilde \var \tr(\rho_0 U_{\tilde\var}\M_0 U_{\tilde\var}^\dagger)4\sin^2\left(\tfrac{\tilde\var}{2}\right),
\end{equation}
(so in notation from \eqref{covariantbound} $g=\frac{\var}{2\pi}$). Without loss, we may restrict ourselves to fully symmetric space and then use the eigenbasis of $U_{\tilde\var}$. Since the mean cost is linear in $\rho$, we may restrict to a pure input state $\ket{\psi^N}$, for which 
\begin{equation}
\label{purediscrete}
 U_{\tilde\var}\ket{\psi^N}=U_{\tilde\var}\left(\sum_{m=0}^Nc_m\ket{m}\right)=\sum_{m=0}^Ne^{im\tilde\var}c_m\ket{m},
 \end{equation}
where $\ket{m}$ is the state with $m$ photons in the upper (sensing) arm. After applying identity $4\sin^2\left(\tfrac{\tilde\var}{2}\right)=2-e^{i\tilde\var}-e^{-i\tilde\var}$ we obtain
\begin{equation}
  \cb=\int \frac{d\tilde\var}{2\pi}\left(\sum_{jk}c^*_jc_ke^{i k\tilde\var}M_{0kj}e^{-i j\tilde\var}\right) (2-e^{i\tilde\var}-e^{-i\tilde\var})=\sum_{jk}c^*_jc_kM_{0kj} (2\delta_{j,k}-\delta_{j+1,k}-\delta_{j-1,k}).
\end{equation}
From condition \eqref{covpovm} $\forall_m M_{0mm}=1$; moreover, without loss of generality, we may assume that $\forall_m c_m\geq 0$ (otherwise, one can always redefine the vectors of the basis, by multiplying them by proper phase factors). Therefore, the minimum value will be obtained for the maximal possible $M_{0m,m+1}$. From positivity condition for $M_0$, $M_{0m,m+1}\leq 1$, where the equality is satisfied if one chooses as $M_0$ the projection onto (unnormalized) vector $\sum_{m=0}^N\ket{m}$, such that: \begin{equation}
\label{contcovmeasu}
\frac{1}{2\pi}\M_{\tilde\var}=\ket{\tilde\var}\bra{\tilde\var},\quad \t{with}\quad \ket{\tilde\var}=\frac{1}{\sqrt{2\pi}}\sum_{m=0}^Ne^{im\tilde\var}\ket{m}.
\end{equation}

After that, the problem simplifies to (introducing $c_{-1}=c_{n+1}=0$ for compact notation):
\begin{equation}
\label{disccost}
  \min_{\ket{\psi^N},\{\M_{\tilde\var}\}}\cb=\min_{\{c_m\}} \sum_{m=0}^{N}(2c_mc_m-c_mc_{m+1}-c_mc_{m-1}),
\end{equation}
which is equivalent to finding the lowest eigenvector of the following tri-diagonal matrix:
\begin{equation}
 \begin{bmatrix} 
    2 & -1& 0 &  0 & &\\
    -1&  2& -1&  0 &\dots&\\
    0 & -1& 2 & -1 & &\\
    0 &  0& -1& 2 & &\\
      &  \vdots& & & \ddots  & -1\\
        &  & & & -1  & 2\\
    \end{bmatrix}.
\end{equation}
The exact solution is known:
\begin{equation}
\label{sinstate}
\ket{\psi^N}=\sqrt{\frac{2}{N+2}}\sum_{m=0}^N \sin\left(\frac{(m+1)\pi}{N+2}\right)\ket{m}
\end{equation}
for which
\begin{equation}
\label{picost}
\cb=2\left(1-\cos\left(\frac{\pi}{N+2}\right)\right)=\frac{\pi^2}{N^2}+\mathcal O \left(\frac{1}{N^3}\right),
\end{equation}
so indeed additional $\pi^2$ factor appears when compared with \eqref{naive}.

While the above provides the exact and complete solution to the problem, some comments are worth mentioning here.

First, while above we have used covariant measurement, the same cost may also be obtained by using the projective measurement, which may be seen as the discrete version of the previously discussed:
\begin{equation}
\label{discrete}
\{M_j\}=\{\ket{\chi_j}\bra{\chi_j}\}_{j=0}^N,\quad \ket{\chi_j}=\frac{1}{\sqrt{N+1}}\sum_{m=0}^N e^{i\frac{2\pi j}{N+1} m}\ket{m},\quad \braket{\chi_j|\chi_k}=\delta_{jk}.
\end{equation}
(note that this is a specific feature of this peculiar model and does not need to be satisfied for arbitrary covariant problems). This choice also has an intuitive interpretation, as this basis may be obtained by applying discrete Fourier transform to the basis with well-defined photons number in upper arm $\{\ket{m}\}_{m=0}^N$. Therefore, it stays in the analog to the problem of the position shift estimation in continuous space, where the shift generator is momentum operator $\hat p=\frac{1}{i}\frac{d}{dx}$, while the optimal measurement is $\{\ket{x}\}$ (which may be obtained by applying Fourier transform to $\{\ket{p}\}$). Nevertheless,  adequacy is not exact, as for the phase $\var$ (unlike for position $x$), no corresponding observable exists.

Second, even if the problem is analytically solvable for finite $N$, it will be useful to see what approximation may be used in the limit $N\to\infty$ (that will allow us to understand better a more general case). Let us introduce $m/N\to \mu $ and consider $\h(\mu)$ and the state \eqref{purediscrete} with $c_m=\frac{1}{\sqrt{N}}\h(m/N)$. Then \eqref{disccost} takes form:
\begin{equation}
 \min_{\ket{\psi^N},\{\M_{\tilde\var}\}}\cb=\min_{\h} \frac{1}{N^2}\cdot\frac{1}{N}\sum_{m=0}^N \h^*(m/N)\frac{2\h(m/N)-\h(m/N+1/N)-\h(m/N-1/N)}{(1/N)^2},
\end{equation}
which, assuming that $\h(\mu)$ does not change very rapidly at distance of order $1/N$, may be approximated by (see \parencite{Imai2009,hayashi2016fourier} for more discussion about exact correspondence in the limit $N\to\infty$):
\begin{equation}
  \min_{\ket{\psi^N},\{\M_k\}}\cb\approx \frac{1}{N^2}\min_{f} \int_0^1 d\mu \h^*(\mu)\left(-\frac{d^2}{d\mu^2}\right)\h(\mu),\quad \h(0)=\h(1)=0,
\end{equation}
so the problem is equivalent to minimizing the kinetic energy of a single particle in the infinite well. The solution is then $\h(\mu)=\sqrt{2}\sin(\mu\pi)$, for which the cost is $\frac{\pi^2}{N^2}$.

Third, it is worth to point the vital difference between the problem of minimizing variance and minimizing the size of confidence interval $\delta_N$ (depended on $N$) of given confidence coefficient $\epsilon$, namely $\min \{\delta_N: p(|\tilde\var-\var|\geq \delta_N/2)\leq \epsilon\}$, discussed broadly in~\parencite{Imai2009}.
In the many repetitions scenario, in the limit of large $k$, for the ML estimator, the distribution of difference $\sqrt{k}(\ml-\var)$ converges to the Gaussian one, which results in exponentially decreasing tail probabilities. Therefore, the solution chosen to minimize the MSE is simultaneously very effective in minimizing the size of a confidence interval. This, however, does not need to be true in the case discussed in this section. The probability of getting outcome $\tilde\var$ where the actual value of the parameter is $\var$ is given as (for optimal covariant measurement \eqref{contcovmeasu}):
\begin{multline}
\label{taildist}
  p(\tilde\var|\var)=\frac{1}{2\pi}\left(\sum_{k=0}^N e^{i m \tilde\var}e^{-i m \var}c_m\right)^2\approx
\frac{1}{2\pi}\left(\int_0^1 d\mu e^{i N\mu (\tilde\var-\var)}\h(\mu)\right)^2=\\
=|f(N(\tilde\var-\var))|^2=\frac{2 \pi (1 + \cos(N(\tilde\var-\var))}{(N^2(\tilde\var-\var)^2 - \pi^2))^2},
\end{multline}
so it decreases only like the forth power of $N(\tilde\var-\var)$. That is -- the strategy using all resources to minimize the quadratic cost (or the one close to quadratic) turns out to have relatively slowly decreasing tail distributions. As a result, it will be highly inefficient for discrimination of confidence interval with a very small risk of error $\epsilon$\footnote{For a fair comparison, it should be noted, that for $\epsilon\approx 1\%$, the solution \eqref{taildist} is rather effective, as the size of the interval is about $5$ times square root of the variance (very similar as in gaussian distribution case). Still, with decreasing $\epsilon$, the size of the corresponding confidence interval increases much faster than for Gaussian-like distribution, and solution \eqref{taildist} starts to be far from optimal~\parencite{Imai2009}}. Therefore, for the optimal usage of all resources, the problem of minimizing variances and the problem of minimizing tail distribution (or the size of confidence interval) need to be discussed independently, and completely different input states turn out to be optimal in each of them.

At last, the broad literature is dedicated to analyzing the relation between the results obtained for the phase estimation in both QFI and Bayesian formalism. The complex analysis of plenty of measurement strategies may be found in \parencite{Berry2009how}. As mentioned in \chref{fisher}, in \parencite{higgins2009demonstrating}, it was analytically proven that the Heisenberg scaling with all resources $N$ may be obtained by a proper sequence of the measurement for $n00n$ states with different $n$ in a different iteration, which was also demonstrated in the experiment \parencite{higgins2007entanglement,higgins2009demonstrating}. In \parencite{Kaftal2014}, the numerical analysis suggests that the factor $\pi^2$ may be exactly obtained for such strategy if one allows for collective measurement on the product of all these $n00n$ states.


\section{Interferometer with a fixed average number of photons}
\label{sec:airy}
The same problem may be discussed with a constraint on the average number of the photons in the sensing arm, which is the constraint for the mean energy in understanding $\braket{\psi|\HH|\psi}=\overline N$, as discussed in \secref{sec:meanen}. Here I recall the results from \parencite{summy1990phase,bandilla1991realistic}.

The reasoning remains almost unchanged compared to the case with a fixed number of photons $N$. The only difference is that in equation \eqref{disccost}, the sum is over all natural indices $\sum_{m=0}^{\infty}$ and the additional constraint appears $\sum_{m=0}^\infty\rho_{mm}m=\overline N$. Similarly, as in the previous case, by approximating discrete variables by continuous one for large $\overline N$ ($m/\overline N\to\mu$), we get:

\begin{equation}
  \min_{\ket{\psi^N},\{\M_{\tilde\var}\}}\cb\approx\frac{1}{\overline N^2}\min_f\int_{0}^{+\infty} d\mu f^*(\mu)\left(-\frac{d^2}{d\mu^2}\right)f(\mu),\quad  \int_0^{+\infty}|f(\mu)|^2=1,\quad f(0)=0,\quad \int_0^{+\infty}|f(\mu)|^2\mu=1.
\end{equation}

The solution may be found using the standard Lagrange multiplier method,
\begin{equation}
-\frac{\partial^2}{\partial \mu^2}f(\mu)+f(\mu)(\lambda_1+\mu\lambda_2)=0 \Rightarrow f(\mu)\propto \t{Ai}\left(\frac{\lambda_1+\lambda_2\mu}{\lambda_2^{2/3}}\right),
\end{equation}
where $\t{Ai}(\cdot)$ is the Airy function of the first kind. After taking into account the conditions, we obtain:
\begin{equation}
\label{eq:g}
f(\mu)= \frac{1}{\t{Ai}'(A_0)}\sqrt{\frac{2|A_0|}{3}}\t{Ai}\left(A_0+\frac{2|A_0|}{3}\mu\right),
\end{equation}
where $A_0\approx -2.34$ is the first zero of the Airy function and $\t{Ai}'(\cdot)$ is its first derivative. 
The corresponding minimal obtainable cost reads:
 \begin{equation}
 \label{summy}
\min_{\ket{\psi^N},\{\M_{\tilde\var}\}}\cb\approx \underbrace{\frac{4 |A_0|^3}{27}}_{c\approx 1.89}.
\end{equation}

One comment should be added here. In the above proof, we restrict ourselves to pure states. Therefore, for completeness, we argue that this cannot be overcome by any mixed input state. For simplicity consider two-rank mixed state $\rho=p_1\ket{\psi_1}\bra{\psi_1}+p_2\ket{\psi_2}\bra{\psi_2}$, where $\tr(\ket{\psi_1}\bra{\psi_1} H)=\overline N_1$, $\tr(\ket{\psi_2}\bra{\psi_2} H)=\overline N_2$, with $p_1\overline N_1+p_2\overline N_2=\overline N$. The minimal cost obtainable for the mixed input state is bigger or equal to the weighted sum of costs optimal for its pure components, therefore:
\begin{equation}
\cb[\rho]\geq p_1\cb[\ket{\psi_1}\bra{\psi_1}]+p_2\cb[\ket{\psi_2}\bra{\psi_2}]\geq p_1\frac{c}{\overline N_1^2}+p_2\frac{c}{\overline N_2^2}\geq \frac{c}{(p_1\overline N_1+p_2\overline N_2)^2}=\frac{c}{\overline N^2},
\end{equation}
where by $\cb[\rho]$ I denoted the minimal variance obtainable for given state $\rho$. The third inequality comes from the convexity property of the $1/x^2$ function (Jensen inequality). The reasoning may be trivially extended for the $\rho$ of higher rank.

\section{\texorpdfstring{$\pi$}--corrected Heisenberg limit}
\label{picorrected}
Let us go back to the fixed number of photons $N$.
Looking at the optimal cost \eqref{picost} and exemplary measurement, which obtains this cost \eqref{discrete}, one may have the impression that the resulting value $\approx\frac{\pi^2}{N^2}$ is somehow related to dividing initial interval $[0,2\pi[$ into $\sim N$ parts corresponding to the measurement outcomes. 

However, this intuition fails: in~\parencite{hayashi2011}, it was proven that, up to the leading term, the optimal cost achievable in parallel strategy does not depend on the size of the initial interval.
More precisely, in the minimax formalism $\min_{\var\in\Theta}\lim_{N\to\infty}N^2\Delta^2\tilde\var=\pi^2$, independently of the size of $\Theta$ (more details about the methods used in this paper will be discussed in \chref{ch:multihl}).

However, three issues remain unclear:
\begin{itemize}
    \item is this results universal (does it works for any unitary generator)?
    \item how the convergence of the above limit depends on the size of $\Theta$?
    \item does the statement also hold in the adaptive scheme?
\end{itemize}
Below I present the results from \parencite{Gorecki2020pi}, which answer all these questions. In the original work, the main thesis has been formulated within Bayesian formalism. Here I present a slightly simplified version formulated within the minimax approach.

\textbf{Theorem 2.} Given a system with unitary evolution $U_\var=e^{i\var\G}$ governed by the bounded generator $\G$ with extreme eigenvalues $\lambda_-$ and $\lambda_+$. No further assumptions about the spectrum of $\G$ are made. Especially the spectrum may be continuous. For more concise notation, I will denote the difference of the extreme eigenvalues by $\lambda=\lambda_+-\lambda_-$. Then, assuming that from the beginning, the parameter is known to be in the interval, $\var\in\Theta=[\var_0-\delta/2,\var_0+\delta/2]$, the optimal minimax cost, obtainable in the most general adaptive scheme with the usage of $N$ quantum gate is bounded by:
\begin{equation}
    \vm\geq \frac{\pi^2}{N^2\lambda^2}\left(1-\frac{8\log(N\lambda\delta)}{N\lambda\delta}\right).
\end{equation}
The logaritmic correction shows how the knowledge gained from the measurement (related to uncertainty of an order of $\sim \frac{1}{N\lambda}$) begins to dominate over the knowledge initially held (uncertainty of order $\sim\delta$) as the amount of resources used $N$ increases. Correction becomes irrelevant when $\frac{1}{N\lambda}\ll \delta$ and asymptotically we obtain $\lim_{N\to\infty}N^2\vm\geq \pi^2/\lambda^2$, which is known to be saturable within the parallel scheme by the analog of the solution from \secref{sec:piinter}. A more detailed discussion will be performed after the derivation of the bound.

\textit{Proof.} The proof contains two major parts. Firstly, I will derive the bound for the Bayesian cost for a specific class of prior distribution $p(\var)$. Next, I will show that, up to small correction, this bound will also be valid for minimax cost for $\var\in\Theta=[\var_0-\delta/2,\var_0+\delta/2]$.

First note that from the Bayesian rule, for any measurement, the Bayesian cost may be bounded from below by the minimal variance of a posteriori distribution $p(\var|x)$
\begin{equation}
\label{minposter}
    \vb=\int d\var \int d x (\tilde\var_x-\var)^2 p(x|\var)p(\var)=\int d\var \int d x (\tilde\var_x-\var)^2 p(\var|x)p(x)\geq \min_x \int d\var(\tilde\var_x-\var)^2 p(\var|x),
\end{equation}
where we assume optimal $\tilde\var_x$, i.e. the one pointing on mean value of $p(\var|x)$.

Let name by $\ket{\psi_\var^N}$ the output state after $N$ acting of gates in the adaptive scheme. According to \secref{sec:projective}, without loss we may restrict to projective rank-one measurements $p(x|\var)=|\braket{\chi|\psi_\var^N}|^2$, so:
\begin{equation}
  p(\var|x)=\frac{|\braket{\chi|\psi_\var^N}|^2p(\var)}{\int |\braket{\chi|\psi_\var^N}|^2 p(\var)d\var}
\end{equation}
 Note that scalar function $f(\var)=\braket{\chi|\psi_\var^N}$ have a finite bandwidth (the support of its Fourier transform $\hat f(\mu)$) of the size of $N(\lambda_+-\lambda_-)$\footnote{note that while $\hat f(\mu)$ plays a similar role as in the previous section, it is not exactly the same object, as here $\mu\in[N\lambda_-,N\lambda_+]$, not $[\lambda_-,\lambda_+]$}. It may be clearly seen, when performing the calculation in the eigenbasis of $\G$, as dependence on $\var$ will be visible only in  terms which are a product of $N$ terms $e^{i\mu_k\var}$ with $\lambda_-\leq \mu_k \leq \lambda_+$:
\begin{multline}
    \braket{\chi|\psi_\var^N}=\\
    \bra{\chi}V_N\left(\int_{\lambda_-}^{\lambda_+} d\mu_Ne^{i\mu_N\var}\ket{\mu_N}\bra{\mu_N}\right)
    \cdots
    V_2\left(\int_{\lambda_-}^{\lambda_+} d\mu_2e^{i\mu_2\var}\ket{\mu_2}\bra{\mu_2}\right)
    V_1\left(\int_{\lambda_-}^{\lambda_+} d\mu_1e^{i\mu_1\var}\ket{\mu_1}\bra{\mu_1}\right)\ket{\psi_{\t{in}}}\\
    =\int_{\lambda_-}^{\lambda_+} d\mu_N\cdots\int_{\lambda_-}^{\lambda_+} d\mu_1 e^{i(\mu_1+\cdots \mu_N)\var}\bra{\chi}V_N\ket{\mu_N}\prod_{k=1}^{N-1} \braket{\mu_{k+1}|V_k|\mu_{k}}\braket{\mu_1|\psi_{in}}=\\
    \int_{N\lambda_-}^{N\lambda_+} d\mu e^{i\mu\var} \underbrace{\int d\boldsymbol{\mu}^\perp \bra{\chi}V_N\ket{\mu_N}\prod_{k=1}^{N-1} \braket{\mu_{k+1}|V_k|\mu_{k}}\braket{\mu_1|\psi}}_{\var-\t{independent part}}
\end{multline}
where $\mu=\mu_1+\cdots \mu_N$, $\int d\boldsymbol{\mu}^\perp$ corresponds to integration over all directions where $\mu=const.$ 

Next, to make the problem analytically solvable, let us consider the class of probability distribution $p(\var)$, which also has finite bandwidth $L$. It was proven \parencite[Lemma 5.1.]{boas1995}, that for any positive function $p_L(\var)$ of the bandwidth $L$ there exists the function $w_{L/2}(\var)$ with bandwidth $L/2$ such that $p_L(\var)=|w_{L/2}(\var)|^2$. Therefore we have
\begin{equation}
    p(\var|x)=\frac{|f(\var)w_{L/2}(\var)|^2}{\int |f(\var)w_{L/2}(\var)|^2d\var}=|g(\var)|^2,\quad \t{with}\quad g(\var):=\frac{f(\var)w_{L/2}(\var)}{\sqrt{\int |f(\var)w_{L/2}(\var)|^2d\var}},
\end{equation}
where $g(\var)$, as a (normalized) product of two functions with bandwidths $N\lambda$ and $L/2$, has a bandwidth $N\lambda+L/2$ (as its Fourier transform is equal to the convolution of Fourier transforms of $f(\var)$ and $w_{L/2}(\var)$).

Therefore RHS of \eqref{minposter} may be bounded from below by minimization over all $p(\var|x)=|g(\var)|^2$ (we may also assume without a loss that its mean value is at $\var=0$, as shifting whole distribution does not change the bandwidth):
\begin{equation}
    \vb\geq \min_{g:\t{supp}\,\hat g\in[N\lambda_--L/4,N\lambda_++L/4]}\int d\var |g(\var)|^2 \var^2,\quad \int d\var |g(\var)|^2=1
\end{equation}
After applying Fourier transform:
\begin{equation}
    \vb\geq \min_{\hat g}\int_{N\lambda_--L/4}^{N\lambda_++L/4} d\mu \left|\frac{d\hat g(\mu)}{d\mu}\right|^2 ,\quad \int_{-N\lambda_--L/4}^{N\lambda_++L/4} d\mu |\hat g(\mu)|^2=1
\end{equation}
so it comes down to finding the minimum energy eigenstate in an infinite potential well and, therefore
\begin{equation}
\label{boundL}
    \vb\geq \frac{\pi^2}{[N\lambda+L/2]^2}.
\end{equation}

Now, let us go to the second part, where I use the above results to derive the bound for minimax cost with $\Theta$ of size $\delta$. It is clear that as $p_L$ has a finite bandwidth, it cannot itself has finite support $\delta$, so it cannot be applied directly from \eqref{bayesmm}. However, one may introduce the finite bandwidth function, for which the probability of finding $\var$ outside of $\Theta$ is sufficiently small that it does not affect the final cost in the leading term. An example of such a function is
\begin{equation}
\label{kaiser}
p_{\alpha,L}(\var)=\mathcal N_\alpha L \t{sinc}^4\left(\pi\alpha \sqrt{(L\var/4\alpha)^2-1}\right),\quad \t{with}\quad \mathcal N_\alpha\approx 4\sqrt{2}\pi^4 \alpha^{7/2}e^{-4\pi\alpha}
\end{equation}
which is proportional to the fourth power of the Fourier transform of the Kaiser window \parencite{kaiser1980on}. The width of the main body of such a function is $8\alpha/L$, while the weight of tails decreases exponentially like $e^{-4\pi\alpha}$.

To show it formally, let me introduce compact notation for minimal Bayesian/minimax cost with given set $\Theta$, probability distribution $p(\var)$ and number of gates $N$:
\begin{equation}
   \mm(\Theta,N):=\inf_{\rho,V_i,\M_i}\sup_{\var\in\Theta}\int d\tilde\var \tr(\rho^N_\var\M_{\tilde\var})(\tilde\var-\var)^2,
\end{equation}

\begin{equation}
   \bay(p(\var),N):=\inf_{\rho,V_i,\M_i}\int p(\var)d\var\int d\tilde\var \tr(\rho^N_\var\M_{\tilde\var})(\tilde\var-\var)^2,
\end{equation}
where $\rho^N_\var$ is the output state obtained by acting on the input state $\rho$ the gate $N$ times with unitaries $V_i$ between, as in the scheme \ref{fig:schemes}(c).

Without loss of generality assume that $\Theta=[-\delta/2,+\delta/2]$ (i.e. $\var_0=0$). Let $R_1$ be the probability, that $\var$ lays outside of $\Theta$ (we consider symmetric $p_L(\var)$):
\begin{equation}
    R_1=2\int_{\delta/2}^{+\infty}d\var p_L(\var).
\end{equation}
Next, we define renormalized cut probability distribution:
\begin{equation}
p^\delta_{L}(\var)=\begin{cases}
\frac{1}{1-R_1}p_L(\var) & \t{for} \,\var\in\Theta\\
0 & \t{for}\,\var\notin\Theta.
\end{cases}
\end{equation}
For such distribution from \eqref{bayesmm}:
\begin{equation}
\label{ineq1}
    \bay(p^\delta_{L},N)\leq \mm([-\delta/2,+\delta/2],N).
\end{equation}
Let $\rho^{N,\delta}_\var,\M^\delta_{\tilde\var}$ be the protocol optimal for $p_L^\delta$. Trivially, for uncut probability $p_L$ this protocol is only suboptimal:
\begin{equation}
\label{ineq2}
    \bay(p_L,N)=\inf_{\rho,V_i,\M_i}\int_{\mathbb R} p_L(\var)d\var \int_{\mathbb R} d\tilde\var\tr(\rho_\var M_{\tilde\var})(\tilde\var-\var)^2\leq
    \int_{\mathbb R} p_L(\var)d\var \int_{\mathbb R} d\tilde\var\tr(\rho^{N,\delta}_\var M^\delta_{\tilde\var})(\tilde\var-\var)^2.
\end{equation}
The last integration may be split between two parts $[-\delta/2,+\delta/2]$ and $\mathbb R\setminus (-\delta/2,+\delta/2)$ 
\begin{equation}
\label{ineq3}
     \frac{1}{1-R_1}\underbrace{\int_{-\delta/2}^{\delta/2} p^\delta_L(\var)d\var \int_{-\delta/2}^{\delta/2} d\tilde\var\tr(\rho^{N,\delta}_\var M^{\delta}_{\tilde\var})(\tilde\var-\var)^2}_{\bay(p^\delta_{L},N)}+
    \underbrace{\int_{\mathbb R\setminus (-\delta/2,+\delta/2)} p_L(\var)d\var \int_{-\delta/2}^{\delta/2} d\tilde\var\tr(\rho^{N,\delta}_\var M^{\delta}_{\tilde\var})(\tilde\var-\var)^2}_{\leq R_2}.
\end{equation}
The first one is simply $\bay(p^\delta_{L},N)$ rescaled by factor $\frac{1}{1-R_1}$. The second part may be bounded from above by assuming the most pessimistic scenario --  each time the true value of parameter $\var$ is to the left (or right) on $\delta/2$, the indication of the estimator is equal $-\delta/2$ (or $+\delta/2$):
\begin{equation}
    R_2=2\int_{\delta/2}^{+\infty}d\var p_L(\var)(\var-\delta/2)^2.
\end{equation}
From \eqref{ineq2}, \eqref{ineq3} we have:
\begin{equation}
    \bay(p_L,N)\leq \frac{1}{1-R_1}\bay(p^\delta_{L},N)+R_2
\end{equation}

Replacing $(1-R_1)$ by one, subtracting $R_2$ and applying \eqref{boundL} and \eqref{ineq1} we get:
\begin{equation}
\label{almost}
    \frac{\pi^2}{(N\lambda+L/2)^2}-R_2 \leq \mm([-\delta/2,+\delta/2],N).
\end{equation}

Next, in \appref{app:window} I prove that for the function \eqref{kaiser} with $L=\frac{8}{\delta}\log(N\lambda\delta)$, $\alpha=\log(N\delta\lambda)$ for $N\lambda\delta>2$:
\begin{equation}
\label{inapp}
\frac{\pi^2}{\lambda^2N^2}\left(1-\frac{8\log(N\lambda\delta)}{N\lambda\delta}\right)\leq \frac{\pi^2}{(\lambda N+L/2)^2}-R_2.
\end{equation}
By applying \eqref{inapp} to \eqref{almost}, we get:
\begin{equation}
\frac{\pi^2}{\lambda^2N^2}\left(1-\frac{8\log(N\lambda\delta)}{N\lambda\delta}\right)\leq \mm([-\delta/2,+\delta/2],N),
\end{equation}
or, in more common notation:
\begin{equation}
\label{result}
    \vm\geq \frac{\pi^2}{\lambda^2N^2}\left(1-\frac{8\log(N\lambda\delta)}{N\lambda\delta}\right),
\end{equation}
what ends the proof. Note, that the function $1-8\log(x)/x$ takes a positive value for $x\geq 26.1$, so only for $N\lambda\delta\geq 26.1$ above bound is informative. $\square$

In \parencite{Gorecki2020pi} the analogous bound has been derived in a similar spirit for the Bayesian cost with rectangular flat prior (the second part of the proof required more calculations):
\begin{equation}
\label{resultbayes}
    \vb\geq \frac{\pi^2}{\lambda^2N^2}\left(1-\sqrt{\frac{8\log(N\lambda\delta)}{N\lambda\delta}}\right).
\end{equation}
One may quickly notice that the bound converge slower than the previous one, which stays consistent with \eqref{bayesmm}.

\subsectionnull{Asymptotic saturability}
The exact solution for the estimating 
unknown phase in the interferometer, namely \eqref{sinstate}, can be used to show asymptotical saturability of \eqref{result}.
Therefore, I will show that the bound may be asymptotically saturated within a parallel scheme.

By equating the upper and lower arms of the interferometer with the eigenstates of $\G$ with extreme eigenvalues, we get the analog of \eqref{sinstate}:
\begin{equation}
\ket{\psi}=\sqrt{\frac{1}{N+2}}\sum_{m=0}^N \sin\left(\frac{(m+1)\pi}{N+2}\right)\ket{N-m}_{\lambda_+}\ket{m}_{\lambda_-},
\end{equation}
where $\ket{N-m}_{\lambda_+}\ket{m}_{\lambda_-}$ is a symmetric state with $m$ particles in state $\ket{\lambda_-}$ and $N-m$ particles in state $\ket{\lambda_+}$.
The only potential problem is that this state cannot distinguish the phases, which differ by multiplication of $2\pi/\lambda$. Hence, it allows only for effective estimation inside the region of size $2\pi/\lambda$. However, unlike in the $N00N$ state case, this is not a severe problem, as the size $2\pi/\lambda$ does not shrink with $N$. Therefore, one may spend at the beginning some sublinear amount of resources (for example $\sqrt{N}$) to discriminate such region (with exponentially decreasing probability of error) without affecting the final cost up to terms of order $1/N^2$.

On the other hand, in a specific case when the spectrum is continuous everywhere between $[\lambda_-,\lambda_+]$, one may use the "continuous" version of the above state, namely 
\begin{equation}
    \ket{\psi_\var^{N}}=\int_{N\lambda_-}^{N\lambda_+}e^{i\mu\var}\sqrt{\frac{2}{\lambda N}}\sin\left(\frac{\mu \pi}{\lambda N}\right)\ket{\mu}d\mu,
\end{equation}
where, by $\ket{\mu}$, we denote an arbitrary combination $\ket{\mu_N}\otimes...\otimes\ket{\mu_1}$ such that $\sum \mu_k=\mu$ (i.e. the one satisfying $U_\var^{\otimes N}\ket{\mu}=e^{i\mu}\ket{\mu}$), with covariant measurement:
\begin{equation}
\label{conticova}
   \ket{\tilde\var}=\frac{1}{\sqrt{2\pi}}\int e^{i\mu\tilde\var}\ket{\mu}d\mu,
\end{equation}
to get immidietly $\vm=\pi^2/N^2\lambda^2$ in single shot experiment.

\subsectionnull{Intuitive interpretation}
\label{sec:intuitive}

Once the bound is derived, I would like to provide a simple intuition for the finite bandwidth prior distribution used during the proof. That would be, in fact, an alternative derivation of the bound \eqref{boundL}, but only for the parallel strategies~\parencite{gorecki2021multiple}.

Let start with the problem, where $\Lambda$ has continuous spectrum $[\lambda_-,\lambda_+]$\footnote{
Note that from an estimation point of view, this is an easier problem than one with a discrete spectrum since we have more space in the choice of the input state. Therefore the lower bound derived for the continuous spectrum is also valid for the discrete one.} and the parameter is entirely unknown $\var\in\mathbb R$ (i.e. $\delta=+\infty$). From the generalization of covariant measurement theorem for non-compact groups \parencite{bogomolow1982minimax}, we can use the optimality of covariant measurement, and the optimal solution is the one discussed above with cost $\vm=\pi^2/N^2\lambda^2$.

 Therefore, assuming that at the beginning one has to disposal $N+N_0$ gates, the minimal achievable minimax cost will be therefore equal:
\begin{equation}
\label{onlyparallel}
    \t{minimax}(\mathbb R,N+N_0,\t{parallel})=\frac{\pi^2}{(N+N_0)^2\lambda^2}.
\end{equation}
On the other hand, we can consider an alternative, sub-optimal strategy:
at first use state $\ket{\psi^{N_0}}$ with the measurement \eqref{conticova} to get an approximated value of the parameter $\tilde\var_{\est}$, and after that spend remaining $N$ gates to estimate a difference $\var-\tilde\var_{\est}$, i.e, use input state

\begin{equation}
\label{parallap}
    \ket{\psi^{N_0+N}}=\ket{\psi^{N_0}}\otimes \ket{\psi^{N}},
\end{equation}
with the measurement:\
\begin{equation}
\ket{\tilde\var_{\est}}\bra{\tilde\var_{\est}}\otimes U_{\tilde\var_{\est}}^{\otimes N}\M_{\tilde\var-\tilde\var_{\est}}U^{\dagger\otimes N}_{\tilde\var_{\est}}.
\end{equation}
The resulting minimax cost will be given as: 
\begin{equation}
\vm =\sup_{\var\in \mathbb R}
\int_{\mathbb R} d\tilde\var 
\int_{\mathbb R} d\tilde\var_{\est}
|\braket{\tilde\var_{\est}-\var|\psi^{N_0}_{0}}|^2 \tr\left(\M_{\tilde\var-\tilde\var_\est}\ket{\psi^N_{\var-\tilde\var_{\est}}}\bra{\psi^N_{\var-\tilde\var_{\est}}}\right)(\var-\tilde\var)^2,
\end{equation}
(where we used $\braket{\tilde\var_{\est}|\psi^{N_0}_{\var}}=\braket{\tilde\var_{\est}-\var|\psi^{N_0}_{0}}$). We see that the cost does not depend on $\var$, so the supremum at the beginning may be omitted. Introducing $\varphi=\var-\tilde\var_{\est}$, $\tilde\varphi=\tilde\var-\tilde\var_{\est}$ we get
\begin{equation}
\vm =
\int_{\mathbb R} d\varphi
\int_{\mathbb R} d\tilde\varphi
|\braket{\varphi|\psi^{N_0}_{0}}|^2 \tr(\M_{\tilde\varphi}\ket{\psi^N_\varphi}\bra{\psi^N_\varphi})(\varphi-\tilde\varphi)^2 ,
\end{equation}
so the formula for minimax cost $\vm$ in this peculiar strategy is, in fact, equal to the formula for the Bayesian cost $\overline{\Delta^2\tilde\varphi}$ with the a priori distribution $p(\varphi)=|\braket{\varphi|\psi^{N_0}_{0}}|^2$! Moreover, for the state:
\begin{equation}
    \ket{\psi_\var^{N_0}}=\int_{N_0\lambda_-}^{N_0\lambda_+}e^{i\mu\var}\hat w_{\lambda N_0}(\mu)\ket{\mu}d\mu,
\end{equation}
the probability density of obtaining results $\tilde\var_\est$ is: 
\begin{equation}
    p(\tilde\var_\est|\var)=|\braket{\tilde\var_\est|\psi}|^2=|w_{\lambda N_0}(\var-\tilde\var_\est)|^2
\end{equation}
Therefore, one get a posteriori distribution $p_{2\lambda N_0}(\varphi)=|w_{\lambda N_0}(\var-\tilde\var_\est)|^2$, with the bandwidth $L=2\lambda N_0$. That it is -- the finite bandwidth probability distribution of the bandwidth $L$ is the one which may be obtained starting from a completely unknown parameter by spending $N_0=L/(2\lambda)$ gates! As this two-step procedure is only sub-optimal, we immediately get:
\begin{equation}
    \t{bayesian}(p_{2\lambda N_0},N,\t{parallel})\geq\t{minimax}(\mathbb R,N+N_0,\t{parallel}), 
\end{equation}
which leads to the same bound formula as \eqref{result}, but here for the parallel strategies only:
\begin{equation}
    \overline{\Delta^2\tilde\varphi}\geq \frac{\pi^2}{[N\lambda+L/2]^2}.
\end{equation}

The reasoning could be performed for the adaptive scheme as well, but then one would need to independently prove \eqref{onlyparallel} for the general adaptive scheme. This makes the first approach more straightforward.

\subsectionnull{Frequency estimation}
\label{sec:frequency}
It is worth noting, that the bound \eqref{result} may be easily generalized for the problem of frequency estimation, where the system of $N_{\t{pr}}$ probes evolve under single-probe Hamiltonian $\omega G$ for total time $T$ and the frequency $\omega$ is to be estimated. Moreover, we assume that one has full freedom in acting on the system and ancilla by any unitary controls during the evolution.

Then the amount of resources is both $N_{\t{pr}}$ and $T$. To use the result \eqref{result} we may divide the total time $T$ to arbitrary small periods of size $t$ and then bound from below the minimal obtainable estimator variance by the analogous problem, where $N=N_{\t{pr}}\cdot \frac{T}{t}$ gates of the form $U_{\omega}=e^{i\omega\oG}$ with $\oG=-t G$, what, after the direct substitution, results with:
\begin{equation}
\label{resultfr}
    \widehat{\Delta^2\tilde\omega}\geq \frac{\pi^2}{N_{\t{pr}}^2\lambda_G^2 T^2 }\left(1-\frac{8\log(N_{\t{pr}}\lambda_G\delta_\omega)}{N_\t{pr}\lambda_G\delta_\omega}\right),
\end{equation}
where $\lambda_G$ is the difference between extreme eigenvalues of $G$ and $\delta_\omega$ is the size of the region where $\omega$ is known to belong.

Such a formalization may be applied for discussing the problem of magnetic field estimation, without the necessity of assuming that $|\omega T|\leq \pi$ or any restriction on how often we use unitary controls. The exemplary problem referring to magnetic field gradient estimation \parencite{altenburg2017estimation} is discussed using this formalism in \appref{app:gradient}.

\section{Average energy constraint}
\label{sec:meanbound}

In this section, I will derive the analogous bound for the problem of the unitary estimation with average energy constraint, discussed in \secref{sec:meanen}. The bound is a consequence of discussions and calculations included in supplementary materials of \parencite{gorecki2021multiple,gorecki2022multiparameter}, but it was not written explicitly anywhere before.

First, note that we cannot simply repeat the main reasoning from \secref{picorrected}. Indeed, having constraints only on average energy, one can easily find an example where RHS in the inequality \eqref{minposter} goes to $0$ for the fixed value of average energy. However, we may make use of the comment made at the end (about an intuitive interpretation) to derive the bound for parallel strategy (with ancilla). For more compact notation, let us assume that the ground state energy is zero $E_g=0$. The problem is defined as follows:
\begin{equation}
   e^{i\var\HH},\quad \braket{\psi|\HH|\psi}\leq E, \quad \var\in[\var_0-\delta/2,\var_0+\delta/2]
\end{equation}
and we want to minimize the MSE of the estimator of $\var$.
Without loss, we assume that the spectrum of $\HH$ is continuous $[0,+\infty[$ (allowing for continuous spectrum may only decrease the minimal obtainable cost, as there is more freedom to choose input state). Then we can use the reasoning mentioned at the end of the previous section. Consider first the problem of estimating a completely unknown parameter $\var\in\mathbb R$, using the state with the average energy $E+E_0$. Then, from the reasoning performed in \secref{sec:airy},
\begin{equation}
    \vm\geq \frac{4|A_0|^3}{27}\frac{1}{(E+E_0)^2}.
\end{equation}
Next, we consider a suboptimal two-step strategy. Consider the input state of the form:
\begin{equation}
   \ket{\psi^{E+E_0}}=\ket{\psi^{\leq E_0}}\otimes\ket{\psi^E},
\end{equation}
where $\ket{\psi^{\leq E_0}}$ is defined as:
\begin{equation}
    \ket{\psi_\var^{\leq E_0}}=\int_0^{E_0}e^{i\mu\var}\hat w_{E_0}(\mu)\ket{\mu}d\mu,
\end{equation}
so for sure $\braket{\psi|\HH|\psi}\leq E_0$. Then we perform a covariant measurement on $e^{i\var H}\ket{\psi^{\leq E_0}}$. The bandwidth of the resulting probability distribution:
\begin{equation}
    p(\tilde\var|\var)=|\braket{\tilde\var|\psi_\var^{\leq E_0}}|^2=|w_{E_0}(\var-\tilde\var)|^2
\end{equation}
is therefore $L=2E_0$. Using the argument that such a two-step strategy is only suboptimal, in an analog to the previous case, we get:
\begin{equation}
    \t{bayesian}(p_{2E_0},E,\t{parallel})\geq \t{minimax}(\mathbb R,E+E_0,\t{parallel})
\end{equation}
\begin{equation}
    \vb\geq \frac{4|A_0|^3}{27}\frac{1}{(E+L/2)^2}.
\end{equation}
Then, choosing as $p_{2E_0}(\var)$ the function \eqref{kaiser}, and repeating reasoning performed in \ref{picorrected}, we obtain:
\begin{equation}
\label{meanbound}
    \vm\geq \frac{4|A_0|^3}{27}\frac{1}{E^2}\left(1-\frac{8\log(E\delta)}{E\delta}\right).
\end{equation}
As mentioned before, the function $1-8\log(x)/x$ takes a positive value for $x\geq 26.1$, so only for $E\delta\geq 26.1$ above bound is informative. For $E\delta\to\infty$ it converge to $\frac{4|A_0|^3}{27}\frac{1}{E^2}$ and, in contrast to existing bounds mentioned in \secref{sec:meanen}, it is asymptotically saturable. 





\chpt{Multiparameter metrology}
\label{ch:multi}

\section{Foundations}

In the previous chapters, I discussed the situation where the quantum channel depends on a single (scalar) unknown parameter we want to measure. On the other hand, many practical metrology problems are related to the estimation of multiple parameters simultaneously, such as vector field sensing  (e.g., magnetic field) \parencite{baumgratz2016}, imaging \parencite{tsang2016quantum}, multiple-arm interferometry \parencite{humphreys2013quantum,Gessner2018} or waveform estimation \parencite{Tsang2011, Berry2013},
which makes multi-parameter metrology an object of intensive interest in recent years~\parencite{matsumoto2002new,kolenderski2008,genoni2013optimal,ragy2016,yuan2016,Liu2017,Kura2018,Nichols2018,ge2018distributed}. Let me formulate the problem more concretely.

Let $\ch_{\bvar}(\cdot)$ be the quantum channel depending on the vector of the parameters $\bvar=[\var_1,...,\var_p]^T$, and so does the results of the measurement performed on the output state $p(x|\bvar)=\tr(\M_x\rho_\bvar)$.

The aim is, therefore, to estimate the values $\bvar$, by probing $x$. We introduce the cost function $\cost(\bvar,\tilde\bvar)$ which is to be minimized:
\begin{equation}
    \cost(\bvar,\tilde\bvar)=\sum_x \cost(\bvar,\tilde\bvar(x))p(x|\bvar),
\end{equation}
which, for a typical problem, for the estimator indication close to the true value may be well approximated by the quadratic function:
\begin{equation}
    \cost(\bvar,\tilde\bvar)\approx (\bvar-\tilde\bvar)^T \C (\bvar-\tilde\bvar),
\end{equation}
where $C$ is the Hessian of the cost function $\C_{ij}=\partial_{i}\partial_{j}\cost(\bvar,\tilde\bvar)\big|_{\tilde\var=\bvar}$ (which, in general, may depend on the exact value of $\bvar$). Here $\partial_i$ is the shortcut for $\frac{\partial}{\partial\var_i}$. It is convenient to write down the average cost as:
\begin{equation}
    \ptr(\C\Sigma), \quad \t{where} \quad \Sigma:=\sum_x p(x|\bvar)(\tilde\bvar(x)-\bvar)(\tilde\bvar(x)-\bvar)^T,
\end{equation}
where $\ptr(\cdot)$ denote trace over $p$-dimensional space of parameters (I use the small letter $\ptr(\cdot)$ to distinguish form the trace over physical Hilbert space $\tr(\cdot)$).
If the estimator is unbiased at $\bvar$, i.e. $\sum_x p(x|\bvar)\tilde\bvar(x)=\bvar$, $\Sigma$ is the covariance matrix of the $\tilde\bvar$. Especially if the cost matrix is identity $C=\openone$, the cost is simply the sum o variances of the estimators $\sum_{i=1}^p\Delta^2\tilde\var$.

\section{Fisher information matrix and multiparameter Cram\'er-Rao bound}
\label{sec:multifisher}
Let me start with the scenario where the experiment is repeated many times so that we may use Fisher information formalism. 

Given probability distribution $p(x|\bvar)$. By locally unbiased estimator of the component $\var_i$, we understand the one which does not only properly changes with arising $\var_i$, but is also insensitive for changes of the remaining parameters $\var_{i\neq j}$. For any estimator locally unbiased at $\bvar$:
\begin{equation}
\label{multilu}
    \forall_j\sum_x p(x|\bvar)\tilde\var_j(x)=\var_j,\quad\forall_{j,i}\sum_x \frac{\partial p(x|\bvar)}{\partial \var_i}\tilde\var_j(x)=\delta_{ij}
\end{equation}
the covariance matrix is bounded from below by multiparameter Cram\'er-Rao bound
\begin{equation}
\label{multiclas}
    \Sigma\geq \bF^{-1}, \quad \bF_{ij}=\sum \frac{1}{p(x|\bvar)}\frac{\partial p(x|\bvar)}{\partial \var_i}\frac{\partial p(x|\bvar)}{\partial \var_j},
\end{equation}
which may be derived by using the Cauchy-Schwarz inequality (by matrix inequality, we understand that their difference is positive semidefinite $\Sigma-\bF^{-1}\geq 0$). Note that diagonal matrix element $\bF_{ii}$ is single-parameter Fisher information for $\var_i$. However, from \eqref{multiclas}, $\Delta^2\tilde\var_i\geq [\bF^{-1}]_{ii}$ (where in general $[\bF^{-1}]_{ii}\geq (\bF_{ii})^{-1}$), as stronger locally unbiasedness condition has been imposed here \eqref{multilu}. Similarly, as in the single-parameter case, the Fisher information matrix scales linearly with the number of repetitions $k$, and in the limit of $k\to\infty$ the corresponding classical Cram\'er-Rao inequality is saturated by maximum likelihood estimator \parencite[CHAPTER 7]{nagaoka2005asymptotic}.

In quantum mechanics, the probability distribution is given by the Born rule $p(x|\bvar)=\tr(\rho_\bvar M_x)$. Here, I present the derivation of quantum Cram\'er-Rao bound, based on direct optimization over locally unbiased observable (based on observation from \parencite{ragy2016}). The advantage of such derivation is that it clearly shows the origin of the measurement incompatibility problem. Moreover, it allows quick improvement to the stronger bounds \parencite{Holevo1982,demkowicz2020multi}.

For compact notation, I will denote the vectors of the operators by properly bounded letters. For fixed $\bvar$ let me defined:
\begin{equation}
\label{bx}
    \bX=\sum_x(\tilde\bvar(x)-\bvar)M_x,
\end{equation}
i.e., $\bX=[X_1,...,X_p]^T$, where $X_i=\sum_x(\tilde\var_i(x)-\var_i)M_x$. We construct a positive semidefinite operator:
\begin{equation}
    0\leq \sum_x [(\tilde\bvar(x)-\bvar)\openone-\bX]M_x[(\tilde\bvar(x)-\bvar)\openone-\bX]^T=\sum_x M_x[(\tilde\bvar(x)-\bvar)][(\tilde\bvar(x)-\bvar)]^T-\bX\bX^T.
\end{equation}
The first inequality is tight if $\{M_x\}$ is a protective measurement. After applying the trace with density matrix $\tr(\rho_\bvar\cdot)$ we obtain:
\begin{equation}
\label{XZ}
    \Sigma\geq \tr(\rho_\var\bX\bX^T)=:Z_\var[\bX]
\end{equation}
Therefore, for any cost function $\C$, the average cost is bounded by:
\begin{equation}
\label{boundz}
    \ptr(\C\Sigma)\geq \ptr(\C Z_\var[\bX]).
\end{equation}
RHS may be directly minimized over all vectors of hermitian matrices $\bX$ satisfying locally unbiased conditions $\tr(\partial_{i}\rho_\bvar X_j)=\delta_{ij}$ (we ignore for a while the fact that all $X_i$ should come from the same measurement $\{M_x\}$). Minimization leads to\footnote{This may be shown by applying the standard Lagrange multipliers method \parencite{ragy2016}. An elegant alternative approach, based on introducing a proper $\rho_\bvar$-dependent scalar product on the space of operators, is presented in \parencite[Section 2.6]{demkowicz2020multi}.} $\bX=\bF_Q^{-1}\bL$, where $\bL=[L_1,...,L_p]^T$ is a vector of symmetric logarithmic derivatives:
\begin{equation}
    \partial_i\rho_\bvar=\frac{1}{2}(L_i\rho_\bvar+\rho_\bvar L_i)
\end{equation}
and $\bF_Q$ is the quantum Fisher information matrix given as:
\begin{equation}
    \bF_Q=\t{Re}\tr(\rho_\bvar \bL\bL^T).
\end{equation}
Therefore, the cost is bounded from below:
\begin{equation}
\label{multicrc}
    \ptr(C\Sigma)\geq \ptr(C\bF_Q^{-1}).
\end{equation}
What may be surprising, the $\bX$ minimizing \eqref{boundz} does not depend on peculiar matrix cost. As the inequality holds for any $\C$, it may be written as matrix inequality:
\begin{equation}
    \Sigma\geq \bF_Q^{-1}.
\end{equation}
Let us discuss the saturability of the above inequality now.
I start with a specific case when the cost matrix $\C$ is rank-one. The problem may be understood that there is only one parameter to be measured\footnote{In general it may be some concrete linear combination of the parameters. However, the cost matrix may be always diagonalized by performing a proper rotation in parameter space (see a discussion in \secref{sec:reparametr}). In the new parametrization, this linear combination is indeed a single parameter.}, but the experimentalist needs to make sure that fluctuations of others do not affect the results (there are called nuisance parameters in this context \parencite{suzuki2020nuisance,suzuki2020quantum}). Then, if only $i^{\t{th}}$ parameter is to be measured, directly from above construction, the optimal measurement will be a projection onto eigenvectors of $X_i=[\bF_Q^{-1}\bL]_i$ (which satifies $\forall_j\tr(\partial_{j}\rho_\bvar X_i)=\delta_{ij}$) and one can immediately see that:
\begin{equation}
    \Delta^2\tilde\var_i=\tr(\rho_\bvar X_i^2)=\tr(\rho_\bvar [\bF_Q^{-1}\bL]_i[\bL^T\bF_Q^{-1}]_i)=\left[\bF_Q^{-1}\tr(\rho_\bvar \bL\bL^T)\bF_Q^{-1}\right]_{ii}=   
    [\bF_Q^{-1}]_{ii},
\end{equation}
where in last step I used the fact that $[\bF_Q^{-1}\t{Im}(\tr(\rho_\bvar \bL\bL^T))\bF_Q^{-1}]_{ii}=0$. This clearly shows that for the rank-one cost matrix, the QCR is always saturable with measurement performed on single copies of the system\footnote{This is a stronger version of the statement from \parencite[Sec. 2.7]{demkowicz2020multi}, where asymptotical saturability was shown via relation with Holevo CR bound (with assumption that we allow collective measurement on many copies of the system).}.

For general cost matrix $\C$, the question about saturability comes down to the question if there exists a single protective measurement $\{M_x\}$, for which all 
$X_i=[\bF_Q^{-1}\bL]_i$ may be written as \eqref{bx}. If so, the classical Fisher information for this measurement is equal to $\bF_Q$, and in the many repetitions scenario, one may use an argument about the asymptotic saturability of classical CR.

Such measurement always exists if all SLDs mutually commute 
\begin{equation}
\label{commute}
\forall_{ij}[L_i,L_j]=0,
\end{equation}
then the such measurement is simply a projection onto a common eigenbasis of all $L_i$. That will be the case for most of the examples discussed in this thesis. However, for completeness, let me also discuss the opposite situation.

If $\exists_{ij}[L_i,L_j]\neq0$, the situation becomes to be more complicated; however, some improvement of \eqref{multicrc} may be performed. Note that matrix $Z_\bvar[\bX]$ in \eqref{XZ} in general may be complex, while covariance matrix $\Sigma$ is always purely real. As $Z_\bvar[\bX]$ is hermitian, its imaginary part is antisymmetric. Therefore, after tracing both sides of \eqref{XZ} with a real positive cost matrix, we completely ignore the information hidden in $\t{Im}Z_\var[\bX]$, as $\ptr(\C\t{Im}Z_\var[\bX])=0$. We can avoid this problem by, instead of minimizing RHS of \eqref{multicrc} directly, performing double minimization:
\begin{equation}
    \ptr(\C\Sigma)\geq \min_{V,\bX}\left[\ptr(\C V): V\geq\ptr(\rho_\var\bX\bX^T)\right],
\end{equation}
with $V$ being symmetric real matrix. Moreover, the optimization over $V$ for fixed $\bX$ may be performed analytically, giving so called Holevo Cram\'er-Rao (HCR) bound:
\begin{equation}
    \ptr(\C\Sigma)\geq \min_{\bX}\left[\ptr(\C Z[\bX])+\|\sqrt{C}\t{Im}Z[\bX]\sqrt{\C}\|_1\right].
\end{equation}
The last inequality has been proven to be asymptotically saturable for many copies of the system if collective measurements on all copies are allowed, i.e., $p(x|\bvar)=\tr(M_x \rho_\bvar^{\otimes k})$ (see \parencite{demkowicz2020multi} for both general proof and simple examples).

From that, we can see that if collective measurements are allowed, the condition for saturability \eqref{commute} may be weakened -- indeed, if SLDs do not commute in general, but commute on state, i.e.
\begin{equation}
\label{weakcommute}
    \tr(\rho_\var[L_i,L_j])=0,
\end{equation}
then for $\bX=\bF_Q^{-1}\bL$, the operator $\t{Im}Z_\bvar[\bX]=0$, so Holevo-Cramer Rao coincidence with standard CR, which means that also the latter one is saturable. Therefore, condition \eqref{weakcommute} is a necessary and sufficient condition for asymptotically saturating standard multiparameter CR bound with collective measurement. Moreover, in \parencite{matsumoto2002new}, it has been proven that collective measurements are necessary only for the general case of mixed states. In contrast, for the pure state, HCR may be saturated with standard local measurements on single copies of the system.


\subsectionnull{Adaptive scheme maximizing QFI}
\label{sec:adamulti}

In the single parameter case, it was shown for the unitary estimation that the optimal adaptive scheme is a sequential one acting of the gates $n$ times, with no necessity of unitary control between (see \secref{sec:unitary}). Moreover, it offers no advantage over the optimal parallel scheme.

However, this no longer applies to multiparameter estimation, where the optimal unitary control is the one that reverses evolution after each step. Moreover, in some cases, it may overcome the optimal parallel scheme \parencite{yuan2016}. The reason for the difference is that in the multiparameter case, the gates with different values of parameter do not necessarily commute $[U_{\bvar_0},U_{\bvar}]\neq 0$, which makes the reversing evolution between each gates something diametrically different from reversing it at the end of the whole procedure. The exemplary problem illustrating this effect will be discussed in \secref{sec:mf}.

The fact, that sequential-adaptive strategy where unitary controls simply reverse the evolution is the best of all those using $N$ gates, was proven in  \parencite{yuan2016} only for a specific case of estimation of the magnetic field sensed by spin-$1/2$ particle; the proof was based on the connection of QFI with Bures metric. Below I present an alternative and fully general proof of this statement (which, best to my knowledge, does not exists in literature). Note, that since any entangled-parallel strategy may be seen as a specific adaptive-sequential one (as discussed in \secref{sec:schemes}), in the proof of optimality I may restrict only to the adaptive-sequential ones.

It should be emphasized that the following reasoning is based only on comparing QFI matrices, so it does not consider the potential incompatibility of the measurements. The question of whether an analogous protocol optimizes the HCR inequality has not been investigated.

\textbf{Theorem 3.} In the multiparameter unitary estimation problem with evolution $U_\bvar$ at point $\bvar=\bvar_0$, 
for any sequential-adaptive strategy $\ket{\psi_\bvar}=V_n (U_\bvar\otimes \openone)...V_1 (U_\bvar\otimes \openone)\ket{\psi}$, there exists an alternative strategy with $\forall_i V_i=U^\dagger_{\bvar_0}\otimes\openone$ and input state $\ket{\psi'}\in\mH_S\otimes\mH_A$, where an ancillary system of the same size as the original one, $\dim(\mH_S)=\dim(\mH_A)$, for which the QFI matrix is bigger or equal to the QFI for the first strategy:
\begin{equation}
  \exists_{\ket{\psi'}\in \mH_S\otimes\mH_A}\quad\bF_Q\Big[(U^\dagger_{\bvar_0}U_\bvar\otimes \openone)^{\otimes n}\ket{\psi'})
  \Big]\geq
    \bF_Q\Big[V_n (U_\bvar\otimes \openone)...V_1 (U_\bvar\otimes \openone)\ket{\psi})\Big].
\end{equation}

\textit{Proof.} For any adaptive strategy, the output state after $n$ steps is given as:
\begin{equation}
  \rho_\bvar=V_n(U_\bvar\otimes\openone) V_{n-1} (U_\bvar\otimes\openone) ... V_1 (U_\bvar\otimes\openone) \rho (U_\bvar\otimes\openone)^\dagger V_1^\dagger...(U_\bvar\otimes\openone)^\dagger V_n^\dagger
\end{equation}
and its QFI matrix is:
\begin{equation}
  \bF_Q=\tr(\rho_\var\bL\bL^T),\quad \partial_{j}\rho_\bvar\big|_{\bvar_0}=\frac{1}{2}(\rho_{\bvar_0}L_j+L_j\rho_{\bvar_0}).
\end{equation}
I introduce $W_i=V_i (U_{\bvar_0}\otimes\openone)$ (so $V_i=W_i (U_{\bvar_0}\otimes\openone)^\dagger$). Then:
\begin{equation}
  \rho_{\bvar}\Big|_{\bvar_0}=W_nW_{n-1}... W_1 \rho W_1^\dagger...W_n^\dagger.
\end{equation}
while its derivatives may be written in the form of the sum of terms when the derivative "hit" $i^{th}$ acting of the channel:

\begin{equation}
  \partial_{j}\rho_\bvar\Big|_{\bvar_0}=\sum_{i=1}^{n} \partial_j W_n...W_{i}U^\dagger_{\bvar_0}U_\bvar\rho^{(i)}U_\bvar^\dagger U_{\bvar_0}W^\dagger_{i}...W^\dagger_n\Big|_{\bvar_0},\quad \t{with}\quad \rho^{(i)}:=W_{i-1}W_{i-2}...W_1\rho W_1^\dagger...W_{i-1}^\dagger.
\end{equation}
We define $L_j^{(i)}$ as SLD  corresponding to $i^{th}$ "hitting":
\begin{equation}
  \partial_jW_n...W_{i}U^\dagger_{\bvar_0}U_\bvar\rho^{(i)}U_\bvar^\dagger U_{\bvar_0}W^\dagger_{i}...W^\dagger_n=\frac{1}{2}(L_j^{(i)}\rho_{\bvar_0}+\rho_{\bvar_0} L_j^{(i)})
\end{equation}
and then full SLD is simply sum of $L_j^{(i)}$:
\begin{equation}
  L_j=\sum_{i=1}^n L_j^{(i)}.
\end{equation}
Now I will show that at least the same QFI matrix may be obtained by preparing the state being a mixture of $\rho^{(i)}$ entangled with additional ancilla of the size $\dim(\mH_{A'})=n$:
\begin{equation}
  \rho'=\frac{1}{n}\sum_{i=1}^n \rho^{(i)}\otimes \ket{i}\bra{i},
\end{equation}
on which we act:
\begin{equation}
  (U^\dagger_{\bvar_0}U_\bvar)^n \otimes\openone
\end{equation}
so
\begin{equation}
  \rho'_{\bvar}=\frac{1}{n}\sum_{i=1}^n[(U^\dagger_{\bvar_0}U_\bvar)^n \otimes\openone] \rho^{(i)}[(U^\dagger_{\bvar_0}U_\bvar)^n \otimes\openone]^\dagger \otimes \ket{i}\bra{i}.
\end{equation}
To show that, I introduce additional matrix $\rho''_{\bvar}$, constructed by acting on $\rho'_{\bvar}$ with unitary  control
\begin{equation}
  \sum_{i=1}^n W_n...W_{i}\otimes \ket{i}\bra{i}.
\end{equation}
As it is $\bvar$-independent unitary, it does not change the QFI (so $\bF_Q'=\bF_Q''$), but the SLDs of $\rho_\bvar$ and $\rho_\bvar''$ will be much simpler to compare. We have:
\begin{multline}
  \partial_{j}\rho''_\bvar\Big|_{\bvar_0}=\frac{1}{n}\sum_{i=1}^n \partial_j W_n...W_{i}
  \left[(U^\dagger_{\bvar_0}U_\bvar)^n \otimes\openone\right]
  \rho^{(i)}
  \left[(U^\dagger_{\bvar_0}U_\bvar)^n \otimes\openone\right]^\dagger
  W^\dagger_{i}...W^\dagger_n\otimes\ket{i}\bra{i}\Big|_{\bvar_0}=\\
  \frac{1}{\cancel{n}}\sum_{i=1}^n \cancel{n}\partial_j W_n...W_{i}
  \left[(U^\dagger_{\bvar_0}U_\bvar) \otimes\openone\right]
  \rho^{(i)}
  \left[(U^\dagger_{\bvar_0}U_\bvar) \otimes\openone\right]^\dagger
  W^\dagger_{i}...W^\dagger_n\otimes\ket{i}\bra{i}\Big|_{\bvar_0},
\end{multline}
so $\rho''_\bvar$ satisfies:
\begin{equation}
  \partial_j\rho''_\bvar\Big|_{\bvar=0}=\frac{1}{2}(L''_j\rho''_0+\rho''_0L''_j)
\end{equation}
with
\begin{equation}
  \bL^{''}=\sum_{i=1}^n n \bL^{(i)}\otimes \ket{i}\bra{i}.
\end{equation}
Therefore:
\begin{equation}
    \bF_Q''=\t{Re}\tr(\rho''_{\bvar_0}\bL^{''}\bL^{''T})=n\sum_{i=1}^n\t{Re}\tr(\rho_{\bvar_0}\bL^{(i)}\bL^{(i)T}) 
\end{equation}
In general, for any hermitian operators:
\begin{equation}
  n\left(\sum^n \bL^{(i)}\bL^{(i)T}\right)-\left( \sum_{i=1}^n \bL^{(i)}\right)\left( \sum_{i=1}^n \bL^{(i)}\right)^T=\frac{1}{2}\sum_{i,j=1}^n\left(\bL^{(i)}-\bL^{(j)}\right)\left(\bL^{(i)}-\bL^{(j)}\right)^T\geq 0,
\end{equation}
which, after applying $\t{Re}\tr(\rho_{\bvar_0}\cdot)$ gives:
\begin{equation}
  \bF''_Q-\bF_Q\geq 0 \Leftrightarrow \bF'_Q-\bF_Q\geq 0.
\end{equation}

To perform the above theoretical construction, one needs an extremely large ancilla. Let me now argue that at least the same QFI matrix may be obtained with the ancilla of the same size as the physical system $\mH_S$. Consider a purification of $\rho'$, namely
$\ket{\psi'}\in \mH_S\otimes \mH_A\otimes \mH_{A'}\otimes \mH_p$ (where $\dim(\mH_p)=\dim(\mH_S\otimes \mH_A\otimes \mH_{A'})$. It is clear that the QFI matrix of $\ket{\psi'_\bvar}=(U_{\bvar_0}U_\bvar)^n \otimes\openone\ket{\psi'}$ is larger or equal to QFI matrix of $\rho_\bvar$, as the second one may be obtained from the first by tracing over $\mH_p$. On the other hand, from Schmidt decomposition:
\begin{equation}
  \ket{\psi'}=\sum_{k=1}^{\dim(\mH_S)}\eta_k \underbrace{\ket{\psi'_{k,S}}}_{\in\mH_S}\otimes \underbrace{\ket{\psi'_{k,A}}}_{\in \mH_A\otimes \mH_{A'}\otimes \mH_p},
\end{equation}
so it is enough to consider $\ket{\psi'}\in \mH_S\otimes \t{span}\{\ket{\psi'_{k,A}}\}_{k=1}^{\dim(\mH_S)}$, what was to be proven. $\square$

It also shows that if, alternatively, $\forall[U^\dagger_{\bvar_0},U^\dagger_{\bvar}]=0$, no unitary feedback is needed, and the optimal controls may be chosen as $\openone$. That also implies that for such cases, the general sequential adaptive scheme offers no advantage over the optimal parallel one.

Indeed, let $\{\ket{i}\}$ be the common eigenbasis of all generators $\G_k$, such that $\G_k\ket{i}=\lambda_i^{(k)}$. Then $n$ sequential acting on the state:
\begin{equation}
  U_\bvar^n \sum_{i=1}^d c_i \ket{i}=\sum_{i=1}^d e^{in\sum_k \var_k\lambda_i^{(k)}}c_i \ket{i}
\end{equation}
is equivalent to:
\begin{equation}
  U_\bvar^{\otimes n} \sum_{i=1}^d c_i \ket{i}^{\otimes n}=\sum_{i=1}^d e^{in\sum_k \var_k\lambda_i^{(k)}}c_i \ket{i}^{\otimes n}.
\end{equation}

\subsectionnull{General theorem for noisy multiparameter estimation within QFI formalism}

The “Hamiltonian-not-in-Kraus-span” (HNKS) condition \eqref{HNKS} may be easily generalized for a multiparameter case. It was first done for the problem of continuous evolution given by master Lindblad equation \parencite{Gorecki2020}, but methodology may be easily adapted to finite-step gates:
\begin{equation}
 \ch_{\bvar}(\rho)=\sum_{k=0}^r K_k(\bvar) \rho K_k^\dagger(\bvar).
\end{equation}
For each parameter we define corresponding Hamiltonian $\tilde \G_i$ and extended Kraus-span $\mathcal S_i$ (where compered to $\mathcal S$, it additionally contain all remaining Hamiltonian $\tilde \G_{j\neq i}$):
\begin{equation}
\tilde \G_i=-i\sum_k K_k^\dagger \partial_i K_k,\quad \mathcal S_i=\t{span}_{\mathbb H}\{\tilde \G_{j\neq i}, K_j^\dagger K_k\, \t{for all},\ j,k\},
\end{equation}
where  $\t{span}_{\mathbb H}\{\cdot\}$ represents all Hermitian operators that are linear combinations of operators in $\{\cdot\}$.
Then Heisenberg scaling in estimation off all $p$ parameters is possible iff $\forall_i \tilde \G_i\notin \mathcal S_i$. When this condition is satisfied, each parameter may be measured separately by constructing quantum error correction protocol as in single-parameter case \parencite{zhou2021asymptotic}, where remaining Hamiltonians $\tilde \G_{j\neq i}$ are treated as the noise we want to protect again. Otherwise, if the condition is violated, it means that there exists some linear combinations of parameter $\var_i$, which cannot be estimated with HS precision.

While if and only if conditions for obtaining HS are known, the general asymptotically saturable bound is not known exactly. The methodology used to derive the saturable bounds in single-parameter estimation, presented in \secref{sec:singlenoisy}, cannot be trivially extended for multiparameter case, mainly due to the difficulties connected with inversing QFI matrix and probe incompatibility. See \parencite{albarelli2022probe} for recent progress in this direction.

\section{Bayesian and minimax approaches}
\label{sec:bayminmulti}
Similarly, as in the single parameter case, we may formulate formulas for Bayesian cost:
\begin{equation}
\label{multibayes}
   \cb=\int p(\bvar) d\bvar \int d\tilde\bvar \tr(\rho_\bvar\M_{\tilde\bvar})\cost(\bvar,\tilde \bvar),
\end{equation}
and and minimax cost:
\begin{equation}
\label{multiminimax}
   \cm=\sup_{\bvar\in\Theta}\int d\tilde\bvar \tr(\rho_\bvar\M_{\tilde\bvar})\cost(\bvar,\tilde \bvar).
\end{equation}
Again, if $\t{supp}\, p(\bvar)\in\Theta$, the minimax cost bigger than Bayesian one:
\begin{equation}
   \cm\geq \cb.
\end{equation}

In the limit of many copies $k$, if collective measurement is allowed, in general, Bayesian and minimax formulas may be related to Holevo Cram\'er-Rao bound (so also with standard Cram\'er-Rao bound if $\tr(\rho_\bvar[L_i,L_j])=0$), however, the formal reasoning is much more complicated and proven fully formally only for specific cases 
\parencite{gill1995,gill2008,guta2006local,guta2008optimal}.

\subsectionnull{Covariant problem}
\label{sec:covadapt}

In this section, we recall the theorem introduced and proven in \parencite{chiribella2008memory}, saying that for the covariant problem of the group element estimation, the optimal strategy may be found as the parallel covariant one, i.e., there is no advantage in considering a more general sequential adaptive scheme. The proof from \parencite{chiribella2008memory} is based on the formalism of quantum combs introduced in \parencite{chiribella2008architecture}. For completeness, I recall here the proof, restricting mathematical formalism only to what is strictly necessary.

Consider a problem where the family of channels depending on the parameter is a unitary representation of a compact Lie group $G$, such that $\ch_g(\rho)=U_g\rho U^\dagger_g$ and the aim is to estimate the group element $g\in G$. Then we say, that the estimation problem is covariant if it satisfies the conditions stated in \secref{sec:covariant}, namely:
\begin{itemize}
    \item the cost function is invariant under the action of the group $\cost(g_1,g_2)=\cost(hg_1,hg_2)$.
    \item (required in Bayesian formalism) a priori distribution is invariant under the action of the group $p(g)dg=p(hg)d(hg)$. For simplicity of the notation, I assume that $dg$ is the normalized Haar measure, $\int_G dg=1$.
\end{itemize}
We can now formulate the stronger version of \eqref{covariantbound}.

\textbf{Theorem 4.} For the covariant group estimation problem, as the optimal strategy using $N$ gates one may always choose the parallel one with covariant measurement, i.e. the one of the form:
\begin{equation}
    \M_g=({U_{\tilde g}}^{\otimes N}\otimes\openone)\M_e ({U^\dagger_{\tilde g}}^{\otimes N}\otimes\openone),\quad 
    \int d\tilde g ({U_{\tilde g}}^{\otimes N}\otimes\openone)\M_e ({U^\dagger_{\tilde g}}^{\otimes N}\otimes\openone)=\openone.
\end{equation}
It cannot be overcome even by the most general sequential-adaptive scheme, namely:
\begin{multline}
\begin{drcases*}
\inf_{\rho, \{V_i\}, \M_{\tilde\bvar}}\cm=
\inf_{\rho, \{V_i\}, \M_{\tilde\bvar}}\sup_{g\in G}\int d\tilde\bvar \tr(\rho_g^N \M_{\tilde g})\cost(g,\tilde g) \\
\inf_{\rho, \{V_i\}, \M_{\tilde\bvar}}\cb=
\inf_{\rho, \{V_i\}, \M_{\tilde\bvar}}\int d\tilde\bvar \tr(\rho_g^N \M_{\tilde g})\cost(g,\tilde g)
\end{drcases*}=\\
 \inf_{\rho\in \lin(\mathcal H_S^{\otimes N}\otimes \mathcal H_A), \M_e}\int d\tilde g \tr(
    \rho ({U^\dagger_{\tilde g}}^{\otimes N}\otimes\openone)\M_e (U_{\tilde g}^{\otimes N}\otimes\openone))\cost(e,\tilde g),
\end{multline}
where in LHS $\rho^N_g=V_N\circ (U_g\otimes\openone)\circ...V_1\circ (U_g\otimes\openone)\circ \rho$.

\textit{Proof.} Consider an arbitrary quantum channel:
\begin{equation}
   \ch: \lin(\mathcal H_{\t{in}}) \to \lin(\mathcal H_{\t{out}}),
\end{equation}
where in principle $\mathcal H_{\t{in}}$ and $\mathcal H_{\t{out}}$ may correspond to the same physical system, yet in this formalism, they will be treated as separated Hilbert spaces.
From Choi-Jemiołkowski isomorphism, the channel is uniquely defined by its Choi matrix:
\begin{equation}
    E=\t{Choi}(\ch):=\ch\otimes \openone(\kett{\Omega}\bbra{\Omega})
\end{equation}
where $\kett{\Omega}=\sum\ket{i}\ket{i}\in\mathcal H_{{\t{in}}}\otimes\mathcal H_{{\t{in}}}$. Indeed, by direct calculation, one may see:
\begin{equation}
    \tr_{\t{in}}(E(\openone\otimes \rho^T))=\tr_{\t{in}}(\sum_{ij}\ch(\ket{i}\bra{j}) \otimes\ket{i}\bra{j}\rho^T)=\sum_{ij}\ch(\ket{i}\bra{j})\rho_{ji}^T=\ch(\rho).
\end{equation}

Moreover, to introduce consistent notation, also density matrix $\rho$ itself as well as each POVM element $M_i$ will be seen as quantum channels, respectively:
\begin{equation}
   \begin{split}
       &\rho: \mathbb C \to \mathcal H_{\t{in}}: 1\mapsto \rho,\\
       &\M_i: \mathcal H_{\t{out}} \to \mathbb C: \rho\mapsto \tr(M_i\rho).
   \end{split}
\end{equation}
Especially, Choi matrix of $\rho$ is simply $\rho$, while Choi matrix of $M_k$ is its transposition, as
\begin{equation}
   M_k\otimes\openone(\kett{\Omega}\bbra{\Omega})=\sum_{ij}\tr(\M_k\ket{i}\bra{j})\ket{i}\bra{j}=M^T.
\end{equation}
By $\EV{}$ I will denote the Choi matrix of the unitary channel $V$, i.e. $\EV{}=(V\otimes \openone)\kett{\Omega}\bbra{\Omega}(V^\dagger\otimes \openone)$.

Next, let us derive the formula for the Choi matrix of the composition of the channels. Consider now two channels, for which the output of the first one is the input of the second:
\begin{equation}
\begin{split}
    \ch_1: \lin(\mathcal H_a)\to \lin(\mathcal H_b),\\
    \ch_2: \lin(\mathcal H_b)\to \lin(\mathcal H_c).
    \end{split}
\end{equation}
By direct calculation, we derive the formula for the Choi matrix of their composition:
\begin{multline}
    \ch_2(\ch_1(\rho))=\tr_{b}(E_2(\openone_{c}\otimes [\tr_{a}(E_1(\openone_{b}\otimes \rho^T))]^T))=\\
    \tr_{a}\left(
    \tr_{b}\left(
    (E_2\otimes \openone_{a})\left[(\openone_{c}\otimes E_1)(\openone_{c}\otimes \openone_{b}\otimes \rho^T)\right]^{T}
    \right)
    \right)=\\
     \tr_{a}\left(
     \underbrace{
    \tr_{b}\left(
    (E_2\otimes \openone_{a})(\openone_{c}\otimes E_1^{T_{b}})\right)}_{E_{2,1}}
    (\openone_{c}\otimes \rho^T)
        \right)
\end{multline}
where in the last step, we used the fact that trace over subspace $\tr_{a}$ is invariant for partial transposition on this subspace, so $\tr_{a}([E_1(\openone_{b}\otimes \rho^T)]^T)=\tr_{a}([[E_1(\openone_{b}\otimes \rho^T)]^T]^{T_{a}})=\tr_{a}([E_1(\openone_{a}\otimes \rho^T)]^{T_{b}})=\tr_{a}([E_1^{T_{b}}(\openone_{a}\otimes \rho^T)])$.

We can go further: consider the channels with an arbitrary number of input and output Hilberts spaces, with the assumption that some of the output of $\ch_1$ is the input of $\ch_2$, for example:
\begin{equation}
    \ch_1: \lin(\mathcal H_a\otimes \mathcal H_b)\to \lin(\mathcal H_c\otimes \mathcal H_d\otimes \mathcal H_f)
\end{equation}
\begin{equation}
    \ch_2: \lin(\mathcal H_f\otimes \mathcal H_g)\to \lin(\mathcal H_h)
\end{equation}
Then Choi matrix of their composition $(\openone_{c,d}\otimes\ch_2) \circ(\ch_1\otimes\openone_{g})$ is given as:
\begin{equation}
    E_{2,1}=\tr_f((\openone_{c,d}\otimes E_2) (E_1\otimes\openone_{g})^{T_f}).
\end{equation}
which we call link product $E_2*E_1$. Note that, as partial trace $\tr_f(\cdot)$ is invariant for partial transposition $\cdot^{T_f}$, the link product is commutative $E_2*E_1=E_1*E_2$.

Note that if two channels do not have any common input-output, the Choi matrix of their composition is simply a tensor product of their Choi matrices $E_1*E_2=E_1\otimes E_2$.

We are now ready to use this formalism to describe the adaptive scheme \ref{fig:schemes}(c). Here $\mH_{\t{in}}\simeq\mH_{\t{out}}\simeq \mH_S\otimes \mH_A$. Moreover, we incorporate the $N^{th}$ unitary gate $V_N$ into the measurement to get a more compact notation. The probability of output $x$ is then given as:
\begin{equation}
\begin{split}
p(x|\bvar)&=\\
&=
\tr(M_x((\ch\otimes \openone)\circ V_{N-1} \circ (\ch\otimes \openone)\circ ... V_{N-2}...(\ch\otimes \openone)\circ\rho))\\
  &=
  M_x^T*(E*\kett{\Omega}\bbra{\Omega})*\EV{N-1}*...\EV{1}*(E*\kett{\Omega}\bbra{\Omega})*\rho\\
  &\overset{(1)}{=}
   (E*\kett{\Omega}\bbra{\Omega}...*E*\kett{\Omega}\bbra{\Omega})*(M_x^T*\EV{N-1}*...*\EV{1}*\rho)\\
   &\overset{(2)}{=}
   (E\otimes \kett{\Omega}\bbra{\Omega})^{\otimes N}*(M_x^T\otimes \EV{N-1}\otimes...\otimes \EV{1}\otimes\rho)\\
   &\overset{(3)}{=}
   \tr((E\otimes \kett{\Omega}\bbra{\Omega})^{\otimes N}(M_x^T\otimes \EV{N-1}\otimes...\otimes \EV{1}\otimes \rho)^T)\\
    &=
   \tr((E\otimes \kett{\Omega}\bbra{\Omega})^{\otimes N}(M_x\otimes \EV{N-1}^T\otimes...\otimes \EV{1}^T\otimes \rho^T)),
   \end{split}
\end{equation}
where in $(1)$, I used the commutative of link product $*$, in $(2)$, the fact that I link only channels with no common input-output space, while in $(3)$ applied the composition formula.

Finally, we may perform the partial trace of all $N$ ancillae to get a generalized Born rule for the probability of getting measurement result $x$:
\begin{equation}
   p(x|\bvar)=\tr(DP_x),\quad \t{with} \quad D=E^{\otimes N},\, P_x=\tr_{\mH_A^{\otimes N}}\left((\openone\otimes \kett{\Omega}\bbra{\Omega})^{\otimes N}(M_x\otimes... \EV{i}^T...\otimes \rho^T)\right),
\end{equation}
where $P_x$ is called a tester, including all information about the measurement strategy.

From definition of POVM, the set of testers $\{P_x\}$ satisfies $\sum_x P_x=\openone\otimes \Xi^{(N)}$, while $\Xi^{(N)}$ is called normalization operator. These allow us for define $\tilde P_x$, $\tilde D$:
\begin{equation}
\begin{split}
    P_x=\left(\openone\otimes \sqrt{\Xi^{(N)}}\right)\tilde P_x \left(\openone\otimes \sqrt{\Xi^{(N)}}\right),\\
    \tilde D=\left(\openone\otimes \sqrt{\Xi^{(N)}}\right)D \left(\openone\otimes \sqrt{\Xi^{(N)}}\right),
    \end{split}
\end{equation}
such that $p(x)=\tr(\tilde P_x\tilde D)$, where $\{\tilde P_x\}$ form POVM\footnote{Note, that we cannot formally write $\tilde P_x=(\openone\otimes \sqrt{\Xi^{(N)}})^{-1}P_x (\openone\otimes \sqrt{\Xi^{(N)}})^{-1}$, as in general $\openone\otimes \sqrt{\Xi^{(N)}}$ may has some zero eigenvalues. Still, as each $P_x\geq 0$, any zero-eigenvector of $\openone\otimes \sqrt{\Xi^{(N)}}$ is simultaneously zero-eigenvector of all $P_x\geq 0$, so proper $\tilde P_x$ always exists. In the case where $\openone\otimes \sqrt{\Xi^{(N)}}$ has some zero eigenvalues, formally, $\{\tilde P_x\}$ does not form a POVM on $(\mH_{\t{out}}\otimes\mH_{\t{in}})^{\otimes N}$, but it may be easily extended to it, adding as the last element the identity on $\t{ker}(\openone\otimes \sqrt{\Xi^{(N)}})$. That will not affect any further reasoning.}
and $\tilde D$ is the state (by construction is semi-positive and one-trace) on (in general abstract) Hilbert space $(\mH_{\t{out}}\otimes\mH_{\t{in}})^{\otimes N}$ (where $\dim(\mH_{\t{in}})=\dim(\mH_{\t{out}})=\dim(\mH_{S})$).

Now consider the case when the channel is given by unitary representation of the group $\ch_g(\rho)=U_g \rho U_g^\dagger$, so 
$E_g=(U_g\otimes\openone)\kett{\Omega}\bbra{\Omega}(U_g^\dagger\otimes\openone)$. In the same way, as in \secref{sec:covariant}, we justify that the optimal tester is a covariant one. Namely, for any tester $\{P'_g\}$ we may construct the corresponding covariant one by proper averaging:
\begin{equation}
    P_h=\mathcal U_h\left(\int_G dg \mathcal U^\dagger_g P'_{g} \mathcal U_g  \right)\mathcal U^\dagger_h, \quad \t{with}\quad \mathcal U_h=(U_h\otimes\openone)^{\otimes N},
\end{equation}
where $dg$ is the normalized Haar measure.

\begin{equation}
    P_h=(U_h\otimes\openone)^{\otimes N}P_e(U^\dagger_h\otimes\openone)^{\otimes N}.
\end{equation}

Next, we show that it may be realized in a parallel scheme with ancilla. Note that:
\begin{equation}
\label{crucialstep}
    [\openone\otimes \Xi^{(N)},(U_h\otimes\openone)^{\otimes N}]=
    \left[\int_G h(\mu)d\mu P_h,(U_h\otimes\openone)^{\otimes N}\right]=0
\end{equation}
Therefore the state $\tilde D$ may be written in the form:
\begin{multline}
    \tilde D=\left(\openone\otimes \sqrt{\Xi^{(N)}}\right)(U_h\otimes\openone)^{\otimes N}\kett{\Omega}\bbra{\Omega}(U^\dagger_h\otimes\openone)^{\otimes N}\left(\openone\otimes \sqrt{\Xi^{(N)}}\right)=\\
    (U_h\otimes\openone)^{\otimes N}\left(\openone\otimes \sqrt{\Xi^{(N)}}\right)\kett{\Omega}\bbra{\Omega}\left(\openone\otimes \sqrt{\Xi^{(N)}}\right)(U^\dagger_h\otimes\openone)^{\otimes N},
\end{multline}
so the optimal cost may be obtained by using the input state $\left(\openone\otimes \sqrt{\Xi^{(N)}}\right)\kett{\Omega}\bbra{\Omega}\left(\openone\otimes \sqrt{\Xi^{(N)}}\right)\in (\mH_S\otimes\mH_A)^{\otimes N}$ (with $\dim(\mH_S)=\dim(\mH_A)$) in the parallel strategy \figref{fig:schemes}(b), what was to be proven. $\square$

The natural question is if the above reasoning could be generalized for the noisy case. We can introduce the class of specific noisy channels for which the proof is still valid. Let us modify the unitary channel by adding classical fluctuation of the parameter itself, namely $\ch_{g}(\rho)=\int d h p(h) U_{gh}\rho U_{gh}^\dagger$ with arbitrary probability distribution $p(h)$. That covers, for example, the common problem of the phase-shift estimation with the occurrence of dephasing noise.
Note that adding such noise does not affect the crucial step of the proof \eqref{crucialstep}, so indeed the statement is still valid. However, this is not a case for arbitrary noise, so the proof cannot be trivially extended for the general case.

\chpt{Heisenberg limit in multiparameter metrology}
\label{ch:multihl}

\section{Introduction}

In this chapter, I will discuss the problem of the Heisenberg limit in multiparameter metrology, presenting the results from \parencite{gorecki2022multiparameter,gorecki2021multiple}. I will focus mainly on two aspects.

The first one is the potential advantage of estimating $p$ parameters jointly, when compared to estimating them separately, spending in total the same amount of resources (i.e., usage of quantum gates). The problem has been extensively discussed in the literature (mainly within QFI formalism). However, there were some doubts about the interpretation of the results\footnote{An alternative approach to quantifying the incompatibility between parameters without discussion the problem of diving resources, has been proposed in \parencite{ragy2016,albarelli2022probe}. However, it does not answer the problems I refer to, so it will not be discussed broader in this thesis.}.

The central ambiguity concerns the issue of resource allocation between parameters when they are measured separately. It has been assumed in many works that, having $n$ gates available in a single implementation, they should be immediately divided between $p$ parameters~\parencite{humphreys2013quantum,yuan2016, baumgratz2016,YINGWANG2018250,goldberg2020}. In the presence of Heisenberg scaling, it would lead to:
\begin{equation}
\sum_{i=1}^p\Delta^2\tilde\var_i\propto p\times \frac{1}{k}\times \frac{1}{(n/p)^2}=\frac{p^3}{kn^2}.
\end{equation}
Based on these assumptions, a significant advantage of the strategy, where the parameter are measured jointly, was postulated (typically of order $\sim p$).

However, as QCR bound is guaranteed to be saturable only in the limit $k\to\infty$, so to use this formalism, we need to assume large $k$. Therefore it is much more efficient to divide the number of repetition $k$ between parameters, keeping $n$ gates in every single trial~\parencite{yousefjani2017estimating,chen2019,ho2020}:
\begin{equation}
\sum_{i=1}^p\Delta^2\tilde\var_i\propto p\times \frac{1}{k/p}\times \frac{1}{n^2}=\frac{p^2}{kn^2},
\end{equation}
which leads to $p$ times smaller final cost -- only due to more reasonable resources division, without the necessity of measuring parameters jointly. That makes questionable previously postulated advantages of a joint strategy.

The highly nontrivial question is how the situation looks when only the total amount of resources $N=kn$ is restricted. Then, due to quadratic scaling with $n$ and linear with $k$, the optimal choice would be $n=N,k=1$ -- but this cannot be fully analyzed within QFI formalism. In this chapter, I will discuss it in the minimax approach.

The second aspect I want to analyze is the advantage of a sequential adaptive scheme over a parallel one -- again, I will compare the results obtained in both paradigms.

\section{Asymptotic equivalence of the global and local minimax cost}

I will compare the minimal cost obtainable within two paradigms. In the first one, all $N$ resources (quantum gates) are used in a fully optimal way (which I will analyze within minimax formalism). In the second, there is a restriction that only $n$ gates may be used in a single trial, while the experiment is repeated a large number of times $k$ (which may be effectively described within QFI formalism).

While the first one demands defining a finite-sized set, in which the procedure is designed to work well $\Theta$, the second is strictly focused on a single point $\bvar_0$. To make a reasonable comparison, in a single parameter case in \chref{ch:baymin}, I have chosen $\Theta=[\var_0-\delta/2,\var_0+\delta/2]$ and I have written the exact bounds for finite region size $\delta$, showing, that its impact decreases with increasing $N$.

Here I use an analogous idea, but I will not investigate the rate of convergence. Instead, I focus only on asymptotic results. A simple approach, which captures the essential problem, has been proposed in~\parencite{hayashi2011} (originally for single parameter case), so-called local asymptotic minimax cost. Let us define $\delta-$neighborhood of $\bvar_0$ as $\Theta(\bvar_0,\delta)=\{\bvar: \forall_i \var_i \in [\bvar_{0i}-\delta/2,\bvar_{0i}+\delta/2]\}$. Consider the sequence of triples: (initial state, $N$ unitary controls and measurement), for each possible $N$, namely $(\rho^N, \{V_i^N\},\{M^N_{\tilde\bvar}\})$. Then for $N$-gates protocol the output state is given as $\rho_\bvar^N=V^N_N\circ (\ch_\bvar\otimes\openone)\circ...V^N_1\circ (\ch_\bvar\otimes\openone)\circ\rho^N$. By local asymptotic minimax cost around point $\bvar_0$ with cost matrix $C_{\bvar_0}$ we define:
\begin{equation}
\label{aslocmini}
   \inf_{\{(\rho^N, \{V_i^N\},\{M^N_{\tilde\bvar}\})\}}\lim_{\delta\to 0}\lim_{N\to\infty}N^2\sup_{\bvar\in\Theta(\bvar_0,\delta)}\int d\tilde\bvar \tr(\rho_\bvar^N \M_{\tilde\bvar}^N)\ptr(C_{\bvar_0}(\tilde\bvar-\bvar)(\tilde\bvar-\bvar)^T),
\end{equation}
Note that the order of taking limits $\lim_{\delta\to 0}\lim_{N\to\infty}$ is crucial here -- for the opposite order, the trivial constant estimator $\tilde\bvar=\bvar_0$ would lead to zero cost, while the current form corresponds to the analysis from  \chref{ch:baymin}, but without the necessity of performing calculations for finite $\delta/N$.
Now, I will generalize the reasoning from ~\parencite{hayashi2011} (originally performed for $U(1)$ estimation) for our case. 

\textbf{Theorem 5.} Given a channel $\ch_g(\rho)=U_g\rho U_g^\dagger$, being a unitary representation of the group $g\in G$. Let $\bvar=[\var_1,...,\var_p]^T$ be the local parametrization of the group element around some point $g_{\bvar_0}$. Let $\cost(g_\bvar,g_{\tilde\bvar})$ be the cost invariant for acting the group, and $C_\bvar$ is its Hessian with respect to variable $\tilde\bvar$, at point $\bvar$ (we also assume some basic regularity properties, which will be given later). Then, for the  most general adaptive scheme, the local asymptotic minimax cost is the same as the global asymptotic minimax cost (while the second one is a covariant problem):
\begin{multline}
\label{theorem}
\forall_{\bvar_0}\inf_{\{(\rho^N, \{V_i^N\},\{M^N_{\tilde\bvar}\})\}}\lim_{\delta\to 0}\lim_{N\to\infty}N^2 \sup_{\bvar\in\Theta (\bvar_0,\delta)}\int d\tilde \bvar \tr(M^N_{\tilde \bvar}\rho^N_{\bvar}) \ptr(C_{\bvar_0}(\tilde\bvar-\bvar)(\tilde\bvar-\bvar)^T)
\\
=\lim_{N\to\infty}N^2 \left(\inf_{(\rho^N, \{V^N_i\},\{M^N_{\tilde g}\})}\sup_{g\in G}\int d\tilde g \tr(\rho^N_{g}M^N_{\tilde g})\cost(g,\tilde g)\right)
\end{multline}

\textit{Proof.}
We start with the observation that the asymptotic minimax cost for $g\in G_\delta \subset G$ is for sure smaller or equal to the asymptotic minimax cost for $g\in G$:
\begin{multline}
\label{theoremp}
\inf_{\{(\rho^N, \{V_i^N\},\{M^N_{\tilde\bvar}\})\}}
\lim_{\delta\to 0}\lim_{N\to\infty}N^2 \sup_{g\in G_\delta}\int d\tilde g \tr(M_{N,\tilde g}\rho_{N,g}) \cost(g,\tilde g)
\\
\leq\lim_{N\to\infty}N^2 \left(\inf_{(\rho^N, \{V^N_i\},\{M^N_{\tilde g}\})}\sup_{g\in G}\int d\tilde g \tr(\rho^N_{g}M^N_{\tilde g})\cost(g,\tilde g)\right),
\end{multline}
where $G_\delta\subset G$ is $\delta-$neighborhood of the group neutral element $e$. I start by showing that they are in fact equal.

Similarly, as in the single parameter case, I introduce the notation for the minimax cost with a finite $\delta, N$:
\begin{equation}
\t{minimax}(G_\delta,N)=\inf_{(\rho^N, \{V^N_i\},\{M^N_{\tilde g}\})}\sup_{g\in G_\delta}\int d\tilde g \tr(\rho^N_{g}M^N_{\tilde g})\cost(g,\tilde g).
\end{equation}
In this notation, RHS of \eqref{theoremp} is simply $\lim_{N\to\infty}N^2\t{minimax}(G,N)$, while LHS may be bounded from below by $\lim_{\delta\to 0}\lim_{N\to\infty}N^2\t{minimax}(G_\delta,N)$ (as taking $\inf$ outside of limit may only increase the value of an objective function). It is also clear that
\begin{equation}
\label{trivial}
\lim_{N\to\infty}N^2\t{minimax}(G_\delta,N)\leq \lim_{N\to\infty}N^2\t{minimax}(G,N).
\end{equation}
What remains to be proven is the weak inequality in the opposite direction:
\begin{equation}
\lim_{N\to\infty}N^2\t{minimax}(G_\delta,N)\overset{?}{\geq}\lim_{N\to\infty}N^2\t{minimax}(G,N).
\end{equation}
I do it by the following reasoning.

Having in total $N$ gates, at first, I perform $\sqrt{N}$ independent single-gate experiments to find an approximated value $g_{\t{est}}$. The probability $p_{\t{err}}(\sqrt{N})$ that the true value does not belong to $\delta-$size neighborhood of the indication of estimator $g\notin g_{\t{est}}G_{\delta}$ (where $g_{\t{est}}G_{\delta}$ is the set $G_{\delta}$ shifted by the action of $g_{\t{est}}$), decreases exponentially with a number of measurements $\sqrt{N}$.

Then I spend remaining $N-\sqrt{N}$ gates to perform optimal estimation strategy for $g\in g_{\t{est}}G_{\delta}$. Since such a two-step strategy is, in general, only suboptimal, I have:
\begin{equation}
\label{prsub}
\t{minimax}(G,N)\leq p_{\t{err}}(\sqrt{N})c_{\max}+(1-p_{\t{err}}(\sqrt{N}))\t{minimax}(g_{\t{est}}G_{\delta},N-\sqrt{N}),
\end{equation}
where $c_{\max}=\max_{g,\tilde g}\cost(g,\tilde g)$. Moreover, due to the symmetry of the whole problem, the RHS does not depend on
$g_{\t{est}}$. After application of $\lim_{N\to\infty}N^2\cdot$ to the both sites, and the use of $\lim_{N\to\infty}\frac{(N-\sqrt{N})^2}{N^2}=1$, we obtain:
\begin{equation}
\lim_{N\to\infty}N^2\t{minimax}(G,N)\leq\lim_{N\to\infty}N^2(G_\delta,N-\sqrt{N})=\lim_{N\to\infty}N^2(G_\delta,N),
\end{equation}
which, together with \eqref{trivial} gives:
\begin{equation}
\lim_{N\to\infty}N^2\t{minimax}(G_\delta,N)= \lim_{N\to\infty}N^2\t{minimax}(G,N).
\end{equation}
Moreover, even if it was formulated for $G_\delta$ around neutral element $e$, due to covariant properties of the problem, it remains valid also for any $g_{\bvar_0}G_\delta$.

Now I will argue that replacing $\cost(g_{\bvar},g_{\tilde\bvar})$ by its Hessian in $\bvar_0$ does not change the minimal obtainable value. To show that we need to justify that for the protocol minimizing LHS of \eqref{theorem}, the probability of the gross error $\tilde\bvar-\bvar$ is negligible. I do it in the following way.

For any protocols sequence $\{(\rho^N, \{V_i^N\},\{M^N_{\tilde\bvar}\})\}$, for any finite $\delta_{\t{f}}$ we introduce following correction. If the result of measurement satisfies $\tilde\bvar\in \Theta(\bvar_0,\delta_{\t{f}})$, we leave it unchanged. However, if any $\tilde\var_i\notin [\var_{0i}-\delta/2,\var_{0i}+\delta/2]$, we change these result from $\tilde\var_i$ to $\var_{0i}$. For any sequence, such a change may only decrease the value of LHS of \eqref{theorem}, so for the optimal one, it does not change. Therefore, for the protocols locally optimal around $\bvar_0$ (in the sense of \eqref{aslocmini}), for any finite $\delta_{\t{f}}$, the probability of getting result outside of $\Theta(\bvar_0,\delta_{\t{f}})$ is negligible.

On the other hand, for sufficiently regular cost function:
\begin{equation}
\forall_\epsilon\exists_{\delta_{\t{f}}}\forall_{\bvar,\tilde\bvar \in \Theta(\bvar_0,\delta_{\t{f}})} |\cost(g_\bvar,g_{\tilde\bvar})-(\bvar-\tilde\bvar)^TC_{\bvar_0}(\bvar-\tilde\bvar)|\leq \epsilon \cdot (\bvar-\tilde\bvar)^TC_{\bvar_0}(\bvar-\tilde\bvar)=\epsilon\cdot\ptr(C_{\bvar_0}(\bvar-\tilde\bvar)(\bvar-\tilde\bvar)^T)
\end{equation}
and from that:
\begin{equation}
\forall_\epsilon\exists_{\delta_{\t{f}}}\forall_{\bvar,\tilde\bvar \in \Theta(\bvar_0,\delta_{\t{f}})}\frac{1}{1+\epsilon }\cost(\bvar,\tilde\bvar)\leq \ptr(C_{\bvar_0}(\bvar-\tilde\bvar)(\bvar-\tilde\bvar)^T)\leq\frac{1}{1-\epsilon}\cost(\bvar,\tilde\bvar).
\end{equation}
The reasoning may be repeated for any $\epsilon$, which proves the statement. $\square$

However, from \secref{sec:bayminmulti}, we know that for the covariant problem, the parallel covariant one is optimal among all (adaptive or not) protocols. We may finally state:
\begin{multline}
\label{proofresult}
\forall_{\bvar_0}\inf_{\{(\rho^N, \{V_i^N\},\{M^N_{\tilde\bvar}\})\}}\lim_{\delta\to 0}\lim_{N\to\infty}N^2 \sup_{\bvar\in\Theta_\delta}\int d\tilde \bvar \tr(M^N_{\tilde \bvar}\rho^N_{\bvar}) \ptr(C(\tilde\bvar-\bvar)(\tilde\bvar-\bvar)^T)
\\
=\lim_{N\to\infty}N^2 \inf_{\rho^N,M^N_e}\int d\tilde g \tr((U^\dagger_{\tilde g}\otimes\openone)^{\otimes N}M_e (U_{\tilde g}\otimes\openone)^{\otimes N}\rho)\cost(e,\tilde g),
\end{multline}

Further, I will be interested mainly in the examples where the Hessian is the identity matrix, so the figure of merit is the sum of MSEs $\ptr(C(\bvar-\tilde\bvar)(\bvar-\tilde\bvar)^T)=\sum_{i=1}^p \Delta^2\tilde\var_i$. I will analyze the optimal achievable precision for the cases where:
\begin{itemize}
    \item all $N$ resources are used optimally (analyzed within minimax formalism) or there is an additional constraint for the maximal amount of resources used in single trial $n$, which is repeated many times $k\to\infty$ (analyzed with the use of quantum Cram\'er-Rao bound).
    \item all parameters are measured jointly (JNT) in a fully optimal way, or from the very beginning, the resources are divided between parameters, which are measured separately (SEP).
\end{itemize}
These give four options, for which the sum of variances $\sum_{i=1}^p\Delta^2\tilde\var_i$ will be labeled as $\jntmm$, $\jntcr$, $\sepmm$, and $\sepcr$ respectively (see also \figref{fig:multi} for a graphical explanation on the example). In a further discussion, I will be interested only in asymptotic values ($N\to\infty$ or $k\to\infty$), which will be denoted by $\simeq$.

\section{Multiple-phase interferometer}
\label{sec:multiphases}

\begin{figure}[h!]
\begin{center}
\includegraphics[width=0.5\textwidth]{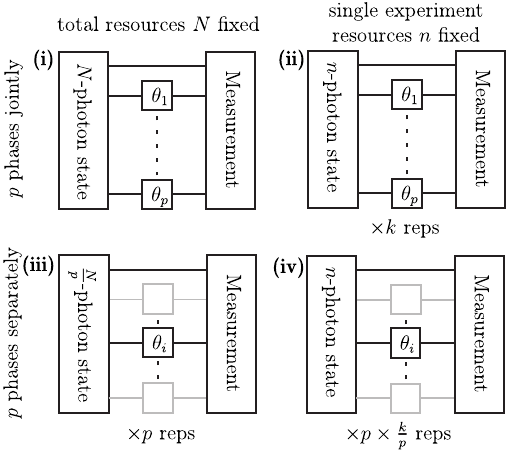}
\caption[Multiple-phase interferometer metrology scheme]{(Figure taken form \parencite{gorecki2021multiple}) $p+1$-arm interferometer with single reference arm and $p$ sensing arms is given. The phase shifts $\var_i$ are estimated by putting as an input an arbitrary entangled state and performing the general measurement involving outputs of all arms. The problem is analyzed in four scenarios: the phases may be measured jointly or separately, where both cases are discussed with the constraint on all resources $N$ or on resources used in a single trial $n$, where the trial is repeated $k$ times. The sum of estimators MSE $\sum_{i=1}^p\Delta^2\tilde\var_i$ in each cases will be denoted as (i) $\jntmm$, (ii) $\jntcr$, (iii) $\sepmm$ and (iv) $\sepcr$.
}\label{fig:multi}
\end{center}
\end{figure}

The first example I would like to discuss is the multiarm interferometer,
with $p$ sensing arms (corresponding to single-photon eigenstates $\ket{i}$ with $i=1,2...,p$) and single reference arm (corresponding to state $\ket{0}$), see \figref{fig:multi}. The single photon evolution is therefore given by:
\begin{equation}
U_\bvar=e^{i\sum_{i=1}^p \var_i\ket{i}\bra{i}}.
\end{equation}

Let me start with the situation where the number of atoms used in a single trial is $n$, while the whole procedure is repeated many times $k$. As all generators mutually commute and therefore $[U_\bvar,U_{\bvar '}]=0$, from discussion in \secref{sec:adamulti} without loss of generality, we may restrict to parallel strategies, as adaptive sequential ones offer no advantage in this case. Moreover, we may restrict to fully symmetric space, where the number of photons in each arm fully characterizes the state. We have, therefore:
\begin{equation}
\ket{\psi^n_\bvar}=\sum_{\boldsymbol{m}:\sum m_i\leq n}c_{\boldsymbol{m}}e^{i\boldsymbol{m}\cdot \bvar}\ket{m_1,...,m_p}
\end{equation}
with $\boldsymbol{m}=[m_1,..,m_p]$, where remaining $n-\sum_{i=1}^pm_i$ atoms are located in reference arm.

In the case where one wants to measure all phases separately, the optimal strategy will use appropriate $n00n$ states (between reference $0^{th}$ and sensing $i^{th}$ arms). In the limit of large $k$, resulting variance for each parameter would be $\frac{1}{k_in^2}$, therefore, getting $k_i=k/p$ one end with:
\begin{equation}
\label{phcrsep}
\sepcr\simeq p\times \frac{1}{k/p}\times \frac{1}{n^2}=\frac{p^2}{kn^2}.
\end{equation}

The optimal strategy of measuring all the phases jointly (within the paradigm of many repetitions) has been discussed in \parencite{humphreys2013quantum}. The optimal state would be then:
\begin{equation}
\label{phcrjntst}
\ket{\psi_\bvar^n}=\beta\ket{0,...,0}+\alpha(e^{in\var_1}\ket{n,..,0}+...+e^{in\var_p}\ket{0,0,...,n}),
\end{equation}
where coefficient minimizing the trace of the inverse of Fisher information are $\alpha=1/\sqrt{p+\sqrt{p}}$ and $\beta=1/\sqrt{1+\sqrt{p}}$, which leads to the cost:
\begin{equation}
\label{phcrjnt}
\jntcr\simeq \frac{1}{k}\tr(\bF_Q^{-1})=\frac{(1+\sqrt{p})^2p}{4kn^2}\overset{p\gg 1}{\approx}\frac{p^2}{4kn^2}
\end{equation}
(CR bound is asymptotically saturable, as all SLDs mutually commute). With an increasing number of phases, the advantage of measuring them jointly converges to a constant factor $ 1/4 $.

However, an interesting conjecture has been made by the Authors of \parencite{humphreys2013quantum} about the consequences of the above results for the situation when only the total amount of photons $N$ is restricted, and one has full freedom in dividing them between $n$ and $k$. They put forward the hypothesis that, as in the first approach, all $p$ phases are measured independently, it would require (in total) $p$ times more trials to saturate CR bound. Then, assuming that there exists some fixed minimal $k_{\min}$ needed for saturating CR, it would lead to $\sepcr\overset{?}{\simeq} p\times \frac{1}{k_{\min}}\times \frac{1}{(N/pk_{\min})^2}=\frac{p^3k_{\min}}{N^2}$ and 
$\jntcr\overset{?}{\simeq}\frac{p^2}{4k_{\min}(N/k_{\min})^2}=\frac{p^2k_{\min}}{4N^2}$. That would indicate the advantage, which scales linearly with $p$. This assumption is, however, unjustified, as CR bound is known only in the limit $k\to\infty$, and the rate of this convergence may depend on a particular model. Therefore, one cannot simply put the same number $k_{\min}$ to both equations to judge the advantage or its lack in terms of all resources used. Such discussion cannot be performed within QFI formalism at all.

To answer the question, I use minimax formalism. I am interested in asymptotic $N\to\infty$, so from \eqref{proofresult} I may restrict to parallel strategies with covariant measurements. If the phases are measured separately, we use the solution from single parameter case \eqref{sinstate}, where only reference $0^{th}$ and sensing $i^{th}$ arms are involved. As for each parameter, we spent $N/p$ photons, it leads to:
\begin{equation}
\label{phmmsep}
\sepmm\simeq p\times \frac{\pi^2}{(N/p)^2}=\frac{p^3\pi^2}{N^2},
\end{equation}
so, comparing to \eqref{phcrsep}, we observe different scaling with the number of parameters, as expected.

For the problem of measuring all phases jointly, the exact solution of the problem is beyond our reach. So instead, I derive a simple lower bound for achievable cost and then propose the state for which the cost is relatively close for the large $N$.

Consider $N-$photonic input state optimal for measuring all phases jointly. Without loss of generality, we may assume that it is symmetric for replacing the sensing arms, so for any sensing arm, the expected number of photons will be $\leq N/p$. Therefore, we can bound from below the variance of every single phase using the bound derived for estimation with mean energy \eqref{meanbound}, having finally:
\begin{equation}
\label{pkmmjnt}
    \jntmm\gtrsim p\times\frac{4|A_0|^3}{27}\frac{1}{(N/p)^2} =\frac{4|A_0|^3}{27}\frac{p^3}{N^2}\approx \frac{1.89 p^3}{N^2}.
\end{equation}
Compared to \eqref{phmmsep}, we see that the maximum potential advantage is again of order $\sim4$. 

That it is -- the fact that for this model, one cannot achieve better scaling with $p$ by measuring all parameters jointly comes directly from the nature of the Heisenberg limit (understood in terms of the average energy). It indicates that if one would like to think in terms of $k_{\min}$, in the joint strategy \eqref{phcrjntst},  it needs to be at least $p$ times bigger than in a separate one.

To check if the bound is close to being saturable, for large $N$ we use the same approximation as in the single-phase case. We replace $m_i\in\{0,1,...,N\}$ with continuous ones $\frac{m_i}{N}\to\mu_i\in[0,1]$ and characterize the state by $p$-dimensional wave function $\hat f(\boldsymbol{\mu})$\footnote{Note the change of convention $f\to \hat f$, compared to original paper \parencite{gorecki2021multiple}, which has been applied here to state in consistency with notation form \chref{ch:baymin}}:
\begin{equation}
\label{eq:contstate}
\ket{\psi_{\bvar}^N}=\int\displaylimits_{\forall \mu_i\geq 0, \sum_i{\mu_i}\leq 1}d\boldsymbol{\mu}\, \hat f(\boldsymbol{\mu})e^{i N\boldsymbol{\mu} \, \bvar}\ket{\mu_1,\mu_2,..\mu_p}.
\end{equation}
Applying covariant measurements:
\begin{equation}
\ket{\tilde\bvar}=\frac{1}{\sqrt{(2\pi/N)^p}}\int \t{d}\boldsymbol{\mu}\, e^{iN\boldsymbol{\mu}\tilde\bvar}\ket{\boldsymbol{\mu}}
\end{equation}
and optimization over the initial state leads to:
\begin{equation}
\jntmm\simeq\min_f \int\displaylimits_{\mathbb R^p} \t{d} \tilde{\bvar}\, |\braket{\tilde{\bvar}|\psi_\bvar^N}|^2 (\tilde\bvar-\bvar)^2  = \frac{1}{N^2}\min_{f}\int\displaylimits_{\mathbb R^p} \t{d} \tilde{\bvar}\,
\left|f(\tilde\bvar)\right|^2\tilde\bvar^2
\end{equation}
where we dropped the irrelevant dependence on $\bvar$. Going back to the $\boldsymbol{\mu}$-representation, the minimization problem takes the following form:
\begin{equation}
\label{eq:jointenergy}
\begin{split}
\jntmm\simeq&\frac{1}{N^2}\min_{f}\int\displaylimits_{\forall \mu_i\geq 0, \sum_i{\mu_i}\leq 1} \t{d}\boldsymbol{\mu}\, \hat f^*(\boldsymbol{\mu})\left(\sum_{k=1}^p-\partial_{\mu_k}^2\right)\hat f(\boldsymbol{\mu}),\\
&{\rm with}\quad \int\displaylimits_{\forall \mu_i\geq 0, \sum_i{\mu_i}\leq 1} \t{d}\boldsymbol{\mu}\, |\hat f(\boldsymbol{\mu})|^2=1,\\
&\hat f(\boldsymbol{\mu})=0\quad \textrm{for } \boldsymbol{\mu} \textrm{ on the boundary } (\mu_i=0 \lor\sum_i \mu_i=1 ).
\end{split}
\end{equation}
This makes the problem equivalent to finding the ground state of a free particle in the $p-$dimensional simplex-shape infinite potential well. While for some simplex shapes, the problem may be solved exactly by mapping to a model of one-dimensional $p$ free fermions \parencite{Krishnamurthy_1982,Turner_1984}, this is not our case (except $p \leq 2$ case \parencite{li1984particle}). The exact solution is beyond our abilities. Still, knowing that the function needs to satisfy boundary consideration and (to minimize kinetic energy) it should be concave inside of the simplex, I propose a simple ansatz, for which the energy may be calculated analytically:
\begin{equation}
\label{eq:state}
\hat f(\boldsymbol{\mu})\propto\left(\prod_{i=1}^p \mu_i\right)^\alpha \left(1-\sum_{i=1}^p \mu_i\right)^\beta.
\end{equation}
Direct calculations \parencite{gorecki2022multiparameter} shows, that for $\alpha = 3/2,\beta = \sqrt{p}$, the cost is equal:
\begin{equation}
\jntmm\simeq\frac{p(1+2\sqrt{p})^2\sqrt{p}(4p+2\sqrt{p}-1)}{(8\sqrt{p}-4)N^2}\overset{p\gg 1}{\approx}\frac{2p^3}{N^2},
\end{equation}
which is indeed close to \eqref{pkmmjnt} and offer constant (asymptotically independent on $p$) advantage over \eqref{phmmsep}.

We see that even if the scaling of the total cost with a number of parameters in separate strategies is different in both paradigms, the advantage of the joint strategy over the separate one is quantitatively the same in both approaches. The advantage in both scenarios comes from the same reason. In joint protocol, the same reference arm is shared by all phases, and therefore more photons remain for sensing arms. As a result, in the limit of large $p$, the amount of atoms in the reference arm (for the optimal state) becomes negligible. Therefore this advantage needs to converge to constant and cannot improve the scaling with $p$.

\section{Magnetic field with unknown direction -- $SU(2)$ estimation}
\label{sec:mf}

As the next example, I would like to discuss estimating three components of the magnetic field sensed by spin-1/2 particles. Similarly, as in the introductory example, I assume the elementary gate is obtained by the field acting sufficiently short in time such that the rotation angle $|\bvar|\leq \pi$, so the problem boils down to the estimation of the rotation. The single atom evolution is therefore given as:
\begin{equation}
U_\bvar=e^{i(\var_1\sigma_x+\var_2\sigma_y+\var_3\sigma_z)/2}.
\end{equation}

Let me start again with the many repetitions scenario, with $n$ gates in each of $k$ trials. 

Note that here, unlike in multiple-phase estimation problem, where each parameter could be measured regardless of the existence of the others, it is no longer the case. There is no subspace for which evolution would depend only on a single parameter. Therefore the values of the remaining two may affect estimation accuracy and should be treated as nuisance parameters, as discussed at the end of \secref{sec:multifisher}. We focus on estimating around the point $\bvar_0$. For simplicity of further formulas, let us choose the coordinates so that $\bvar_0=[0,0,\var_0]$.

Instead of deriving the optimal cost by direct optimization, we bound it from below and then show the example saturating these bounds. As mentioned before, the variance of every single parameter may be bounded by the inverse of the diagonal element of the QFI matrix:
\begin{equation}
\label{first}
    k\Delta^2\tilde\var_i\gtrsim [\bF_Q^{-1}]_{ii}\geq ([\bF_{Q}]_{ii})^{-1}
\end{equation}
while $[\bF_{Q}]_{ii}$ is simply single parameter $\var_i$ QFI, so from \eqref{unitarybound} is equal:
\begin{equation}
 [\bF_Q]_{ii}= 4\Delta^2 [-iU^\dagger_\bvar \partial_i U_\bvar]
\end{equation}
which, after direct maximization of RHS over the state (independently for $i=1,2,3$), gives
\begin{equation}
\label{threein}
 [\bF_Q]_{11}\leq\t{sinc}^2(\var_0),\quad [\bF_Q]_{22}\leq\t{sinc}^2(\var_0),\quad [\bF_Q]_{33}\leq1.
\end{equation}
We see that the non-zero value of $|\bvar_0|$ decreases the value of QFI connected to the estimating components perpendicular to $\bvar$. It may be easily interpreted since as $\bvar$ is oriented in $z$ direction, for all non-zero values of $\var_0$ the changes of $z$ component are connected with changes in the length of the rotation vector, while changes of $x,y$-- with changes of angle of rotation vector. Since $\var_0=2\pi$ corresponds to full rotation, exactly at this point, the infinitesimal changes of the angle of the rotation vector do not change the channel at all. For direct calculations performed in spherical coordinates, see \parencite{yuan2016}.

For a single usage of the gate, the maximally entangled state $\frac{1}{\sqrt{2}}(\ket{\up}_z\ket{0}+\ket{\down}_z\ket{1})$ saturates all three inequalities \eqref{threein}. At first sight, it may seem that this state distinguishes the $z$ direction. However, all maximally entangled states of the qubit are mutually equivalent up to local transformations on the ancilla $\openone \otimes V$ \parencite{kraus2010local} (which does not change the QFI of the output state, as commute with $U_{\bvar}\otimes\openone$). Moreover, the resulting QFI matrix is diagonal, so $[\bF_Q^{-1}]_{ii}=([\bF_{Q}]_{ii})^{-1}$. Together with the fact that SLDs commute on state, \eqref{first} is saturated, therefore for a single quantum gate (with $k$ repetitions), we have:
\begin{equation}
\label{crmfsinglegate}
    \sum_{i=1}^3\Delta^2\tilde\var_i\simeq\frac{1}{k}\left(1+\frac{2}{\t{sinc}^2(|\bvar|)}\right).
\end{equation}

For $n$ usage of gates in a single trial, the situation becomes a bit more complicated.
The states optimal for measuring each of the parameters separately $\frac{1}{\sqrt{2}}(\ket{\up}_x^{\otimes n}\ket{0}+\ket{\down}_x^{\otimes n}\ket{1})$, $\frac{1}{\sqrt{2}}(\ket{\up}_y^{\otimes n}\ket{0}+\ket{\down}_y^{\otimes n}\ket{1})$, $\frac{1}{\sqrt{2}}(\ket{\up}_z^{\otimes n}\ket{0}+\ket{\down}_z^{\otimes n}\ket{1})$ are no longer equivalent. For separate strategy, we therefore have:
\begin{equation}
\label{crsepmf}
    \sepcr\simeq\frac{1}{k/3}\frac{1}{n^2}+\frac{1}{k/3}\frac{1}{n^2\t{sinc}^2(|\bvar|)}+\frac{1}{k/3}\frac{1}{n^2\t{sinc}^2(|\bvar|)}=\frac{3}{kn^2}\left(1+\frac{2}{\t{sinc}^2(|\bvar|}\right).
\end{equation}
To see if this result may be significantly improved by measuring all parameters jointly, note first that $\forall_{i\in\{x,y,z\}} [\bF_Q]_{ii}\leq 4\Delta^2 J_{i}\leq 4\braket{J_i^2}$, which further implies \parencite{kolenderski2008}\footnote{Note a missing $4$ factor in Eq. (10) of \parencite{kolenderski2008}.}:
\begin{equation}
    \jntcr\gtrsim \frac{1}{k}\tr(\bF_Q^{-1})\leq \frac{1}{k}\frac{9}{\tr(\bF_Q)}\leq\frac{1}{k}\frac{9}{4(J_x^2+J_y^2+J_z^2)}\leq\frac{1}{k}\frac{9}{4\cdot n/2(n/2+1)}=\frac{1}{k}\frac{9}{n(n+2)}.
\end{equation}
For finite $|\bvar|$ the solution calculated directly gives \parencite{baumgratz2016}\footnote{Note difference by factor $1/2$ in the definition of the generators in \parencite{baumgratz2016}.}:
\begin{equation}
    \jntcr\simeq\frac{1}{k}\frac{3}{n(n+2)}\left(1+\frac{2}{\t{sinc}^2(|\bvar|)}\right)\overset{|\bvar|\ll 1}{\approx}\frac{9}{kn(n+2)}.
\end{equation}
Compared with \eqref{crsepmf}, one can see that the advantage of the joint strategy over separate one $\frac{n+2}{n}$ disappears with an increasing number of gates used in a single trial.

However, looking at \eqref{crmfsinglegate}, one may easily notice that the parallel results may be significantly improved if sequential adaptive strategy with optimal unitary controls $\forall_i V_i=U^\dagger _{\bvar}$ will be applied. Then, for $n$ usage of the gate, the QFI matrix is multiplied by $n^2$ when compared to single usage, as the output state is the same, while the derivative is multiplied by $n$; that gives \parencite{yuan2016}\footnote{Note difference by a factor $1/2$ in the definition of the generators in \parencite{yuan2016}.}
\begin{equation}
    \sum_{i=1}^3\Delta^2\tilde\var_i\simeq\frac{1}{k}\frac{1}{n^2}\left(1+\frac{2}{\t{sinc}^2(|\bvar|)}\right)\overset{|\bvar|\ll 1}{\approx}\frac{3}{kn^2}.
\end{equation}
We see that within the many repetitions scenario, the sequential adaptive scheme is needed to observe the asymptotic advantage of measuring all components jointly over measuring them separately.

Next, I would like to compare these results with the one obtained for the situation where all $N$ gates are used optimally. To do that, I will use theorem \eqref{proofresult} and the results obtained for the covariant problem. Note that for the covariant cost (invariant for the rotation), the exact form of the Hessian of this function, written in introduced coordinates $x,y,z$ will depend on the values $\bvar$ (which compensates for the fact that also the optimal QFI matrix depends on $\bvar$). Therefore, to make the comparison reasonable and simple, I will choose the cost function for which $\C=\openone$ at $\bvar=0$ and make the comparison around this point.

Note that in this case, the problem of measuring each of the three parameters separately is highly non-trivial, as, unlike in the multiple-phase example, there is no subspace of Hilbert space, for which the evolution would depend on the single parameter. As a result, even when wishing to measure one parameter, to construct a meaningful measurement and estimator, one needs to have some (at least approximate) knowledge of the values of the others. 

It makes the problem of measuring the parameters separately a bit artificial, but in principle, we can formulate the problem where all $N$ is divided between three independent experimenters (who cannot communicate) when each of them is focused on estimating a single component $\var_i$. Then, from the point of view of single parameter estimation, existing the others may only make the full problem harder (see next section for broader explanation), so the total cost for such strategy may be bounded from below analogously as in multiple-phase case:
\begin{equation}
\sepmm \gtrsim 3\times \frac{\pi^2}{(N/3)^2}=\frac{27\pi^2}{N^2}
\end{equation}
(while here, there is no guarantee for saturating).

In the case where all parameters are measured jointly, from \eqref{proofresult}, we know that up to terms $o(1/N^2)$ the optimal strategy is the covariant one, realizable in the parallel scheme with ancilla, without the necessity of applying adaptivity. The problem has been solved in ~\parencite{chiribella2004,chiribella2005su2}, with a covariant cost:
\begin{equation}
\label{su3}
e(\bvar,\tilde\bvar)=6-2\tr(U^{(1)}_\bvar U^{(1)\dagger}_{\tilde\bvar})\overset{|\bvar|,|\tilde\bvar|\ll 1}{\approx}2\sum_{i=1}^3\Delta^2\tilde\var_i,
\end{equation}
where $U^{(1)}_\bvar$ is a rotation matrix of a spin-1 particle. It was shown that the asymptotic minimal cost is $e(\bvar,\tilde\bvar)\simeq8\pi^2/N^2$, which is achievable using initial state:
\begin{equation}
\label{3dopt}
\ket{\psi_{in}}=\sqrt{\frac{2}{N/2+1}}\sum_{j=0(\tfrac{1}{2})}^{N/2-1}\sin\left(\tfrac{(j+1)\pi}{J+1}\right)
\left(\sum_{\alpha=1}^{2j+1}\frac{\ket{j\alpha,m_j=\alpha}}{\sqrt{2j+1}}\right),
\end{equation}
where $\ket{j\alpha,m_j=\alpha}$ are states with a well defined total angular momentum $j$ and its projection onto $z$ direction, while $\alpha$ numerates different subspaces corresponding to equivalent irreducible representations of $SU(2)$. Note that here the degree of freedom connected with different irreducible representations plays the same roles as ancilla, so as a result, the optimal protocol does not need an additional one. After rescaling the cost function by factor $\frac{1}{2}$ we get
\begin{equation}
\label{su2res}
 \jntmm\simeq\frac{4\pi^2}{N^2}.
\end{equation}
Note that these results in the asymptotic limit $N\to\infty$ may also be obtained using Fourier analysis \parencite[Section 12]{hayashi2016fourier}.

Compared to the previous multiple-phase example, we see significant qualitative differences. In the multiple-phase case, the relation between optimal separate and joint protocol within both paradigms was very similar; in all cases, adaptivity was not needed for optimal performance.

In contrast, for magnetic field estimation, there are significant differences between the results obtained in both paradigms. In the many repetitions scenario, the optimal parallel strategy offers asymptotically no advantage over separate protocol, while the adaptive one allows for a decrease in the cost by a factor $3$. Contrarily, for the optimal usage of all resources $N$, already the parallel scheme offers a big advantage, which cannot be further improved by an adaptive one.

A similar analysis may be performed for the problem of estimating only two components of the magnetic field, where the third one is known to be zero, namely $U_\bvar=e^{i(\var_1\sigma_x+\var_2\sigma_y)}$, which in covariant formalism was analyzed in~\parencite{bagan2000,bagan2001}. The results are qualitatively the same as in the example discussed above \parencite{gorecki2022multiparameter}.

Looking at the two examples discussed above, one may ask if the discrepancy between the results obtainable in different paradigms is inextricably linked to the commutativity of the evolution generators. 

From discussions performed in \chref{ch:multi} we know that superiority of adaptive protocols over parallel ones is indeed strictly related to commutativity -- if all generators mutually commute, the advantage cannot appear in any paradigm. Contrary, if generators do not commute, the superiority sequential-adaptive scheme may appears in the many repetitions scenario (while for optimal usage of all $N$ resources, the entangled-parallel is guaranteed to be optimal).

What about the existing advantage of measuring parameters jointly vs. separately? In a moment, I will give an example to show that divergence on this matter between the two paradigms is \textbf{not} clearly related to the issue of commutation of operators. Before I do so, I will introduce some general theories.

\section{General bound for precision in multiparameter unitary evolution}
\label{sec:reparametr}
The problem of the multiparameter estimation with optimal usage of all $N$ gates has known analytical solution for some specific group estimation problems: SU(2)/U(1)~\parencite{bagan2000,bagan2001}, SU(2)~\parencite{chiribella2004,bagan2004su2,chiribella2005su2,hayashi2006su2}, SO(3)~\parencite{hayashi2016fourier}; moreover, the existing of Heisenberg scaling with $N$ has been proven for the general case of SU(d)~\parencite{kahn2007}. However, there is a lack of the universal lower bound. Below I present the universal lower bound \parencite{gorecki2022multiparameter} (unfortunately not tight), being the generalization of \eqref{result}.

Consider a unitary channel dependent on $p$ parameters, which enter linearly via generators $\oG_i$, such that $U_\bvar=e^{i\sum_{i=1}^p\var_i\oG_i}$. First, I want to justify that if one is interested in estimating only one of these parameters $\var_i$, the existence of another $\var_j$ may only make the task more difficult and never easier. Namely, even in the presence of
other generators, the bound \eqref{result} is still valid (but not necessarily saturable). For more compact notation, let me name by $\lambda[\oG]$ the difference of maximal and minimal eigenvalues of operator $\oG$. Let me remind the bound \eqref{result} in this notation:
\begin{equation}
\label{repeat}
    \vm\geq \frac{\pi^2}{\lambda^2[\oG]N^2}\left(1-\frac{8\log(N\lambda[\oG]\delta)}{N\lambda[\oG]\delta}\right).
\end{equation}
Next, note that a single gate may be arbitrarily well approximated as:
\begin{equation}
U_\bvar=e^{i\sum_{i=1}^p\var_i\oG_i}\overset{l\gg 1}{\approx} \left(e^{i\var_i \frac{\oG_i}{l}}e^{i\sum_{j\neq i}\var_j\frac{\oG_j}{l}}\right)^l
\end{equation}
where from Trotter formula~\parencite{COHEN198255} the approximation is exact in the limit $l\to\infty$ (even if $[\oG_i,\oG_j]\neq 0$). Therefore $N$ actions of the unitary $U_\bvar$ may be seen as $l\cdot N$ action of $U_{\var_i}=e^{i\var_i \frac{\oG_i}{l}}$, with unitary controls $V=e^{i\sum_{j\neq i}\var_j\frac{\oG_j}{l}}$ in between. As a result, the product $\lambda[\oG_i/l]\cdot (lN)=\lambda[\oG_i]N$ remains unchanged, so does the bound \eqref{repeat} (including the range of convergence). Therefore, for each parameter;
\begin{equation}
\label{singlesimple}
\vm_i\gtrsim \frac{\pi^2}{\lambda^2[\Lambda_i]}.
\end{equation}
To make use of the above bound in the multiparameter case, let me discuss the concept of re-parametrization.

For given unitary channel $U_\bvar=e^{i\bvar\cdot \bG}$ with cost matrix $\ptr(C(\tilde\bvar-\bvar)(\tilde\bvar-\bvar)^T)$, for real invertible $p\times p$ matrix $A$, we may perform the transformation of parameters, generators and cost matrix:
\begin{equation}
\bvar'=A^{-1}\bvar,\quad\bG'=A^T\bG,\quad C'=A^TCA,    
\end{equation}
such that estimation problem remain unchanged:
\begin{equation}
\bvar'\cdot\bG'=\bvar\cdot\bG,\quad \ptr(C'(\tilde\bvar'-\bvar')(\tilde\bvar'-\bvar')^T)=\ptr(C(\tilde\bvar-\bvar)(\tilde\bvar-\bvar)^T).
\end{equation}
Therefore, even if the problem is initially formulated with $\bvar'$ and arbitrary cost matrix $C'$, we may always reduce it to minimize the sum of variance by choosing $A=\sqrt{C}$, such that for $\bvar$ the cost is simply $\sum_{i=1}^p\Delta^2\tilde\var_i$.

This allows us the bound from below the minimal achievable cost by assuming the most optimistic scenario, i.e., not only do the other parameters not interfere but there is a single initial state and a single optimum measurement for all parameters at the same time:
\begin{equation}
    \jntmm \gtrsim\sum_{i=1}^p \frac{\pi^2}{\lambda^2[\G_i]}.
\end{equation}
However, the parametrization for which the cost matrix is identity is not unique. Therefore, the bound may be potentially improved by optimizing it over orthogonal transformation $O$:
\begin{equation}
\label{lastbound}
\jntmm\geq \sum_{i=1}^p\max_O\frac{\pi^2}{\lambda^2[[O^T\bG]_i]}.
\end{equation}
Note that restricting to orthogonal transformations is crucial, as otherwise, we get cost matrix $C$ with non-zero of diagonal elements, and then $\ptr(C(\tilde\bvar-\bvar)(\tilde\bvar-\bvar)^T)$ cannot be bounded by the sum of \eqref{singlesimple}.

For the problems discussed in \secref{sec:multiphases}, \secref{sec:mf}, the bound is rather far from being saturable.
In multiple-phase case, by direct minimization over $O$, one can find that it leads to $\frac{5\pi^2}{2N^2}$ for $p=2$ (which is slightly tighter than \eqref{pkmmjnt}, but still two times smaller than an exact solution), while for $p\geq 3$ it is even weaker than \eqref{pkmmjnt}. In the case of $SU(2)$, one can notice that application $O$ does not improve the bound at all (as rotating Pauli matrices do not change their eigenvalues), which gives $\frac{3\pi^2}{N^2}$, which is a quarter less than the exact solution.

\section{Magnetic field in two spatially separated points}
At last, I would like to present a simple example for which the bound \eqref{lastbound} will be tight, and also re-parametrization will play an important role. Consider a magnetic field oriented in $z$ direction in two spatially separated points, sensed by spin-1/2 atoms. The Hilbert space of a single atom is spanned by vector $\ket{i,\pm}$, where coordinate $i=1,2$ is the spacial one. We have:
\begin{equation}
\var_1\oG_1+\var_2\oG_2=\var_1 \ket{1}\bra{1}\otimes\sigma_z+\var_2\ket{2}\bra{2}\otimes \sigma_z,
\end{equation}
so $\lambda_1=\lambda_2=1$, what leads to the bound $\jntmm\gtrsim \frac{2\pi^2}{N^2}$.

To make the bound tighter, let us consider reparametrization:
$\var_1'=(\var_1+\var_2)/\sqrt{2}, \var_2'=(\var_1-\var_2)/\sqrt{2}$, for which:
\begin{equation}
\label{matrix}
\begin{bmatrix}
\var_1 & 0 & 0 & 0\\
0 & -\var_1 & 0 & 0\\
0 & 0 & \var_2 & 0\\
0 & 0 & 0 & -\var_2
\end{bmatrix}=
\begin{bmatrix}
(\var'_1+\var'_2)/\sqrt{2} & 0 & 0 & 0\\
0 & -(\var'_1+\var'_2)/\sqrt{2} & 0 & 0\\
0 & 0 & (\var'_1-\var'_2)/\sqrt{2} & 0\\
0 & 0 & 0 & -(\var'_1-\var'_2)/\sqrt{2},
\end{bmatrix}
\end{equation}
so $\lambda_1=\lambda_2=\frac{1}{\sqrt{2}}$ which gives $\jntmm\gtrsim \frac{4\pi^2}{N^2}$. Moreover, this bound turns out to be achievable.

To see that, let as reorder the eigenbasis to be $\{\ket{1,+},\ket{2,-},\ket{2,+},\ket{1-}\}$. Then the matrix \eqref{matrix} takes the specific form:
\begin{equation}
\var'_1\oG'_1+\var'_2\oG'_2=\tilde\var_1\tfrac{1}{\sqrt{2}}\openone\otimes\sigma_z+\tilde\var_2\tfrac{1}{\sqrt{2}}\sigma_z\otimes\openone
\end{equation}
Then, by renaming $\{\ket{1,+},\ket{2,-},\ket{2,+},\ket{1-}\}\to\{\ket{+,+},\ket{+,-},\ket{-,+},\ket{-,-}\}$, on the mathematical level we can identify the Hilbert space of single particle with a 2-dimensional spacial degree of freedom with Hilbert space of 2 independent spin-1/2 particles (with no spacial degree of freedom).

It makes it clear that, therefore, $\tilde\var_1,\tilde\var_2$ may be simultaneously measured optimally with the same input state without interfering with each other. More precisely, we can write the Hilbert space as 
$\mH=\mathbb C^{\otimes 2}\otimes\mathbb C^{\otimes 2}$, so for $N$ copies in the parallel strategy, we would have $\mH^{\otimes N}=(\mathbb C^{\otimes 2}\otimes\mathbb C^{\otimes 2})^{\otimes N}=(\mathbb C^{\otimes 2})^{\otimes N}\otimes(\mathbb C^{\otimes 2})^{\otimes N}$. Therefore the state of the form $\ket{\t{sin}}\otimes\ket{\t{sin}}$ (where $\ket{\t{sin}}$ is the one of the form \eqref{sinstate}) with covariant measurement will saturate the bound:
\begin{equation}
\jntmm\simeq \frac{4\pi^2}{N^2}.
\end{equation}

We see that such a strategy offers the advantage over the one where $\var_1,\var_2$ are measured separately, which leads to:
\begin{equation}
\sepmm \simeq2\times\frac{\pi^2}{(N/2)^2}=\frac{8\pi^2}{N^2}.
\end{equation}
An interesting observation may be concluded if one looks at the problem from a many-repetition scenario perspective. Then for joint strategy, by the same reasoning, one obtains
\begin{equation}
\jntcr\simeq\frac{4}{kn^2}.
\end{equation}
However, for measuring parameters separately:
\begin{equation}
\sepcr\simeq 2\times\frac{1}{k/2}\times\frac{1}{n^2}=\frac{4}{kn^2},
\end{equation}
so, in this case, there is no advantage to applying a more advanced joint strategy! Unlike in previously discussed examples, we observe qualitative differences about existing advantages depending on the paradigm.

For the number of spatial points $p$ bigger than $2$, the above construction cannot be performed, as, simply speaking, we lose the identity $2p=2^p$. The broad discussion about the case $p>2$ may be found in \parencite{gorecki2022multiparameter}.

One more feature should be mentioned here. In all discussed examples in the separate strategy, we measured the initially introduced parameters. Especially in the last example, we were focused on measuring $\var_1,\var_2$, while in principle, one could consider as a separate strategy measuring independently $\var'_1,\var'_2$, or, in general, any linear combinations of them. From a practical point of view, typically, some parametrization is more natural, and in trying to measure their linear combination, we simply lose the simplicity, which is the crucial motivation for a separate strategy (here: restricting the measurements local in space). Still, it is a reasonable comment from a general point of view. However, it turns out that for all discussed examples, this "natural" parametrization was indeed the optimal one for measuring parameters separately, as shown in \appref{app:repar}.

\section{Summary}
The analysis of the examples shows that, in contrast to the estimation of a single unitary parameter, in the multiparameter case, there is no clear and unambiguous correspondence between the results obtained in the many repetitions paradigm (analyzed within CR formalism) and those corresponding to the optimal use of all resources (analyzed using minimax approach (MM)). I analyzed in detail two issues
\begin{itemize}
\item superiority of sequential-adaptive scheme over the entangled-parallel one.
\item the advantage of measuring all parameters jointly, compared to the strategy, where they are measured separately.
\end{itemize}
In the first case, the existence of this superiority is strictly related to the commutativity of evolution generators, which is consistence with theorems from \chref{ch:multi}. If all generators mutually commute, the optimal results may be obtained within a parallel scheme (in both paradigms). However, in non-commuting cases, in CR adaptive scheme may lead to significantly better results, while in MM a parallel scheme is asymptotically sufficient to obtain the optimal result (which I illustrated by the problem of estimation of the three components of the magnetic field).

On the other hand, the divergence of the conclusions obtained in the two paradigms regarding the gain from measuring the parameters jointly is not related to commutativity. I have shown an example of estimating one component of the magnetic field in two spatially separated points (so generators mutually commute), where CR postulates do advantage from measuring parameters jointly, while the MM approach allows for decreasing cost by a factor of $2$.

In general, the conclusions about the usefulness of carrying out a joint measurement made solely with CR may be overly pessimistic. For optimal use of all resources (MM), the advantage may exist for models that do not exist in a many repetitions scenario (CR).

\chpt{Conclusion and outlook}
For single-parameter unitary noiseless estimation, I have shown that the fundamental bound on the accuracy of the estimation with a limited number of the resources $N$ is $\pi$ times larger than would result from naive use of the error propagation formula or exploiting the quantum Fisher information (without considering multiple repetitions). The result considers any evolution generator (including one with a continuous spectrum) and remains valid for arbitrary entangled input states and a more general sequential adaptive scheme. The bound is asymptotically saturable. The rate of convergence depends on the ratio between the generator's spectral width and the size of the region of the parameter's occurrence. The superiority of the adaptive sequential scheme over the parallel one asymptotically disappears with increasing $N$. An analogous analysis of converging to the fundamental bound was performed (only for the parallel strategy) when only the average number of resources is limited (in this case, I am not discussing an adaptive strategy).
In both cases, in the limit of $N\to\infty$, up to the leading order, the problems can be equated with estimating a completely unknown phase shift in a two-arm interferometer.

The analysis of the analogous problem in the presence of noise remains an open question. On the one hand, the QFI formalism offers extensive quantum error-correction tools and protocols, offering hope for recovering Heisenberg scaling. Unfortunately, their successful implementation requires almost exact knowledge of the parameter value from the very beginning. An actual implementation requires sequential tweaking of the circuit settings, considering the knowledge gained so far. There has been no thorough analysis of whether this would allow the MSE to keep scaling as $1/N^2$.

In the multi-parameter case, I compared optimal metrological protocols for cases when the experiment is repeated many times (analyzed within CR formalism) or when all available resources can be accumulated together (analyzed within MM formalism). In the case of the multi-arm interferometer, the results in both cases are similar. However, in the case of the estimation of the three components of the magnetic field, the results are qualitatively different -- CR predicts an asymptotic lack of benefit from measuring parameters jointly in a parallel scheme (however, it postulates a substantial advantage of the sequential adaptive scheme). In contrast, in the MM analysis, a substantial gain is already apparent for the parallel scheme (which cannot be improved by adaptiveness). 
It is known that the divergence in the utility of adaptivity in the two paradigms occurs only when the evolution generators do not commute. However, further analysis shows that the divergence in the benefit of measuring the parameters jointly (versus measuring them separately) is unrelated to the commuting of the generators.
I also derived a simple (not necessarily saturated) lower bound on the achievable MSE. The problem of deriving a general saturable bound for multi-parameter metrology (even without noise) remains open.

\end{mainf}

\begin{appendices}
\multappendices

\chpt{Example: squeezed vacuum states in the interferometry}
\label{examplelosses}
\subsectionnull{Quasi-classical intuition for quantum advantage}
Before discussing realistic noisy problem, let me briefly introduce an intuition standing behind the advantage of applying squeezed vacuum state into the lower arm of the interferometer.

First, assume that the inputs of both arms are a classical light with complex amplitudes $A,B$, which will be later treated as random variables, which expectation values will be denoted as $\langle\cdot\rangle$. Then the difference of intensities between two output arms is equal
\begin{equation}
N_-=(|A|^2-|B|^2)\sin(\var)+2\t{Im}AB^*\cos(\var).
\end{equation}
From error propagation formula (note that the square root of both sides has been taken compared to \eqref{errpro}) gives:
\begin{equation}
\Delta\tilde\var=\frac{\Delta N_-}{\left|\frac{d\langle N_-\rangle}{d\var}\right|},
\end{equation}
so around point $\var\approx 0$
\begin{equation}
\Delta\tilde\var\approx\frac{2\Delta(\t{Im}AB^*)}{\langle|A|^2-|B|^2\rangle}.
\end{equation}
It is reasonable to take as the $B$ simply vacuum state, for which $\langle B\rangle =0$. However, from the principle of quantum mechanics, we know that $B$ has some unavoidable fluctuations satisfying 
\begin{equation}
\label{unsq}
\Delta (\t{Re}B)\Delta (\t{Im}B)\geq \frac{1}{4}.
\end{equation}
Assuming $A$ is purely real and $\langle|A|^2\rangle\gg \langle|B|^2\rangle$, $|\langle A\rangle|\gg \{|\langle B\rangle|,|\Delta A|, |\Delta B|\}$, we can write:
\begin{equation}
\Delta\tilde\var\approx\frac{2A\Delta(\t{Im}B)}{\langle|A|^2\rangle}.
\end{equation}
From the above formula, we can see that it is profitable to slightly decrease $\Delta (\t{Im}B)$ -- but not too much, as from \eqref{unsq} we can see that too strong squeezing $\Delta (\t{Im}B)$ would result in large $\Delta (\t{Re}B)$ and, consequently, violating conditions $\langle|A|^2\rangle\gg \langle|B|^2\rangle$, $|\langle A\rangle|\gg |\Delta B|$. For slightly squeezed vacuum $\Delta(\t{Im}B)=\frac{1}{2}e^{-r}$ in the lower arm and coherent state with average $N$ photons in the upper one, we recover the famous formula 
\begin{equation}
\Delta\tilde\var\approx\frac{e^{-r}}{\sqrt{N}}.
\end{equation}

Of course above reasoning is not strict, as, during freely jumping between classical amplitudes and quantum uncertainty relation, it does not take into account all features connected with non-commuting of quantum amplitude operators. Still, it gives solid intuition that putting in the lower arm the proper squeezed state of the vacuum should significantly increase the precision. Formal derivation requires performing whole calculations within quantum mechanic formalism.

\subsectionnull{Interferometer with losses}

Let us now go to the noisy case. Consider two arms interferometer \figref{fig:interf} with symmetric photon losses, i.e. for each arm only $\eta<1$ of the signal go further, while $1-\eta$ part is loss. For such a model the bound \eqref{nohs} gives \parencite[Supplemental Material, Erasure]{demkowicz2014using}:
\begin{equation}
\label{eq:losses}
 \min_{\{K_k\},\beta=0}\|\alpha\|=\frac{\eta}{4(1-\eta)}\Longrightarrow \Delta^2\tilde\var\geq \frac{1}{kF_Q}\geq\frac{1}{kn}\frac{1-\eta}{\eta},
\end{equation}
where $n$ is the number of photons used in a single iteration.

It turns out that this bound may be saturated by putting to the upper arm the coherent state with the average value of photons equal $\NN$ and the vacuum states weakly squeezed by a factor $e^{-2r}$ to the lower one \parencite{caves1981quantum}, followed by simply estimating the value of phaseshift from the difference of photon counting in both counter\footnote{Strictly speaking, as this strategy involves an indefinite number of photons with fixed mean value $\NN$, the bound \eqref{eq:losses} does not apply here directly (as it was derived with the assumption of a well-defined number of gates used). Still, as we consider the case when the variance of the number of photons is much smaller than the square of its mean number, we do not expect any significant effects (contrary to the ones from \secref{sec:infinite}) coming from this fact.}.

The expectation value of this difference is then
\begin{equation}
    \braket{N_-}=\tfrac{\eta}{2}\cos(\var)(\NN-\sinh^2r),
\end{equation}
so, unlike in the case of $n00n$ states or similar ones, the range of effectiveness of such strategy is finite and does not shrink with increasing $\NN$. From the error propagation formula we have:
\begin{equation}
    \Delta^2\tilde\var=\frac{\braket{\Delta^2N_-}}{\left|\frac{d\braket{N_-}}{d\var}\right|^2},
\end{equation}
with
\begin{equation}
    \left|\frac{d\braket{N_-}}{d\var}\right|^2=\frac{\eta^2}{4}\sin^2(\var)(\NN-\sinh^2(r))^2,
\end{equation}
\begin{multline}
    \braket{\Delta^2N_-}=\frac{\eta^2}{4}\Big(\cos^2(\var)(\NN+\tfrac{1}{2}\sinh^2(2r))+\\
    \sin^2(\var)\left(\NN(\cos^2\phi e^{-2r}+\sin^2\phi e^{+2r})+\sinh^2r\right)+\tfrac{1-\eta}{\eta}(\NN+\sinh^2r)\Big)
\end{multline}   
where $\phi$ is the angle between the phase of coherent input and squeezing direction, which should be set $\phi=0$ for optimal performance. In asymptotic limit, choosing $r$ such as $\frac{\eta}{1-\eta}\ll \sinh^2 r\ll \NN$ we may approximate above by:
\begin{equation}
    \Delta^2\tilde\var\approx \frac{\cos^2\var +\frac{1-\eta}{\eta}}{\sin^2\var \NN}.
\end{equation}
Clearly, the optimal point is at $\var=\pi/2$, when it saturates the bound \eqref{eq:losses}. However, the MSE does not change significantly for $\cos^2(\var)\ll \frac{1-\eta}{\eta}$, so indeed, the strategy works well in a finite-sized region.

Such a method has been used in the LIGO gravitational wave detector \parencite{ligo2011gravitational,aasi2013enhanced}. In practice the procedure implementing squeezed vacuum to the lower arm significantly increases the noise, so the condition $\frac{\eta}{1-\eta}\ll \sinh^2 r$ may be not satisfied. Moreover, to talk about quantum advantage, one would compare the resulting variance with the one which would be obtained for the coherent light without these additional losses or -- equivalently -- for coherent light with the same \textbf{output energy} (which is in practice true limitation, given by photons counters abilities). In a small neighborhood of $\var=\pi/2$ the formula for MSE may be approximated as:
\begin{equation}
   \Delta^2\tilde\var\approx \frac{\eta e^{-2r}+(1-\eta)}{\eta \NN}.
\end{equation}
Therefore, the measure of quantum advantage is the value of the nominator in the above equation. In \parencite{ligo2011gravitational} for squeezing rate $e^{-2r}=0.1$ ($10\t{dB}$) and detection efficiency $\eta=0.62$ decreasing the MSE by a factor $0.47$ ($3.5\t{dB}$) has been observed, in accordance with the above formula.

In \parencite{aasi2013enhanced} for squeezing rate $e^{-2r}=0.93$ ($10.3\t{dB}$) and detection efficiency $\eta=0.44$ was obtained, which, according to the above formula, should lead to decreasing MSE by $0.6$ ($2.2\t{dB}$). The observed effect was minimally weaker $2.15\t{dB}$, which has been identified as the result of normal fluctuation of the phase $\phi$ of order $37\pm6 \t{mrad}$.

\chpt{Derivation of \eqref{inapp}}
\label{app:window}

Consider probability distribution
\begin{equation}
p_{\alpha,L}(\var)=\mathcal N_\alpha L {\rm sinc}^4\left(\pi\alpha\sqrt{(L\var/4\alpha)^2-1}\right)=\mathcal N_\alpha L\frac{{\rm sinh}^4\left(\pi\alpha\sqrt{1-(L\var/4\alpha)^2}\right)}{\left(\pi\alpha\sqrt{1-(L\var/4\alpha)^2}\right)^4},
\end{equation}
where $L=2N_0$ is the bandwidth, $\alpha$ determines the shape and $\mathcal N_\alpha$ may be bounded by:
\begin{equation}
\label{Nbound}
\mathcal N_\alpha\lesssim4\sqrt{2}\pi^4\alpha^{7/2}e^{-4\pi\alpha},
\end{equation}
where the bound is tight for big $\alpha$~\parencite{Gorecki2020pi}. As shown in \parencite{Gorecki2020pi} only exponentially small (with $\alpha$) part lays outside of region $[-4\alpha/L,4\alpha/L]$ and, therefore, for our purpose we choose $\delta/2=4\alpha/L$.

For such a function, I will show that for properly chosen $L$, inequality \eqref{inapp}
\begin{equation}
\label{toprove}
\frac{\pi^2}{(\lambda N+L/2)^2}-R_2\overset{?}{\geq} \frac{\pi^2}{\lambda^2N^2}\left(1-\frac{8\log(N\lambda\delta)}{N\lambda\delta}\right) 
\end{equation}
holds. Using general identity $\frac{1}{(1+x)^2}=1-2x+\frac{3x^2+x^3}{(1+x)^2}$ we may rewrite LHS of \eqref{toprove} as:
\begin{equation}
\frac{\pi^2}{(\lambda N+L/2)^2}-R_2=
\underbrace{\frac{\pi^2}{\lambda^2N^2}\left(1-\frac{L}{\lambda N}\right)}_{B_1}+\underbrace{\frac{\pi^2}{\lambda^2N^2}\frac{3(L/2\lambda N)^2+2(L/2\lambda N)^3}{(1+L/2\lambda N)^2}-R_2}_{B_2}.
\end{equation}
Next, I will show that for proper $L$, $B_1$ is equal to RHS of \eqref{toprove}, while  $B_2$ is strictly positive for big $N\lambda \delta$.

First I bound from above $R_2$:
\begin{multline}
\label{R2b}
R_2= 2\mathcal N_{\alpha} \int^{+\infty}_{4\alpha/L} \t{d}\var\, L {\rm sinc}^4\left(\pi\alpha\sqrt{(L\var/4\alpha)^2-1}\right)(\var+\delta/2)^2
\\
=2\mathcal N_{\alpha} \frac{4\alpha}{L}\int^{+\infty}_{1}\t{d}x\, L {\rm sinc}^4\left(\pi\alpha\sqrt{x^2-1}\right)(x+1)^2\left(\frac{4\alpha}{L}\right)^2\leq\\
\leq 2\mathcal N_{\alpha} L\left(\frac{4\alpha}{L}\right)^3\Big( \int_1^2\t{d}x\, (x+1)^2+\frac{1}{\pi^4\alpha^4}\int_2^{+\infty}\t{d}x\, \frac{(x+1)^2}{(x^2-1)^2}\Big)
\\
=2\mathcal N_{\alpha} L\left(\frac{4\alpha}{L}\right)^3\Big(\frac{19}{3}+\frac{1}{\pi^4\alpha^4}\Big)\overset{\alpha>1/2}{\leq}14\mathcal N_\alpha L\left(\frac{4\alpha}{L}\right)^3,
\end{multline}
where inequalities ${\rm sinc}(x)\leq 1$ and ${\rm sinc}(x)\leq 1/x$ were used.

Now, choosing $L=\frac{8}{\delta}\log(N\delta\lambda)$ (so $\alpha=\log(N\delta\lambda)$) we obtain:
\begin{equation}
B_1=\frac{\pi^2}{\lambda^2N^2}\left(1-\frac{8\log(N\lambda\delta)}{N\lambda\delta}\right)
\end{equation}
and, introducing $y=N\delta \lambda$ and $z(y)=4\log(y)/y$ for more compact notation,
\begin{equation}
L^2\cdot B_2=\pi^2(2z)^2\frac{3z^2+2z^3}{(1+z)^2}-14\mathcal N_{\log(y)} \left(4\log(y)\right)^3.
\end{equation}
By numerical calculation, it may be shown that the above is positive for $y\geq 2$, which was to be shown.

\chpt{Gradient estimation}
\label{app:gradient}
Here I address the problem raised in  \parencite{altenburg2017estimation}, where the authors have analyzed the problem of magnetic field gradient estimation within QFI and Cram\'er-Rao formalism, leaving an open question, of whether the Bayesian (or minimax) approach would lead to a tighter bound.

Consider ${N_{\t{pr}}}$ atoms of spin $\sfrac{1}{2}$ distributed along the $x$-axis, such that for each atom $x_i\in [x_0,x_0+L]$. The evolution of the system is given as\footnote{Note that, comparing to notation form \parencite{altenburg2017estimation}, $G$ was replacing by $\omega$, $t$ by $T$ and subscript "pr" has been added to the number of probes $N_{\t{pr}}$, to keep notation consistency with the rest of thesis.}
\begin{equation}
U_\omega=\exp(-i T \gamma (H_0+\omega H_\omega)),\quad \t{where}\quad H_0:=B_0 J_z,\quad H_\omega:=\frac{1}{2}\sum_{i=1}^{N_{\t{pr}}}(x_i-x_0)\sigma_z^{(i)}.
\end{equation}
Following \parencite[Sec. V]{altenburg2017estimation}, I consider a situation, where the system is under collective phase noise (strong fluctuation of $B_0$), so looking for optimal input state I restrict to decoherence-free states, i.e. the ones of the form:
\begin{equation}
\rho=\sum_{k=0}^{N_{\t{pr}}}p_k \rho_k,
\end{equation}
where each $\rho_k$ has well defined $J_z$. 

As $J_z$ acts trivially on these states, without changing the problem we may shift the generator $H_\omega$:
\begin{equation}
H_\omega':=H_\omega+L/2\cdot J_z=\frac{1}{2}\sum_{i=1}^{N_{\t{pr}}}(x_i-(x_0+L/2))\sigma_z^{(i)}.
\end{equation}

Now we may formulate a new abstract problem: estimate $\omega$ for the evolution $U'_{\omega}=\exp(-it\gamma \omega G')$ with \textbf{no restriction on the state}. This problem is obviously easier than the original one -- therefore any lower bound on the precision obtainable in this problem will remain valid in the original one (even if it may be not saturable). This is, however, exactly the problem discussed in \secref{sec:frequency}. Therefore, setting 
\begin{equation}
\oG_\omega=-\hbar \gamma \frac{1}{2}\sum_{i=1}^{N_{\t{pr}}}(x_i-(x_0+L/2))\sigma_z\quad \Rightarrow \lambda_{\omega}=(L/2)\hbar \gamma
\end{equation}
from \eqref{resultfr} we have:
\begin{equation}
\widehat{\Delta^2\tilde \omega}\overset{{N_{\t{pr}}}T\gg 1}{\gtrsim}\frac{4\pi^2}{({N_{\t{pr}}}t\hbar \gamma L)^2},
\end{equation}
which may be saturated letting the state evolve for time $t$, using as the input state the analogue of \eqref{sinstate}, namely:
\begin{equation}
\begin{split}
\label{singrad}
\ket{\psi^{N_{\t{pr}}}}=&\sqrt{\frac{2}{{N_{\t{pr}}}+2}}\sum_{m=0}^{N_{\t{pr}}} \sin\left(\frac{(m+1)\pi}{{N_{\t{pr}}}+2}\right)\cdot\\
\frac{1}{\sqrt{2}}\Big(&\ket{\up:m,\down:{N_{\t{pr}}}/2-m}_{x_0}\otimes\ket{\up:{N_{\t{pr}}}/2-m,\down:m}_{x_0+L}\\
&+\ket{\up:{N_{\t{pr}}}/2-m,\down:m}_{x_0}\otimes\ket{\up:m,\down:{N_{\t{pr}}}/2-m}_{x_0+L}\Big),
\end{split}
\end{equation}
where $\ket{\up:m,\down:k}_{x}$ is the state with $m+k$ atoms located at $x$, while $m$ of them is oriented up and $k$ of them is oriented down.

Note, that this state belongs to the class of decoherence-free states (as $J\ket{\psi^{N_{\t{pr}}}}=0$), so it is simultaneously the optimal solution of the initial problem. As expected, we see, that an additional $\pi^2$ comparing to the analysis based on QFI has appeared.

It is worth stressing, that the strategy saturating the bound is not unique. Especially, alternatively to using state \eqref{singrad} and evolving for time $t$, one may use the analogue of $N00N$ state (which maximizes QFI):
\begin{equation}
\begin{split}
\ket{\psi^{N_{\t{pr}}}}=&\\
\frac{1}{\sqrt{2}}\Big(&\ket{\up:{N_{\t{pr}}}/2,\down:{N_{\t{pr}}}/2}_{x_0}\otimes\ket{\up:{N_{\t{pr}}}/2,\down:{N_{\t{pr}}}/2}_{x_0+L}\\
&+\ket{\up:{N_{\t{pr}}}/2,\down:{N_{\t{pr}}}/2}_{x_0}\otimes\ket{\up:{N_{\t{pr}}}/2,\down:{N_{\t{pr}}}/2}_{x_0+L}\Big),
\end{split}
\end{equation}
and performed the sequence of measurements which different $t_i$, such that $\sum_i t_i=T$ (numerical results suggest the possibility of obtaining optimal cost in such an approach \parencite{Kaftal2014}).

\chpt{Broader discussion -- reparametrization}
\label{app:repar}
Here I recall the reasoning from \parencite{gorecki2022multiparameter}. Let $\bvar=[\var_1,...,\var_p]^T$ be the parametrization for which the cost matrix is identity $C=\openone$. Consider the extended procedure of measuring parameters separately with usage totally $N$ gates, $\t{SEP}+$. Namely:
\begin{itemize}
    \item first apply the linear transformation $\bvar'=A^{-1}\bvar.$
    \item then estimate each of $\var_i'$, using $N_i$ gates, such that $\sum_{i=1}^p N_i=N$.
\end{itemize}
From resulting $\tilde\bvar'$ reconstruct $\tilde\bvar=A\tilde\bvar'$. We want to minimze $\sum_{i=1}^p \Delta^2\tilde\var_i$ over transformation $A$, resources' distribution $\sum_{i=1}^p N_i=N$ and single sequential-adaptive protocol for each $\tilde\var_i'$. I will derive a simple lower bound for the final precision obtainable in such a procedure.

Note that if all parameters $\var_i$ are measured separately with the usage of different probes, the resulting covariance matrix will be diagonal (as indications of different estimators come from uncorrelated measurements results):
\begin{equation}
\Sigma'=\t{diag}(\Delta^2\tilde\var_1',...,\Delta^2\tilde\var_p').
\end{equation}
Therefore, the final cost, which is to be minimized may be written as follows:
\begin{equation}
\sum_{i=1}^p \Delta^2\tilde\var_i=\tr(\Sigma)=
\tr(AA^T\Sigma')=\sum_{i=1}^p [AA^T]_{ii}\Delta^2\tilde\var_i.
\end{equation}
Using \eqref{singlesimple} we may write:
\begin{equation}
\sepmmp\gtrsim \min_{A,N_j}\sum_{j=1}^p\frac{\pi^2}{N_j^2}\frac{[A^TA]_{jj}}{\lambda^2([A^T\bG]_j)}
\geq
\min_{A,N_j}\sum_{j=1}^p\frac{\pi^2}{N_j^2}\left(\min_i\frac{[A^TA]_{ii}}{\lambda^2([A^T\bG]_i)}\right)
= \frac{p^3\pi^2}{N^2}\min_{A,i} \frac{[A^{T}A]_{ii}}{\lambda^2([A^T\bG]_i)}.
\end{equation}
For any fixed $i$, RHS depends on on $i^{th}$ column of $A$, the minimization is equivalent to minimization over single vector $\boldsymbol{a}$:
\begin{equation}
\min_{A,i}\frac{[A^{T}A]_{ii}}{\lambda^2([A^T\bG]_i)}=\min_{\boldsymbol{a}}\frac{|\boldsymbol{a}|^2}{\lambda^2[\boldsymbol{a}\bG]}=\min_{\boldsymbol{a}:|\boldsymbol{a}|^2=1}\frac{1}{\lambda^2[\boldsymbol{a}\bG]}=\frac{1}{\max\limits_{\boldsymbol{a}:|\boldsymbol{a}|^2=1}\lambda^2[\boldsymbol{a}\bG]},
\end{equation}
so we get:
\begin{equation}
\label{usemm}
\sepmmp\gtrsim \frac{p^3\pi^2}{N^2}\frac{1}{\max\limits_{\boldsymbol{a}:|\boldsymbol{a}|^2=1}\lambda^2[\boldsymbol{a}\bG]}.
\end{equation}
Similarly, in the many repetitions scenario:
\begin{equation}
\label{usecr}
\sepcrp\gtrsim \frac{p^2}{kn^2}\frac{1}{\max\limits_{\boldsymbol{a}:|\boldsymbol{a}|^2=1}\lambda^2[\boldsymbol{a}\bG]}.
\end{equation}
After applying the examples discussed in \chref{ch:multihl}, it shows, that the parametrization, which was used, was indeed the one minimizing the cost obtainable in a separate strategy.

\end{appendices}


\clearpage
\phantomsection
\addtocontents{toc}{\protect\setcounter{tocdepth}{1}}
\addcontentsline{toc}{chapter}{BIBLIOGRAPHY}

\printbibliography


\end{document}